\PassOptionsToPackage{dvipsnames}{xcolor}
\documentclass[authorversion,acmsmall,nonacm,screen]{acmart}

\acmJournal{PACMPL}
\acmVolume{1}
\acmNumber{CONF} %
\acmArticle{}
\acmYear{2021}
\acmMonth{1}
\acmDOI{} %
\startPage{1}

\setcopyright{none}

\usepackage[scaled=0.8]{beramono}
\usepackage{listings,array,varwidth}
\usepackage{mathpartir}
\usepackage{bcprules}
\usepackage{mathtools}
\usepackage{makecell}
\usepackage{diagbox}
\usepackage{float}
\usepackage{varioref}
\usepackage{xfrac}
\usepackage{enumitem}
\usepackage[skip=2.5pt]{caption}
\usepackage{slashbox}
\usepackage{cleveref}
\usepackage{tikz}
\usepackage{cancel}
\usepackage{bm}
\usepackage{mathtools}
\usepackage{lipsum}
\usepackage{wrapfig}
\usepackage{soul}
\usepackage{xspace}
\usepackage{mdframed}
\usepackage[normalem]{ulem}
\usepackage{cancel}
\usepackage{scalerel}
\usepackage{multirow}
\usepackage[export]{adjustbox}
\usepackage{xifthen}
\usepackage{tikz} %
\usepackage{xcolor,colortbl}
\usepackage{tikz-cd}
\usepackage{comment}
\usepackage{amsmath}
\usepackage{longfbox}
\usepackage{arydshln}

\colorlet{lightred}{red!30} %

\usepackage{comment}

\usetikzlibrary{arrows}

\makeatletter %
\def\arcr{\@arraycr}
\makeatother

\newcommand{\showDOI}[1]{\unskip}

\newtheoremstyle{mystyle}
  {1pt}    %
  {1pt}    %
  {\itshape}  %
  {\parindent}    %
  {} %
  {.}      %
  {5pt}    %
  {\thmname{#1}\thmnumber{ #2}\thmnote{ (#3)}} %

\theoremstyle{mystyle}
\newtheorem{re-theorem}{Theorem}[section]
\newtheorem{re-lemma}[re-theorem]{Lemma}
\newtheorem{re-definition}[re-theorem]{Definition}

\newcommand{\figref}[1]{Fig.~\ref{#1}}
\newcommand{\secref}[1]{Section~\ref{#1}} %
\newcommand{\thmref}[1]{Theorem~\ref{#1}}
\newcommand{\lemref}[1]{Lemma~\ref{#1}}

\newcommand{\Specsharp}{%
  {\settoheight{\dimen0}{C}Spec\kern-.05em \resizebox{!}{\dimen0}{\raisebox{\depth}{\#}}}}
\newcommand{\Csharp}{%
  {\settoheight{\dimen0}{C}C\kern-.05em \resizebox{!}{\dimen0}{\raisebox{\depth}{\#}}}}

\newcommand{\fun}[1]{\operatorname{#1}}
\newcommand{\DOM}{\fun{dom}}

\definecolor{blue-violet}{rgb}{0.54, 0.17, 0.89}
\definecolor{ccomment}{HTML}{006400}
\definecolor{depmap}{HTML}{00007B}
\definecolor{dark-cyan}{HTML}{135579}
\definecolor{magenta}{HTML}{a8264f}

\lstdefinelanguage{PolyRT}%
{morekeywords={abstract,%
      case,catch,char,class,%
      def,else,extends,final,finally,for,%
      if,import,implicit,%
      match,module,%
      new,null,%
      object,override,%
      package,private,protected,public,%
      for,public,return,super,%
      this,throw,trait,try,type,%
      val,var,%
      with,while,%
      yield,%
      let,end,%
      in,fun,alloc,%
      assert,%
    },%
  mathescape=true,%
  sensitive,%
  keywordstyle={\color{dark-cyan}\bf\ttfamily},%
  commentstyle=\color{dark-cyan},%
  escapebegin=\color{dark-cyan},%
  morecomment=[l]//,%
  morecomment=[s]{/*}{*/},%
  morecomment=[s][\color{dark-cyan}]{@}{\ },%
  morestring=[b]",%
  morestring=[b]',%
  showstringspaces=false%
}[keywords,comments,strings]%

\newcommand{\code}[1]{\lstinline[basicstyle=\footnotesize\ttfamily] {#1}\xspace}

\newcommand{\fr}{\textsf{fr}}

\newcommand{\langb}{\ensuremath{\lambda_B}\xspace}

\newcommand{\ts}[1][]{\ensuremath{\ifthenelse{\isempty{#1}}{\,\vdash\,}{\,\vdash_{#1}\,}}}

\newcommand{\maybelang}{\ensuremath{\lambda^{\vardiamondsuit}}\xspace}
\newcommand{\maybelange}{\ensuremath{\lambda_{\EPS}^{\vardiamondsuit}}\xspace}
\newcommand{\polylang}{\ensuremath{\mathsf{F}_{<:}^{\vardiamondsuit}}\xspace}
\newcommand{\X}{\ensuremath{\mathsf{\lambda^{\vardiamondsuit}_B}}\xspace}
\newcommand{\Xsub}{\ensuremath{\mathsf{\lambda^{\vardiamondsuit}_{B<:}}}\xspace}

\newcommand{\Type}[1]{\ensuremath{\mathsf{#1}}}
\newcommand{\Var}{\Type{Var}}

\newcommand{\TRef}{\Type{Ref}}

\newcommand{\tref}{\text{\textsf{ref}}}

\newcommand{\TUnit}{\Type{Unit}}

\newcommand{\ty}[2][]{\ensuremath{\ifthenelse{\isempty{#1}}{#2}{#2^{\,#1}}}}

\newcommand{\flt}{\ensuremath{\varphi}}

\newcommand{\cx}[2][]{\ensuremath{\ifthenelse{\isempty{#1}}{#2}{#2^{\,#1}}}}
\newcommand{\csx}[3][]{\ensuremath{\ifthenelse{\isempty{#1}}{#3\mid #2}{{\color{gray!50}[}#3\mid#2{\color{gray!50}]}^{\,#1}}}}
\providecommand{\G}{G} %
\renewcommand{\G}[1][]{\cx[#1]{\Gamma}}

\newcommand{\cdsx}[4][]{\ensuremath{\ifthenelse{\isempty{#1}}{#3\mid #2\has #4}{{\color{gray!50}[}#3\mid#2{\color{gray!50}]}^{\,#1}\has #4 }}}

\newcommand{\QFresh}{\ensuremath{\vardiamondsuit}}
\newcommand{\qbot}{\ensuremath{\varnothing}}
\newcommand{\qfresh}{\ensuremath{\vardiamondsuit}}
\newcommand{\subq}{\ensuremath{\subseteq}}
\newcommand{\qlub}{\ensuremath{\cup}}

\newsavebox{\SMALLSTAR}
\savebox{\SMALLSTAR}{\(\raisebox{.25ex}{\(\qfresh\)}\)}
\newcommand{\starred}[1]{\ensuremath{\mathord{\scalerel*{\usebox{\SMALLSTAR}}{*}}#1}}
\newsavebox{\OVRLP}
\savebox{\OVRLP}{$\raisebox{.37ex}[0pt][0pt]{$\mathrlap{\hspace{.415ex}\scaleobj{.5}{\vardiamondsuit}}$}\cap$}

\newcommand{\qsat}[1]{\ensuremath{#1\mathord{*}}}

\newcommand{\reaches}{\ensuremath{\mathrel{\leadsto}}}

\newfboxstyle{mybox}{padding=2pt,margin=2pt,baseline-skip=false}
\newcommand{\BOX}[1]{\lfbox[mybox]{\strut \footnotesize \ensuremath{#1}}}

\newcommand{\FV}{\ensuremath{\operatorname{fv}}}

\newcommand{\vgap}{\vspace{7pt}}

\colorlet{mute}{teal}
\colorlet{eff}{magenta}
\newcommand{\mute}[1]{{\color{mute}#1}}

\definecolor{light-gray}{gray}{0.92}
\definecolor{dark-gray}{gray}{0.5}
\definecolor{light-yellow}{HTML}{FFFACD}

\colorlet{lightred}{red!30}

\newcommand{\FX}[1]{\ensuremath{{\color{eff}#1}}}
\newcommand{\EPS}[1][]{\ifthenelse{\isempty{#1}}{\FX{\bm{\varepsilon}}}{\FX{\bm{\varepsilon_{#1}}}}}
\newcommand{\EPSS}[1][]{\ifthenelse{\isempty{#1}}{\FX{\qsat{\bm{\varepsilon}}}}{\FX{\qsat{\bm{\varepsilon_{#1}}}}}}
\newcommand{\EPSPR}[1][]{\ifthenelse{\isempty{#1}}{\FX{\bm{\varepsilon'}}}{\FX{\bm{\varepsilon'_{#1}}}}}
\newcommand{\EPSSPR}[1][]{\ifthenelse{\isempty{#1}}{\FX{\qsat{\bm{\varepsilon'}}}}{\FX{\qsat{\bm{\varepsilon'_{#1}}}}}}
\newcommand{\PURE}{\FX{\boldsymbol{\varnothing}}}
\newcommand{\EFFSEQ}{\ensuremath{\mathbin{\FX{\boldsymbol{\rhd}}}}}

\newcommand{\DEP}[1][]{\ensuremath{\ifthenelse{\isempty{#1}}{\mute{\delta}}{\mute{\delta_{#1}}}}}

\newcommand{\HDEP}[1][]{\ensuremath{\ifthenelse{\isempty{#1}}{\mute{\mathsf{h}}}{\mute{\mathsf{h}_{#1}}}}}
\newcommand{\SDEP}[1][]{\ensuremath{\ifthenelse{\isempty{#1}}{\mute{\mathsf{s}}}{\mute{\mathsf{s}_{#1}}}}}

\newcommand{\has}{\ensuremath{\mathbin{\bullet}}}

\newcommand{\trackset}[1]{^{\texttt{\{#1\}}}}

\newcommand{\track}[0]{^{\varnothing}}
\newcommand{\trackvar}[1]{^{\texttt{#1}}}
\newcommand{\trackfresh}{^{\texttt{\ensuremath\vardiamondsuit}}}

\newcommand{\hole}[1]{\ensuremath{[\,#1\,]}}
\newcommand{\CX}[3][black]{\ensuremath{{\color{#1}#2\ifthenelse{\isempty{#3}}{}{\hole{{\color{black}#3}}}}}}

\newcommand{\bfparagraphX}[1]{\paragraph{\textbf{#1}}}
\newcommand{\bfparagraph}[1]{\vspace{-0.4\baselineskip}\paragraph{\textbf{#1}}}

\newcommand\restr[2]{{%
\left.\kern-\nulldelimiterspace %
#1 %
\vphantom{\big|} %
\right|_{#2} %
}}

\lstdefinelanguage{DOT}%
{morekeywords={val,new},%
  sensitive,%
  morecomment=[l]//,%
  morecomment=[s]{/*}{*/},%
  morestring=[b]",%
  morestring=[b]',%
  showstringspaces=false%
}[keywords,comments,strings]%

\newlength{\trulemargin}
\newlength{\trulewidth}
\newlength{\srulewidth}
\newenvironment{trules}{$\vspace{0.5em}\ba{p{\trulemargin}@{~}p{\trulewidth}@{~}p{\trulemargin}}}{\ea$}
\newenvironment{srules}{$\vspace{0.5em}\ba{p{\trulemargin}@{~}p{\srulewidth}}}{\ea$}

\newcommand{\ba}{\begin{array}}
\newcommand{\ea}{\end{array}}

\newcommand{\ei}{\end{array}}
\newcommand{\bcases}{\left\{\begin{array}{ll}}
\newcommand{\ecases}{\end{array}\right.}

\newcommand{\eg}{{\em e.g.}\xspace}
\newcommand{\ie}{{\em i.e.}\xspace}

\newcommand{\judgement}[2]{{\textsf{\textbf{#1}}} \hfill #2}

\newcommand{\synbracket}[1]{[\![{#1}]\!]}
\newcommand{\DEF}{\stackrel{{\rm def}}{=}}
\newcommand{\equiva}{\ensuremath{\mathrel{\approx_{\text{ctx}}}}}
\newcommand{\equivlog}{\ensuremath{\mathrel{\approx_{\text{log}}}}}
\newcommand{\carrow}{\Rrightarrow}

\newcommand{\fltp}{\ensuremath{\varphi'}}

\newcommand{\extends}{\ensuremath{\ensuremath{;}}}

\newcommand{\GP}[1][]{\cx[#1]{\Gamma'}}

\newcommand{\BMt}[2]{\mathcal{E}\synbracket{#2}_{{#1}}}

\newcommand{\DEPS}[3]{\ensuremath{{#1}{\rightarrow{\!\scalebox{0.7}[0.7]{{$#2$}}}}}{~#3}}
\newcommand{\LEFFS}[1]{\ensuremath{\stackrel{\!\scalebox{0.7}[0.7]{{$#1$}}}{\rightarrow}}}

\newcommand{\csxs}[2][]{\ensuremath{\ifthenelse{\isempty{#1}}{\mid #2}{{\color{gray!50}[}#2{\color{gray!50}]}^{\,#1}}}}

\newcommand{\State}[2]{\ensuremath{{#1}, \ {#2}}}

\newcommand{\xreaches}[1]{\ensuremath{\reaches_{#1}}}%

\newcommand{\tabs}[2]{\ensuremath{({#1})^{#2}}}
\newcommand{\cll}[2]{\ensuremath{\langle {#1}, \ {#2} \rangle }}
\newcommand{\cl}[3]{\ensuremath{\langle {#1}, \ \tabs{#2}{#3} \rangle }}

\newcommand{\eval}{\Downarrow}

\newcommand{\config}[3]{\ensuremath{{#1},\ \State{#2}{#3}}}
\newcommand{\pconfig}[2]{\ensuremath{{#1},\ {#2}}}

\newcommand{\WFV}[2]{\mathtt{WF}(\ensuremath{{#1}, {#2}})}

\newcommand{\UV}[1]{V\synbracket{{#1}}}
\newcommand{\Utt}[1]{E\synbracket{#1}}
\newcommand{\Ut}[2]{E\synbracket{#2}_{#1}}
\newcommand{\UTC}[3]{\ensuremath{(\State{#1}{#2}, \ {#3})}}
\newcommand{\UG}[1]{G\synbracket{{#1}}}
\newcommand{\STCPA}[3]{\ensuremath{{#1} \sqsubseteq{\!\scalebox{0.7}[0.7]{$#2$}}~ {#3}}}
\newcommand{\STCPB}[3]{\ensuremath{{#1} \equiv{\!\scalebox{0.7}[0.7]{$#2$}}~ {#3}}}
\newcommand{\STC}[2]{\ensuremath{{#1} \sqsubseteq {#2}}}

\newcommand{\WFS}[2]{{#1}:{#2}}
\newcommand{\WFRS}[3]{({#1}, {#2}):{#3}}

\newcommand{\W}{\text{W}}

\newcommand{\WFQA}{\ensuremath{\text{Inv}_{o}}}
\newcommand{\WFGA}[3]{\ensuremath{\WFQA(\Gamma,#1,#2,#3)}}
\newcommand{\WFQB}{\ensuremath{\text{Inv}_{S}}}
\newcommand{\WFGB}[3]{\ensuremath{\WFQB(#1,#2,#3)}}

\newcommand{\VE}{\ensuremath{\hat{H}}}
\newcommand{\BVT}[2]{\mathcal{V}\synbracket{{#1, #2}}}

\newcommand{\BTC}[4]{\ensuremath{(\State{#1}{#2},{#3}, {#4})}}
\newcommand{\BSTCP}[4]{\ensuremath{{#1} \equiv_{({#2},{#3})} {#4}}}
\newcommand{\BSTCPS}[3]{\ensuremath{{#1} \equiv_{{#2}} {#3}}}

\newcommand{\STFN}{\text{WR}}
\newcommand{\STF}[3]{\ensuremath{\STFN({#1}, {#2}, {#3})}}  %

\newcommand{\varslocs}[2]{\synbracket{{#2}}_{#1}}
\newcommand{\vallocs}[1]{\ensuremath{\mathnormal{L}({#1})}}
\newcommand{\qsatl}[1]{\ensuremath{#1{\mathord{*}}}}
\newcommand{\lls}[2]{{#2}^{\mathord{*}}_{{#1}}}  %
\newcommand{\llsvars}[3]{\synbracket{#3}_{\scalebox{0.6}[0.6]{$#2$}}^{\qsatl{{\scalebox{0.6}[0.6]{$#1$}}}}}
\newcommand{\frlocs}[2]{\ensuremath{\overline{\DOM(#2)}}}

\newcommand{\ccup}[1]{\ensuremath{\cup_{{#1}}}}

\newcommand{\ttrue}{\text{true}}
\newcommand{\tfalse}{\text{false}}

\newsavebox{\CFUNION}
\savebox{\CFUNION}{$\raisebox{.37ex}[0pt][0pt]{$\mathrlap{\hspace{.415ex}\scaleobj{.5}{\vardiamondsuit}}$}\cup$}

\definecolor{cyan}{rgb}{0.0, 0.72, 0.92}
\definecolor{light-pink}{HTML}{faf2f7}
\definecolor{light-gray}{gray}{0.92}

\newfboxstyle{teallf}{padding=0pt,margin-top=0pt,margin-bottom=0pt,baseline-skip=false,border-color=teal!12,background-color=teal!12}
\newfboxstyle{teallfsuper}{padding=0pt,margin-top=-2pt,margin-bottom=-2pt,baseline-skip=false,border-color=teal!12,background-color=teal!12}
\newfboxstyle{lpinklf}{padding=0pt,margin=1pt,baseline-skip=false,border-color=light-pink,background-color=light-pink}
\newfboxstyle{lgraylf}{padding=0pt,margin=1pt,baseline-skip=false,border-color=light-gray,background-color=light-gray}
\newfboxstyle{lgraylfsub}{padding=-0pt,margin-top=-2pt,margin-bottom=-2pt,baseline-skip=false,border-color=light-gray,background-color=light-gray}

\newcommand{\HLMath}[1]{\lfbox[teallf]{\strut \footnotesize \ensuremath{#1}}}
\newcommand{\HLMathSUPER}[1]{\lfbox[teallfsuper]{\strut \footnotesize \ensuremath{#1}}}
\newcommand{\HLMathP}[1]{\lfbox[lpinklf]{\strut \footnotesize\ensuremath{#1}}}
\newcommand{\HLMathG}[1]{\lfbox[lgraylf]{\strut \footnotesize\ensuremath{#1}}}
\newcommand{\HLMathGSUB}[1]{\lfbox[lgraylfsub]{\strut \footnotesize\ensuremath{#1}}}

\newcommand{\QF}{\ensuremath{q}}
\newcommand{\qminus}{\ensuremath{\!\setminus\!}}
\newcommand{\NFQ}[1]{\ensuremath{\tilde{#1}}}

\newcommand{\QSelf}{\ensuremath{\varspadesuit}}

\newcommand*\circled[1]{\protect\tikz[baseline=(char.base)]{
            \protect \node[shape=circle,draw,inner sep=1pt,color=red, font=\scriptsize] (char) {#1};}}

\newcommand*\circledx[1]{\protect\tikz[baseline=(char.base)]{
            \protect \node[shape=circle,draw,inner sep=1pt,color=red, font=\scriptsize] (char) {#1};}}

\newcommand{\NUM}[2]{\ensuremath{{\circledx{#1}}}~{#2}}

\newcommand*\rect[1]{\tikz[baseline=(char.base)]{
            \node[shape=rectangle,draw,inner sep=1pt,color=red, font=\scriptsize] (char) {\ensuremath{#1}};}}

\newcommand{\RSTEP}[1]{\ensuremath{\stackrel{{\rect{#1}}}{~<:~}}} 

\newcommand{\WFP}[3]{\ensuremath{\mathtt{WF}(#1_{#2}^{#3})}}

\newfboxstyle{oplf}{padding=1.5pt,margin=1pt,baseline-skip=false,border-style=dashed,border-color=gray}
\newcommand{\op}[1]{\lfbox[oplf]{\ensuremath{#1}}}

\newcommand{\prog}[1]{\ensuremath{\text{Prog}_{#1}}}

\newcolumntype{R}[2]{%
  >{\begin{turn}{#1}\begin{minipage}{\ensuremath{#2}}\small\raggedright\hspace{0pt}}l%
  <{\end{minipage}\end{turn}}%
}

\newcommand{\val}{\ensuremath{\mathbb{V}}}
\newcommand{\bval}{\ensuremath{\hat{\mathbb{V}}}}

\newsavebox{\SMALLF}
\savebox{\SMALLF}{\(\raisebox{0.3ex}{\(\QSelf\)}\)}
\newcommand{\fstarred}[1]{\ensuremath{\mathord{\scalerel*{\usebox{\SMALLSTAR}\usebox{\SMALLF}}{*}}#1}}

\renewcommand{\L}{\ensuremath{\mathbb{L}}}

\newcommand{\wrt}{{\em w.r.t.}\xspace}

\let\emptyset\varnothing

\newlength{\myskip}
\makeatletter%
\begin{document}

\title[Modeling Reachability Types with Logical Relations]{Modeling Reachability Types with Logical Relations: Semantic Type Soundness, Termination, Effect Safety, and Equational Theory}\thanks{This article is an extended version of \citet{bao2025logrel}.}

\author{Yuyan Bao}
\orcid{0000-0002-3832-3134}             %
\affiliation{
  \institution{Augusta University}            %
  \city{Augusta}
  \country{USA}                    %
}
\email{yubao@augusta.edu}          %

\author{Songlin Jia}
\orcid{0009-0008-2526-0438}             %
\affiliation{
  \institution{Purdue University}            %
  \city{West Lafayette}
  \country{USA}                    %
}
\email{jia137@purdue.edu}          %

\author{Guannan Wei}
\orcid{0000-0002-3150-2033}             %
\affiliation{
  \institution{Tufts University}            %
  \city{Medford and Somerville}
  \country{USA}                    %
}
\email{guannan.wei@tufts.edu}          %
\authornote{Work completed while at Purdue University.}

\author{Oliver Bra\v{c}evac}
\orcid{0000-0003-3569-4869}             %
\affiliation{
  \institution{EPFL}            %
 \city{Lausanne}
  \country{Switzerland}                    %
}
\email{oliver.bracevac@epfl.ch}          %
\authornote{Work completed while at Purdue University.}

\author{Tiark Rompf}
\orcid{0000-0002-2068-3238}             %
\affiliation{
  \institution{Purdue University}            %
  \city{West Lafayette}
  \country{USA}                    %
}
\email{tiark@purdue.edu}          %

\lstMakeShortInline[keywordstyle=,%
  flexiblecolumns=false,%
  language=Scala,
  basewidth={0.56em, 0.52em},%
  mathescape=false,%
  basicstyle=\ttfamily]@

\begin{abstract}
Reachability types are a recent proposal to bring 
Rust-style reasoning about memory properties to 
higher-level languages, with a focus on 
higher-order functions, parametric 
types, and shared mutable state -- features that
are only partially supported by current
techniques as employed in Rust.
While prior work has established key type soundness
results for reachability types using the usual
syntactic techniques of progress and preservation,
stronger metatheoretic
properties have so far been unexplored.
This paper presents an alternative
semantic model of reachability types using
logical relations, providing a framework
in which we study key properties of interest:
(1) semantic type soundness, including
of not syntactically well-typed code fragments,
(2) termination, especially in the presence 
of higher-order mutable references, 
(3) effect safety, especially the absence
of observable mutation, and, finally,
(4) program
equivalence, especially reordering of 
non-interfering expressions for 
parallelization or compiler optimization.
 \end{abstract}

\maketitle

\vspace{-1ex}
\section{Introduction}

Reachability types are a recent proposal to bring 
Rust-style reasoning about memory properties to 
higher-level languages. Instead of Rust's rigid
``shared XOR mutable'' model, reachability types
emphasize \emph{tracking} of resources such as
mutable references using type qualifiers, which
enables richer programming patterns, \eg, based
on sharing captured mutable data, in languages
that make pervasive use of higher order
functions and type-level abstraction.

In prior work, key type soundness results for reachability
types have been established~\cite{DBLP:journals/pacmpl/BaoWBJHR21,wei2023polymorphic} using the usual
syntactic techniques of progress and preservation 
\cite{DBLP:journals/iandc/WrightF94} with respect
to (\wrt) a small-step, substitution-based, operational semantics.
These results assure us that reachability
information is preserved through evaluation, serving, \eg, 
as an informal justification that expressions with 
non-overlapping reachability can be safely
evaluated in a different order or in parallel. 
Likewise, prior work~\cite{DBLP:journals/pacmpl/BaoWBJHR21,wei2023polymorphic} has proposed that reachability 
types could be used as the basis for effect systems
to track (the absence of) store modifications
in a fine-grained way, \eg, to parallelize overlapping
reads as long as there are no conflicting writes.
However, no end-to-end formal proof of such properties
exists so far, and in general, stronger metatheoretic 
properties beyond basic type safety have  
been left unexplored.

We address this gap by presenting
an alternative semantic model of reachability types
using the technique of logical relations~\cite{plotkin_lambda_1973,DBLP:journals/jsyml/Tait67,logical-approach,DBLP:conf/popl/AhmedDR09,DBLP:conf/ppdp/BentonKBH07},
providing a framework in which to study key properties
of interest such as semantic type soundness, termination,
effect safety, and contextual equivalence.

\vspace{-1ex}
\paragraph*{\textbf{Bottom-up Design of Logical Relations}}

We construct logical relations (LR) for a family of reachability 
type systems with increasing sets of features. Our starting point is
the standard LR for the simply-typed $\lambda$-calculus with first-order
references, which we augment with internal invariants to track the
flow of store locations, without restricting expressiveness 
(\secref{sec:motiv}). From these added
invariants, we ``rediscover'' reachability types by propagating
certain choices to qualifiers in the user-facing type system
(\secref{sec:unary_base}). Then, we scale up the set of features, 
from first-order mutable references to higher-order mutable references (\secref{sec:unary_nested}), 
from tracking all mentioned references uniformly to distinguishing 
reads and writes (\secref{sec:effects}), and from a unary LR
to a binary LR (\secref{sec:direct-lr}) to support equational reasoning (\secref{sec:equiv}).

\vspace{-1ex}
\paragraph*{\textbf{Semantic Type Soundness}}

The most immediate result from the LR model is semantic
type soundness. As~\citet{logical-approach} have argued, %
semantic type soundness is a stronger notion than syntactic
type soundness, so our semantic type soundness result has significance
beyond the existing syntactic results
\cite{DBLP:journals/pacmpl/BaoWBJHR21,wei2023polymorphic}. 
First, semantic type soundness does not require terms to be 
\emph{syntactically} well-typed, just to \emph{behave} in a well-typed
way, \eg, disregarding ill-typed branches that are never taken. 
Thus, it provides a foundation for studying potentially unsafe features 
and interactions with the outside world.
Second, unlike prior results~\cite{DBLP:journals/pacmpl/BaoWBJHR21,wei2023polymorphic}
that were established \wrt a small-step 
operational semantics based on substitution,
we adopt a big-step operational semantics based on
closures and environments. Thus, our results 
map more closely to realistic language implementations.

\vspace{-1ex}
\paragraph*{\textbf{Termination}}

The LR model actually proves a stronger property than
the absence of type errors, namely that evaluation of 
all well-typed terms \emph{must terminate} with a 
well-typed value.
This termination result is interesting in addition to
soundness. First, the reachability type systems in
\secref{sec:unary_nested} and beyond contain several
non-trivial features such as nested mutable references
and self references in function qualifiers, whose termination
properties are non-obvious. Second, our LR model
is defined by simple structural recursion on types,
with a store typing invariant that precludes cycles 
in the store.
This is in contrast to other recent work, which
proposed a form of transfinite step indexing
to prove termination for a feature-rich language
with higher-order state~\cite{DBLP:journals/pacmpl/SpiesKD21}. 
In comparison, our model achieves a similar result
but is entirely elementary and requires 
neither step-indexes, nor advanced set-theoretic 
concepts, nor classical reasoning.
More broadly, the termination result is both reassuring, 
in the sense that reachability types can support flexible 
mutation in places where termination is required
(\eg, dependent type systems) but also points out
a non-obvious limitation in prior 
work~\cite{DBLP:journals/pacmpl/BaoWBJHR21,wei2023polymorphic},
namely the inability to construct cyclic heap structures
such as doubly-linked lists. 
We leave it to future 
work to study this limitation and possible remedies
further. Recently, promising steps in this direction 
were made by \citet{deng2025completecyclereachabilitytypes}.

\vspace{-1ex}
\paragraph*{\textbf{Effect Safety}}

The built-in store invariants of the LR model guarantee
that an expression can only observe or modify reachable
store locations. Thus, reachability provides a natural
upper bound on potential side effects. Specifically, 
the latent effect of a function is limited to the
store locations reachable from the function and any
locations reachable from the provided argument.
In \secref{sec:effects}, we extend our LR model with
an effect system that tracks \emph{write effects}
on sets of variables in addition to reachability. 
With this added
precision of distinguishing pure observations from 
effectful write operations, we are able to prove 
an effect safety theorem, stating that all store 
changes are covered by write effects, \ie, pure
expressions never cause any observable effects.
While similar extensions have been proposed in
prior work~\cite{DBLP:journals/pacmpl/BaoWBJHR21,oopsla23}, 
no end-to-end proofs of effect safety have been provided so far. 
And in fact the desired properties are non-trivial to capture 
using progress and preservation, since effects are ``performed''
during reduction and not preserved in the types
of values.

\vspace{-1ex}
\paragraph*{\textbf{Program Equivalence}}

We extend our unary LR to a binary LR model in \secref{sec:direct-lr}
to support equational reasoning, and we prove key equivalences in
\secref{sec:equiv}. 
This equational reasoning framework is significant as
it provides a foundation for parallelization and
for a variety of effect-based compiler
optimizations in the style of~\citet{DBLP:conf/ppdp/BentonKBH07}
or~\citet{DBLP:journals/iandc/BirkedalJST16}.
First, we prove a reordering theorem for non-interfering
expressions which justifies parallel execution.
Second, we prove $\beta$-equivalence for functions 
applied to pure and non-interfering arguments.
This result shows that the evaluation semantics
used here is consistent with substitution (specifically,
substitution with values), despite
not using substitution internally. It also justifies
function inlining as a compiler optimization, \eg,
in the context of \citet{oopsla23}'s compiler IR
based on reachability types for dependency analysis.
We formulate equivalences for the
calculus with and without effect qualifiers 
(\secref{sec:effects}) and show that write effects 
enable a more precise notion of non-interference
than pure reachability.

\bfparagraph{Contributions} In summary, this paper makes the following contributions:
\begin{itemize}[leftmargin=2em]
    
    \item We review the standard LR model for simply-typed $\lambda$-calculus with first-order mutable references, and discuss how adding certain store invariants leads to reachability types (\secref{sec:motiv}).
    
    \item We present a unary logical relation that establishes semantic 
    type soundness as well as termination for types with reachability qualifiers
    (\secref{sec:unary_base}).
    
    \item We extend the calculus to support higher-order mutable
    references and show that the system remains sound and terminating 
    (\secref{sec:unary_nested}).
    
    \item We extend the calculus with an effect system including an effect
    safety invariant, guaranteeing that pure expressions have no observable side effects
    (\secref{sec:effects}).
    
    \item We extend the model to a binary logical relation to support relational reasoning \wrt the observational equivalence of two programs
    (\secref{sec:direct-lr}).
    
    \item We prove a reordering theorem for non-interfering expressions and $\beta$-equivalence for functions applied to pure and non-interfering arguments
    (\secref{sec:equiv}).
\end{itemize}

The formal results in this paper have all been mechanized in Rocq.
The developments are available online at \url{https://github.com/tiarkrompf/reachability}.

\section{Motivation} \label{sec:motiv}

\subsection{Brief Review of Reachability Types}\label{sec:overview}
Reachability types~\cite{DBLP:journals/pacmpl/BaoWBJHR21,wei2023polymorphic}
are a recent proposal to bring
Rust-style reasoning about memory properties to
higher-level languages, specifically reasoning about
sharing and the absence of sharing: separation.
Reachability types are based on four key ideas:
\begin{enumerate}[label=\textbf{(\arabic*)}, leftmargin=*, wide]
  \item \textbf{Tracking reachable variables in type qualifiers:} Types are of the form $\ty[p]{T}$, 
  where $p$ is a \emph{reachability qualifier}, a set of variables that may additionally include the \emph{freshness marker} $\QFresh$.
  \begin{lstlisting}
  val x = new Ref(0)       // : Ref[Int]$\trackvar{x}$ in context [ x: (Ref Int)$\trackfresh$ ]
  val y = x                // : Ref[Int]$\trackvar{y}$ in context [ y: (Ref Int)$\trackvar{x}$, x: (Ref Int)$\trackfresh$ ]
\end{lstlisting}
The value of expression @new Ref(0)@ is \emph{fresh}: it must be tracked, but is not bound to a variable. 
The typing context $x: \ty[\QFresh]{(\TRef ~\Type{Int})}$ means @x@ reaches a fresh value.
The type system keeps reachability sets minimal in type assignment, \ie, in the example, we assign the one-step reachability 
set @y@ to the type of variable @y@. The complete reachability set @y,x@~\footnote{For readability, we often drop the set notation for qualifiers
and write them down as comma-separated lists of atoms.} can be retrieved by computing transitive closures \wrt the typing context~\cite{wei2023polymorphic}.

Functions can reach all tracked values they close over. 
Thus, the reachability qualifier of a function includes the set of its free variables, consistent with the interpretation as a closure record:
 \begin{lstlisting}
  def incr() = { x += 1 }  // : (() => Int)$\trackvar{x}$ in context [ y: (Ref Int)$\trackvar{x}$, x: (Ref Int)$\trackfresh$ ]
\end{lstlisting}

\item \textbf{Contextual freshness for function arguments:} Reachability types can implement Rust-style borrowing that grants unique access to 
  a store location while temporarily disabling other access paths. %
  The following code defines a combinator @borrow@:~\footnote{Reachability types is a dependent type system, allowing a function's result type to mention its argument.} %
\begin{lstlisting}
  // : ((z: (Ref Int)$\trackfresh$) => ((Int => Int)$\trackfresh$ => Unit)$\trackvar{z}$)$\trackvar{\qbot}$
  def borrow(z: (Ref Int)$\trackfresh$)(f: (Int => Int)$\trackfresh$) = { z := f(!z) }  
\end{lstlisting}
A fresh argument type means that the locations reachable from the argument and the function must be mutually disjoint, so that the function cannot \emph{observe} any potential overlap with variables in the environment. This extends naturally to curried arguments such as @z@, @f@ above:
\begin{lstlisting}
  val d = new Ref(0)
  borrow(x)(y => y + !d)     // ok: argument type (Int => Int)$\trackvar{d}$ is fresh relative to borrow(x)
  borrow(x)(y => incr())     // type error: argument (Int => Int)$\trackvar{incr}$ overlaps with borrow(x) via x
\end{lstlisting}
The first function call typechecks as the arguments are separate and fresh relative to the partially applied function @borrow(x)@, %
but the second does not, as both the passed closure and the partially
applied function reach @x@.

  \item \textbf{Self references to approximate reachable data for escaping values:} 
  The type system uses \emph{self-references} in function types to track reachable values that escape from their defining scope. Our self references are similar to \texttt{this} pointers in OO languages, as formalized, \eg, in the DOT family of type systems~\cite{DBLP:conf/birthday/AminGORS16,DBLP:conf/oopsla/RompfA16}.
  \begin{lstlisting}
  def cell(init: Int) = {  // : Int => ($\mu$f.() => (Ref Int)$\trackvar{f}$)$\trackfresh$
    val c = new Ref(init)    // : (Ref Int)$\trackvar{c}$
    () => c                  // : (() => (Ref Int)$\trackvar{c}$)$\trackvar{c}$
  }                        // : ($\mu$f.() => (Ref Int)$\trackvar{f}$)$\trackfresh$    introduce self-ref to avoid local variable $c$
  val z = cell(0)          // : (() => (Ref Int)$\trackvar{z}$)$\trackvar{z}$    instantiate self-ref $f$ with bound name $z$
\end{lstlisting}
The above function @cell@ returns a closure that captures and leaks the internal reference @c@.
Once the closure escapes its defining scope, the variable @c@ is no longer in scope, and thus cannot be mentioned in its type, 
as doing so would make the type meaningless to clients.
\footnote{The return type can only mention
what is in scope, and needs to avoid any variable names going out of scope. 
This is called the ``avoidance problem'' in the context of ML modules and 
other languages with restricted forms of dependent types (\eg, DOT~\cite{DBLP:conf/birthday/AminGORS16,DBLP:conf/oopsla/RompfA16}),
where substitution is not always permitted.} 
Thus, we use self-references in its type instead.
The escaping closure is a fresh value 
before being bound to @z@. 
Its return type uses a self-reference introduced by the $\mu$-notation (similar to DOT) to express that the returned value overlaps with the closure itself.
Like @this@ pointers in OO languages, self-references provide
a way to tie hidden internal data to an externally visible identity,
allowing us to track it throughout the interaction with the caller.

   \item \textbf{Tracking effects:} 
   Reachability qualifiers provide a capability to reason about
   effects. Specifically, a function's qualifier is as an upper bound on what the function can observe or modify from the context. 
   Effects can be tracked even more precisely if the type system is extended with a reachability-sensitive effect system~\cite{DBLP:journals/pacmpl/BaoWBJHR21,oopsla23,oopsla23supplement}. This enables, \eg, distinguishing 
   reads and writes on \emph{the same} store reference. 
   In this paper, we focus on tracking store modifications as \emph{write effects}~(\secref{sec:effects}). This allows us to distinguish pure from effectful code.
   This distinction also provides additional precision for program equivalence~(\secref{sec:equiv}).
   Consider the following example, where variables @x@ and @y@ are aliased.
   \begin{lstlisting}
    val x = new Ref(5); val y = x; val z = new Ref(1)
    y := y + !z            // :  @wr(y) in context [z: Ref[Int]$\trackfresh$, y: Ref[Int]$\trackvar{x}$, x: Ref[Int]$\trackfresh$]
   \end{lstlisting}
  Incrementing the content of @y@ by the value referenced by @z@ induces a write effect (@≠wr@) on variable @y@,
  and transitively on @x@ too, but not on @z@.  
  Thus, such an effect system can be used to show that values (\eg, the referent of @z@) are preserved over a potential effectful computation~\cite{oopsla23supplement}. 
\end{enumerate}

\bfparagraph{Towards a Semantic Model of Reachability Types} 
Having seen key examples of reachability types, we will now recap the 
structure of semantic type soundness via the standard $\lambda$-calculus 
LR definitions, and then proceed to ``rediscover'' reachability types from semantic 
store properties.

\subsection{Semantic Soundness of \langb{}}

After introducing the standard LR model for simply-typed $\lambda$-calculus (STLC), 
we will add high-level semantic store invariants in \secref{sec:motiv_limitation}
that track
the flow of store-allocated values without restricting the set
of typeable terms. Setting the scene for \secref{sec:unary_base}, we show how 
these semantic invariants
naturally lead to the idea of exposing the corresponding
information through user-facing type qualifiers, 
so that the flow of store-allocated
values can be tracked and controlled by the programmer.
We thus rediscover reachability types from first principles, 
and justify them in a semantic rather than syntactic way.

\bfparagraph{The \langb{} Language}
\begin{figure}[t]\footnotesize
    \begin{mdframed}
        \judgement{Syntax}{\BOX{\langb}}\vspace{-8pt}
        \begin{minipage}[t]{\linewidth}
        \[\begin{array}{l@{\quad}l@{\quad}l@{\quad}l}
                S,T,U,V & ::= & \ty{Bool} \mid  T \to U \mid \TRef~ T                                                  & \text{Types}               \\
                c       & :=  & \ttrue \mid \tfalse                                                           & \text{Constants}       \\
                t       & ::= & c \mid x \mid \lambda x. t \mid t_1~t_2 \mid \tref~t \mid\ !~t  \mid t_1 \coloneqq t_2 \mid t_1;t_2  & \text{Terms}               \\
                v       & ::= & c \mid \ell \mid \cll{H}{\lambda x. t}                      & \text{Values}              \\
                \Gamma  & ::= & \varnothing\mid \Gamma, \ x : T                                                       & \text{Typing Environment} \\
                H       & ::= & \varnothing \mid H \extends (x, v)                                                           & \text{Value Environment}   \\
                \sigma  & :=  & \varnothing \mid \sigma \extends (\ell, v)                                                             & \text{Store} \\
            \end{array}\]\\
        \end{minipage}%
        \vspace{-16pt}
        \\ 
        \judgement{Typing Rules}{\BOX{\G \ts t: T}} \vspace{-2pt} 
        \begin{minipage}[t]{.20\linewidth}\vspace{0pt}
            \infrule[t-cst]{
            }{
                \G \ts c : \ty{Bool}
            }          
        \end{minipage}%
        \begin{minipage}[t]{.03\linewidth}
            \hspace{1pt}%
        \end{minipage}%
        \begin{minipage}[t]{.20\linewidth}\vspace{0pt}
            \infrule[t-var]{
                \G(x) = T
            }{
                \G \ts x : T
            }
        \end{minipage}%
        \begin{minipage}[t]{.06\linewidth}
            \hspace{2pt}%
        \end{minipage}%
        \begin{minipage}[t]{.25\linewidth}\vspace{0pt}
            \infrule[t-ref]{
                \G \ts t : \ty{Bool}
            }{
                \G \ts \tref~t : \TRef~\ty{Bool}
            }
        \end{minipage}%
        \begin{minipage}[t]{.03\linewidth}
            \hspace{1pt}%
        \end{minipage}%
        \begin{minipage}[t]{.23\linewidth}\vspace{0pt}
            \infrule[t-$!$]{ %
                \G \ts t : \TRef~\ty{Bool}
            }{
                \G \ts !t : \ty{Bool}
            }
          \end{minipage}%
        \\ \\
        \begin{minipage}[t]{.23\linewidth}\vspace{0pt}    
                 \infrule[t-$:=$]{
                \G \ts t_1 : \TRef~\ty{Bool} \\
                \G \ts t_2 : \ty{Bool}
            }{
                \G \ts t_1 \coloneqq t_2 : \ty{Bool}
            }
        \end{minipage}%
        \begin{minipage}[t]{.24\linewidth}\vspace{0pt}
         \infrule[t-seq]{
        \G \ts t_1 : \ty{Bool}\\
        \G \ts  t_2 : \ty{Bool} \\
      }{
        \G \ts t_1;t_2 : \ty{Bool} \\
      }
        \end{minipage}%
        \begin{minipage}[t]{.29\linewidth}\vspace{0pt}
            \infrule[t-abs]{
                \\
                \G, \, x: T \ts t : U
            }{
                \G \ts \lambda x.t : T \to U
            }
        \end{minipage}%
        \begin{minipage}[t]{.24\linewidth}\vspace{0pt}
            \infrule[t-app]{
                \G \ts t_1:  T \to U \\ \G \ts t_2: T
            }{
                \G \ts t_1~t_2 : U
            }
        \end{minipage} \\ 
        \judgement{Big-Step Semantics}{\BOX{\config{t}{H}{\sigma} \eval \pconfig{v}{\sigma'}}} \vspace{-5pt} \\
        \begin{minipage}[t]{.25\linewidth}\vspace{0pt}
            \infrule[e-cst]{ \\
            }{
                \config{c}{H}{\sigma} \, \eval \, \pconfig{c}{\sigma}
            }
        \end{minipage}%
        \begin{minipage}[t]{.03\linewidth}
            \hspace{1pt}%
        \end{minipage}%
        \begin{minipage}[t]{.36\linewidth}\vspace{0pt}    
            \infrule[e-\(!\)]{
                \config{t}{H}{\sigma} \, \eval \, \pconfig{\ell}{\sigma'}  \quad \sigma'(\ell) = v
            }{
                \config{!t}{H}{\sigma} \, \eval \, \pconfig{v}{\sigma'}
            }
        \end{minipage}
        \begin{minipage}[t]{.03\linewidth}
            \hspace{1pt}%
        \end{minipage}%
        \begin{minipage} [t]{.35\linewidth}\vspace{0pt}       
            \infrule[e-abs]{ \\
            }{
                \config{\lambda x.t}{H}{\sigma} \, \eval \, \pconfig{\cll{H}{\lambda x.t}}{\sigma}
            }
        \end{minipage}%
        \\\\ 
        \begin{minipage}[t]{.30\linewidth}\vspace{0pt}
            \infrule[e-ref]{
                \config{t}{H}{\sigma} \, \eval \, \pconfig{v}{\sigma'}  \\  \ell \not\in \DOM(\sigma')
            }{
                \config{\tref~t}{H}{\sigma}\, \eval \, \pconfig{\ell}{\sigma'\extends(\ell, v)}
            }
        \end{minipage}%
        \begin{minipage}[t]{.32\linewidth}\vspace{0pt} 
            \infrule[e-seq]
            {\config{t_1}{H}{\sigma} \, \eval \, \pconfig{b_1}{\sigma'} \\
                \config{t_2}{H}{\sigma'} \, \eval \, \pconfig{b_2}{\sigma''}}
            {\config{t_1;t_2}{H}{\sigma} \, \eval \, \pconfig{b_1 \land b_2}{\sigma''}}
        \end{minipage}%
        \hfill
        \begin{minipage}[t]{.36\linewidth}\vspace{0pt}    
            \infrule[e-\(:=\)]{
            \config{t_1}{H}{\sigma} \, \eval \, \pconfig{\ell}{\sigma'} \\  \config{t_2}{H}{\sigma'} \eval \pconfig{v}{\sigma''}
            }{
            \config{t_1 := t_2}{H}{\sigma} \, \eval \, \pconfig{\ttrue}{\sigma''[\ell \mapsto v]}
            }
            
        \end{minipage} \\\\
        \begin{minipage}[t]{.23\linewidth}\vspace{0pt}    
            \infrule[e-var]{
                H(x)  =  v
            }{
                \config{x}{H}{\sigma} \, \eval \, \pconfig{v}{\sigma}
            }
            
        \end{minipage}
        \begin{minipage}[t]{.03\linewidth}
            \hspace{1pt}%
        \end{minipage}%
        \begin{minipage}[t]{.75\linewidth}\vspace{0pt}
            \infrule[e-app]{
                \config{t_1}{H}{\sigma} \, \eval \, \pconfig{\cll{H'}{\lambda x.t}}{\sigma'} \quad
                \config{t_2}{H}{\sigma'} \, \eval \, \pconfig{v}{\sigma''} \quad
                \config{t}{H' \extends(x, v)}{\sigma''} \, \eval \, \pconfig{v}{\sigma'''}
            }{
                \config{t_1t_2}{H}{\sigma} \, \eval \, \pconfig{v}{\sigma'''}
            }
        \end{minipage}
    \end{mdframed}
    \caption{The syntax, static and dynamic semantics of \langb{}. We use type \ty{Bool} as a representative type for primitive types, \eg, TUnit and TInt.
    Thus, assignments (rule \textsc{T-:=}) return type \ty{Bool}.}\label{fig:lambda}
    \vspace{-4ex}
\end{figure} \figref{fig:lambda} shows the syntax and typing rules for \langb{}, STLC with first-order store-allocated mutable references (restricted to hold Boolean values).
The definitions are mostly standard, and can be found, e.g., in the popular TAPL textbook~\cite{10.5555/1076265}.
We use a big-step semantics with a value environment $H$ and a store $\sigma$,
where $H$ is a partial function that maps variables to values, and
$\sigma$ is a partial function that maps locations to values.
We write $\config{t}{H}{\sigma} \eval \pconfig{v}{\sigma'}$  to denote that term $t$ is evaluated to value $v$, resulting in a store transition from $\sigma$ to $\sigma'$.
The value environment is immutable, as is standard. 
Evaluating each sub-term in a sequence $t_1;t_2$ term yields a Boolean value, and the overall result conjuncts the sub-terms' resulting values.~\footnote{
To verify that reordering two terms (rules in \secref{sec:reorder}) yields equivalent results, 
we have to choose a commutative operation to combine the results. 
The most natural would be to combine the two results into a pair, 
but this would require either including pairs in the base language, 
or using a $\lambda$-encoding. 
Instead, we chose to allow each sub-term to yield a Boolean value, as it captures the essential property with less complexity than those alternatives approaches.
}
The definition of dynamic semantics shown in \figref{fig:lambda} is standard. 

Following~\cite{Jeremy,DBLP:conf/popl/ErnstOC06,DBLP:conf/popl/AminR17,DBLP:conf/ecoop/WangR17}, in the proof of semantic type soundness, we extend the big-step semantics $\eval$ to a total evaluation function by adding a numeric fuel value and explicit \text{timeout} and \text{error} results. Note, however, that while the operational semantics is step-indexed in this way, the logical relations we define are not, and could be
defined just as well with a partial (big-step or small-step) evaluation semantics.

\bfparagraph{Modeling Stores}
Considering the restriction to first-order references here, the store layouts are always ``flat'', \ie, free of cycles.
To specify well-typed stores, we use a store typing $\Sigma$ that maps from locations to semantic types.
A semantic type is a set of values ($\val$).
Observing the semantics, a store always grows monotonically during the course of evaluation, which is modeled by the store typing extension relation, written $\STC{\Sigma}{\Sigma'}$. 
Store typing extension satisfies reflexivity and transitivity.
Given a store $\sigma$, we write $\WFS{\sigma}{\Sigma}$ to mean that $\sigma$ is well-formed \wrt the store typing $\Sigma$. 
\figref{fig:lambda_lr} (top, ignoring the highlighted formulas) summarizes the definitions of these notations.

\bfparagraph{Value Interpretation of Types and Term Interpretation}
\begin{figure*}[t]\footnotesize
    \begin{mdframed}[innertopmargin=0pt, innerbottommargin=3pt, leftmargin=2pt, rightmargin=2pt]
        \judgement{\HLMathG{\textsf{\textbf{Interpretation of Value Reachability}}}\phantom{lifiers}}{\BOX{\langb}}\vspace{-1.5ex}\\
    $
      \begin{array}{l@{\ \,}c@{\ \,}l@{\qquad\qquad\qquad\qquad\ \ }r}

        \HLMathG{\vallocs{c} = \qbot}    \qquad
        \HLMathG{\vallocs{\ell} = \{\ell\}}   \qquad
        \HLMathG{\vallocs{\cll{H}{\lambda x.t}} = \vallocs{\FV(\lambda x.t)}_H} \qquad 
        \HLMathG{\vallocs{q}_{H} = \bigcup_{x \in q} L(H(x))}
      \end{array}
    $\\
        \judgement{Semantic Interpretation of Types, Terms, and Typing Contexts}\vspace{-1ex}
        \begin{footnotesize}
        \begin{mathpar}
                \begin{array}{@{}r@{\hspace{1ex}}l@{\hspace{1ex}} l}
                    \Sigma  & ::= & \varnothing\mid \Sigma, \ell : \val \\
                    \STC{\Sigma}{\Sigma'} &  \DEF &  \forall \,\ell. \, \ell \in \DOM(\Sigma) \Rightarrow  (\ell \in \DOM(\Sigma') \land \Sigma(\ell) = \Sigma'(\ell)) \\
                    \HLMathG{\STCPA{\Sigma}{\L}{\Sigma'}} & \DEF & \HLMathG{\L \subq \DOM(\Sigma) \land \L \subq \DOM(\Sigma') \, \land \, (\forall\, \ell \in \L.\, \Sigma(\ell) = \Sigma'(\ell'))} \\
                   \sigma :\Sigma & \DEF & \DOM(\sigma) = \DOM(\Sigma) \, \land  \, (\forall \ell \in \DOM(\sigma). \, \sigma(\ell) \in \Sigma(\ell)) \\
                    \HLMathG{\DEPS{\sigma}{L}{\sigma'}} & \DEF & \HLMathG{\forall \ell \in \DOM(\sigma). \ \ell \not\in L \Rightarrow \sigma(\ell) = \sigma'(\ell) }                                                                                                                                                                                                 \\
                    \\
                   
                    \UV{\Type{Bool}}                    & =    & \{ \UTC{H}{\Sigma}{c} \}                                                                                                                                                                                                                                   \\

                    \UV{\TRef~ \Type{Bool}}             & =    & \{ \UTC{H}{\Sigma}{\ell} \mid  \ell \in \DOM(\Sigma) \, \land  \,(\forall\, \Sigma'.\ (v \in \Sigma(\ell) \iff (H, \Sigma', v) \in \UV{\Type{Bool}})) \}                    \\
                    \\

                    \UV{T \to U}          & =    & \{ \UTC{H}{\Sigma}{\cll{H'}{\lambda x. t}} \mid   \forall \, v, \, \sigma', \, \Sigma'. \, \circledx{1}\STCPA{\Sigma}{\HLMathGSUB{\vallocs{\cll{H'}{\lambda x. t}}}}{\Sigma'} \, \land \, \WFS{\sigma'}{\Sigma'} \, \land \\
                                                        &      & \quad {\circledx{2}}{\HLMathG{\vallocs{v} \subq \vallocs{\cll{H}{\lambda x.t}} \cup  \overline{\vallocs{\cll{H}{\lambda x.t}}}}} \, \land \, \UTC{H}{\Sigma'}{v}\in \UV{T}\, \Rightarrow  \\
                                                        &      & \quad\quad \quad \exists \, \sigma'', \Sigma'', v'. \, \config{t}{H'\extends (x;v)}{\sigma'} \, \eval \pconfig{v'}{\sigma''} \, \land \, \WFS{\sigma''}{\Sigma''} \, \land \, \STC{\Sigma'}{\Sigma''} \, \, \land \, \UTC{H}{\Sigma''}{v'} \in \UV{U} \, \land \\
                                                        &      & \quad \quad\quad \quad \circledx{3}\HLMathG{\vallocs{v'} \subq \vallocs{\cll{H}{\lambda x.t}} \cup \vallocs{v}  \cup \, \frlocs{\Sigma''}{\Sigma'}} \, \land \, \circledx{4}\HLMathG{\DEPS{\sigma'}{(\vallocs{\cll{H}{\lambda x.t}} \cup \vallocs{v})}{\sigma''}} \}           \\
                    \\

                    \Utt{T}                      & =    & \{ \UTC{H}{\Sigma}{t} \mid \, \forall \, \sigma. \, \WFS{\sigma}{\Sigma} \, \land \, \exists \,\sigma', \Sigma', v'.\,  \config{t}{H}{\sigma} \eval \pconfig{v'}{\sigma'} \,  \land \, \WFS{\sigma'}{\Sigma'} \, \land \, \STC{\Sigma}{\Sigma'} \, \land \\
                                                        &      & \quad  \UTC{H}{\Sigma'}{v'} \in \UV{T} \, \land \, \HLMathG{\vallocs{v'} \subq \vallocs{\FV(t)}_H  \, \cup \, \frlocs{\Sigma'}{\Sigma}} \, \land \, \HLMathG{\DEPS{\sigma'}{\vallocs{\FV(t)}_H}{\sigma''}}  \}
                    \\
                    \\                                                                                  
                    \UG{\G[\HLMathGSUB{\flt}]}                             & =    & \{ (H, \Sigma) \mid \DOM(H) = \DOM(\G) \, \land \, \HLMathG{\flt \subq \DOM(\G)} \, \land \, (\forall \, x, T. \, \G(x) = T \, \land \, \HLMathG{x \in \flt} \Rightarrow  \\
                                                                            & & \quad  \UTC{H}{\Sigma}{H(x)} \in \UV{T})\}
                \end{array}\vspace{-1.5ex}%
        \end{mathpar}%
        \textsf{\textbf{Semantic Typing Judgment}} \qquad  $\G \models t: T \quad \DEF \quad \forall\, (H, \Sigma) \in \UG{\G[\HLMathG{\FV(t)}]}. \, \UTC{H}{\Sigma}{t} \in \Utt{T}$
    \end{footnotesize}
    \end{mdframed}%
    \caption{The value interpretation of types, terms, and typing context interpretation for \langb{}.
    Formula $\circledx{2}$ is a tautology here, but we will see that our reachability qualifiers can specify 
    how argument values nontrivially interact with function values in \secref{sec:unary_base}.
    The highlighted formulas demonstrate that the knowledge about locations reachable from given values allows the proof to track properties of an effectful operation, \ie, properties of locations unreachable during an operation will be preserved.
    }
    \label{fig:lambda_lr}
    \vspace{-4ex}
\end{figure*}

 \figref{fig:lambda_lr} (middle) defines the value interpretation of types, as well as the term and typing context interpretations, ignoring the highlighted formulas for now.
The value interpretation of type ${T}$, written as $\UV{T}$,
is a set of triples of form $\UTC{H}{\Sigma}{v}$,
where $H$ is a value environment, $\Sigma$ is a store typing, and $v$ is a value.

The values of type \Type{Bool} are $\ttrue$ and $\tfalse$;
the values of reference type $\TRef ~\Type{Bool}$ are store locations
that store a value of type $\Type{Bool}$.
The set of values of function type $T \to U$ \wrt store typing $\Sigma$ is defined based on its computational behavior.
It says that for all future stores $\WFS{\sigma'}{\Sigma'}$, where $\STC{\Sigma}{\Sigma'}$,  %
for any value $v$ of type $T$ \wrt store typing $\Sigma'$,
if evaluating the function body in the extended value environment with store $\sigma'$
results in a value of type $U$, a store $\sigma''$, and a store typing $\Sigma''$, such that
$\sigma'':\Sigma''$ and $\STC{\Sigma'}{\Sigma''}$,
then we conclude that
the function values are valid at type $T \to U$ \wrt store typing $\Sigma$.
Note that we inline the term interpretation to characterize the function body's computational behavior.

A term $t$ has type $T$ \wrt store typing $\Sigma$ if for all $\sigma:\Sigma$,
term $t$ evaluates to a value of type $T$, a final store $\sigma'$ and a store typing $\Sigma'$, such that
$\sigma':\Sigma'$ and $\STC{\Sigma}{\Sigma'}$.

\bfparagraph{The Compatibility Lemmas, Fundamental Theorem and Adequacy}
The semantic typing judgment $\G \models t: T$ is defined in \figref{fig:lambda_lr} (bottom, ignoring the highlighted formula).
It means that term $t$ inhabits the term interpretation of type $T$
under any value environment $H$ and store typing $\Sigma$ that satisfies the
semantic interpretation of the typing context $\UG{\G}$, as defined in \figref{fig:lambda_lr} (bottom).
The semantic typing rules have the same shape as the syntactic ones, but turning the symbol $\ts$ (in \figref{fig:lambda}) into the symbol $\models$.
Each of the semantic typing rules is a compatibility lemma.

\begin{theorem}(Fundamental Theorem  of Unary Logical Relations)\label{thm:lambda_ufp}
  Every syntactically well-typed term is semantically well-typed, \ie,
  if $\G \ts t: T$, then
  $\G \models t : T$.
\end{theorem}

The following theorem entails that a semantically typed term always terminates.
The termination property is defined as part of the semantic typing judgment.
\begin{theorem}(Adequacy of Unary Logical Relations)\label{thm:lambda-adeq}
  Every closed semantically well-typed term $t$ is safe: if
  $\emptyset \models\ t : \Type{T} $, then $\exists ~v, \sigma.\, \config{t}{\emptyset}{\emptyset} \, \eval \, \pconfig{v}{\sigma}$.
\end{theorem}

The fundamental theorem (\thmref{thm:lambda_ufp}) follows immediately from the compatibility lemmas, and 
\thmref{thm:lambda-adeq} (Adequacy) and semantic type safety follow from the fundamental theorem (\thmref{thm:lambda_ufp}) and the definition of semantic term interpretation $\G \models t : T$ and $\Utt{T}$.

\subsection{Rediscovering Reachability Types}\label{sec:motiv_limitation}
\begin{comment}
\begin{wrapfigure}{r}{0.29\textwidth}
  \vspace{-4ex}
\begin{lstlisting}
def m(init: int) {
  val x = new Ref(init) 
  val y = new Ref(init)
  (() => x = !y; y += 1, 
   () => assert (!x <= !y))
}
\end{lstlisting}
\caption{Safe use of unsafe feature encapsulated in function \texttt{m}.}
\label{fig:redis}
\vspace{-3ex}
\end{wrapfigure}
\end{comment}
\begin{wrapfigure}{r}{0.30\textwidth}
  \vspace{-4ex}
\begin{lstlisting}
def m() {  
  val x = new Ref(1)   
  val y = new Ref(2)
  def f () = { y += 1 }
  f()
  assert (!x == 1)
}
\end{lstlisting}%
\caption{Safe use of unsafe feature encapsulated in function \texttt{m}.\protect\footnotemark}
\label{fig:redis}
\vspace{-4ex}
\end{wrapfigure}%
\footnotetext{The code that omits the assertion statement can be encoded as a sequence of $\lambda$-abstractions and applications in our systems.} 
\noindent The logical definition allows us to establish strong theoretical
results such as semantic type soundness and termination but remains
fairly limited in describing program behavior in other ways. 
Firstly, it does not allow us to prove the assertion true on the right, 
as the logical definition does not say anything about preserving
store \emph{values} across certain operations. 
Secondly, the assertion statement gets stuck if @!x == 1@ is false, thus it is an unsafe feature~\cite{logical-approach},
although the example uses it safely, which cannot be proven using the given logical definitions.~\footnote{Those are not the general motivations behind reachability types, but rather to give an explanation for some requirements of an effective logical relations model. Please refer to prior works~\cite{DBLP:journals/pacmpl/BaoWBJHR21,wei2023polymorphic} for general motivation.}
\footnote{See an informal justification of the safe use of the assertion statement in \figref{fig:redis}, using our binary logical relations in \secref{sec:binary}, 
but our goal is not primarily to verify safe use of general unsafe code like~\citet{DBLP:journals/pacmpl/0002JKD18}'s work.} 

The highlighted formulas in \figref{fig:lambda_lr} illustrate that information about reachable locations from given values
can give us useful semantic store invariants.
We first introduce function $\vallocs{v}$ (\figref{fig:lambda_lr}, top) that computes reachable locations from value $v$.
In \figref{fig:redis}, %
function @m@ reaches what its free variables reach; \ie, $\qbot$. %

With the knowledge of reachable locations, instead of using the standard definition of store typing extension, we introduce \emph{relational store typing} (RST), 
written as $\STCPA{\Sigma}{\L}{\Sigma'}$,
which parameterizes the definition of store typing extension with reachable locations $\L$.
RST is symmetric, allowing $\Sigma'$ to be defined from $\Sigma$, such that they agree on $\L$, 
leaving the remainder unspecified.\footnote{Starting from \secref{sec:unary_base}, we adopt the notation $\STCPB{\Sigma}{L}{\Sigma'}$ to reflect the symmetric nature of the RST.}
The standard definition of store typing extension $\Sigma \sqsubseteq \Sigma'$ can be defined as $\STCPA{\Sigma}{\DOM(\Sigma)}{\Sigma'}$.

\bfparagraph{The Time Travelling Property} 
Now the reasoning can time travel between a past
store typing and another (possible future) one back and forth, as long as they agree on reachable locations, \ie, formula $\circledx{1}$,
which is characterized by the following lemma:
\begin{lemma}[Time Travelling]\label{lem:valt_store_change}
    If $(H, \Sigma, v) \in \UV{T}$, 
    and $\STCPA{\Sigma}{\vallocs{v}}{\Sigma'}$,
    then $(H, \Sigma', v) \in \UV{T}$. 
\end{lemma}%
The lemma allows us to establish that the triple $(H, \Sigma', v)$ is a valid element of $\UV{T}$ if we know $(H, \Sigma, v)$ is valid as long as the two store typings agree on the locations reachable from value $v$.

In the interpretation of function types,
formula $\circledx{1}$ states that types are only preserved \wrt reachable locations from function values,
which gives us a degree of \emph{local reasoning}, 
as $\Sigma'$ does not assume anything from $\Sigma$ that is not reachable from the function value.
Under this assumption together with the argument value being semantically well-typed in $\Sigma'$, 
the safe reduction of the function body implies that functions can only \emph{access}
locations reachable from their arguments and themselves.
This characterizes a separation logic~\cite{DBLP:conf/lics/Reynolds02,DBLP:conf/csl/OHearnRY01} style of reasoning, 
but without enforcing exclusiveness of ownership, uniqueness, or immutability.

In the example of \figref{fig:redis}, 
formula $\circledx{1}$ implies no store locations are needed in the pre-store (specified by $\Sigma$), 
characterizing proper encapsulation.
In function @m@, function @f@ reaches what @y@ reaches, which is separate from what @x@ reaches.
Thus, we can prove the assertion true by formula $\circledx{4}$.

In addition, with the reachability information, the value interpretation of function types can also know that
an argument is either separate from the function, or has some overlap with the function (formula $\circledx{2}$).
This is a tautology here, but we will see that our reachability qualifiers can specify how argument values nontrivially interact with function values in \secref{sec:unary_base}. 
The return value may further reach fresh locations (formula $\circledx{3}$).
In the proof context, we know fresh locations are part of $\Sigma''$, 
so it is redundant to use $\Sigma'' - \Sigma'$ here. 
A similar refinement is adapted to the term interpretation.
Now, for $\langb$, 
the semantic typing judgment for open terms $t$ can be localized to $t$'s free variables, as formulated by the highlighted formula in \figref{fig:lambda_lr} (bottom).

Having added these semantic store invariants to the logical 
relation, why don't we expose them to the programmer as features of 
the type system? This is precisely the idea of reachability types.

\section{Semantic Type Soundness of \X{}}
\label{sec:unary_base}
In this section, we first present the reachability type system \X{} that tracks aliasing and expresses values reachable from a given expression's result,
define its value interpretation of types and terms, and present the semantic typing rules and show semantic type soundness.
Then we present $\Xsub{}$ (in \secref{sec:subtyping}) that extends \X{} with subtyping,
and $\maybelang$ (in \secref{sec:unary_nested}) that extends \X{} to support higher-order mutable references. 
Finally, we present $\maybelange$ (in \secref{sec:effects}) that extends $\maybelang$ with observable write effect qualifiers, 
demonstrating a stronger value preservation property than what is defined for $\maybelang$.

\subsection{The \X{} Language} \label{sec:syntaxb}
\figref{fig:base} shows the syntax of \X{}, where the differences from $\langb{}$ (in \figref{fig:lambda}) are highlighted.
Terms follow \langb{}, except that $\lambda$ terms are in the form of $\tabs{\lambda x.t}{\QF}$, carrying a reachability qualifier $\QF$ that includes the function's captured variables. 
\X{} adds reachability qualifiers to its type language and allows dependent function types, which are in the form of \((x: \ty[s]{T}) \to \ty[r]{U}\), where both argument
and return types are qualified. The codomain \(\ty[r]{U}\) may depend on the argument \(x\) in its qualifier $r$. %
To avoid dealing with deep substitution, we prohibit references to $x$ from
inside $U$. This restriction is purely technical; some prior works~\cite{wei2023polymorphic} do allow deep references, but this choice does not appear to impact expressiveness in any significant way~\cite{jia2024escape} as deep references can also be modeled indirectly via chains of argument and self-references.
\begin{figure}\footnotesize
  \begin{mdframed}
      \judgement{Syntax}{\BOX{\X}}\vspace{-8pt}
      \[\begin{array}{l@{\quad}l@{\quad}l@{\quad}l}
              c       & :=  & \ttrue \mid \tfalse                                                           & \text{Literals}       \\
              t         & ::= & c \mid x \mid \tabs{\lambda x.t}{\HLMathSUPER{\QF}} \mid t_1~t_2 \mid \tref~t \mid\ !~t \mid t \coloneqq t \mid t_1;t_2 & \text{Terms}                   \\

              r,s     & \in & \mathcal{P}_{\mathsf{fin}}(\Var \uplus \{ \QFresh \} \uplus \{ {\QSelf }\})                                                              & \text{Function Domain/Codomain Qualifiers} \\
              p,o     & \in & \mathcal{P}_{\mathsf{fin}}(\Var \uplus \{ \QFresh \})                                                              & \text{Reachability Qualifiers} \\
              \flt, \QF & \in & \mathcal{P}_{\mathsf{fin}}(\Var)                                                                                   & \text{Observations}            \\
              S,T,U,V   & ::= &  \ty{Bool} \mid (\HLMath{x:}\ty[\HLMathSUPER{s}]{T}) \to \ty[\HLMathSUPER{r}]{U} \mid \TRef~\ty[\HLMathSUPER{p}]{T}             & \text{Types}                   \\
              \Gamma    & ::= & \varnothing\mid \Gamma, \, x : \ty[\HLMathSUPER{p}]{T}                                                                  & \text{Typing Environments}     \\
              v         & ::= & c \mid \ell \mid \cl{H}{\lambda x.t}{\HLMathSUPER{\QF}}                                    & \text{Values}                  
          \end{array}\]
  \end{mdframed}
  \caption{The syntax of \X. Differences from \langb{} (in \figref{fig:lambda}) are highlighted.}
  \label{fig:base}
  \vspace{-3ex}
\end{figure}

Following prior works on reachability types~\cite{DBLP:journals/pacmpl/BaoWBJHR21,wei2023polymorphic}, types in \X{} are of the form $\ty[p]{T}$, where $p$ is a \emph{reachability qualifier}, a finite set of 
variables that may include the distinct freshness marker $\QFresh$, representing the values reachable from the given 
expression's result.

For simplicity, we introduce a self-reference marker $\QSelf$,
rather than formalizing self-references $\lambda f$ as a binder of a $\lambda$ term (as in prior works~\cite{DBLP:journals/pacmpl/BaoWBJHR21,wei2023polymorphic}).
The self-reference marker $\QSelf$ may appear in the qualifiers of a function's argument and its result, often expressed by the symbols $s$ and $r$ as shown in \figref{fig:base}.
If present, it denotes that the corresponding value may reach what the function reaches. 
This simplified formalization does not change the use of self-references in prior works~\cite{DBLP:journals/pacmpl/BaoWBJHR21,wei2023polymorphic}, 
but frees us from engaging in reasoning about recursion within the logical definitions, which could otherwise be done using well-established techniques 
such as step-indexing~\cite{ahmed2004semantics,DBLP:conf/popl/AhmedDR09,DBLP:journals/toplas/AppelM01}.

Mutable reference types $\TRef~\ty[p]{T}$ track the known aliases of the value held by the reference.
We use \emph{observation} \(\flt\), which is a finite set of variables,
to specify which variables in the typing
context \(\Gamma\) are observable, where the typing context assigns qualified typing
assumptions to variables.

\subsection{Store Typing and Semantic Typing Context}
\begin{figure*}\footnotesize
    \begin{mdframed}
        \judgement{Interpretation of Value Reachability\phantom{liftel}}{\BOX{\X{}}}\\
    $
      \begin{array}{l@{\ \,}c@{\ \,}l@{\qquad\qquad\qquad\qquad\ \ }r}
        \vallocs{c} = \qbot    \qquad
        \vallocs{\ell} = \{\ell\}   \qquad
        \vallocs{\cl{H}{\lambda x.t}{\QF}} = \varslocs{H}{\QF} \qquad 
        \varslocs{H}{\QF}  =  \bigcup_{x \in q} L(H(x))  
      \end{array}
    $\vspace{-1ex}\\
    \judgement{Variable Reachability and Qualifier Saturation}{\BOX{{\color{gray}\G\vdash}\,x \reaches x}\ \BOX{{\color{gray}\G\vdash}\, \qsat{q}}}\vspace{-1ex}
    $$
      \begin{array}{ll@{\quad\qquad}ll}
        \text{Reachability Relation} & {\color{gray}\G\vdash}\, x \reaches y \Leftrightarrow  x : T^{q,y} \in \Gamma   &
        \text{Variable Saturation}   & {\color{gray}\G\vdash}\, \qsat{x} := \left\{\, y \mid x \reaches^* y\, \right\}   \\[1.1ex]
        \text{Qualifier Saturation}  & {\color{gray}\G\vdash}\, \qsat{q} :=  \bigcup_{x\in q} \qsat{x} &
      \end{array}$$\vspace{-1.5ex}\\
       \judgement{Unary Logical Relations}{}\vspace{-1ex}
        \begin{mathpar}
                \begin{array}{@{\hspace{-1ex}}r@{\hspace{1ex}}l@{\hspace{1ex}} l}
                  \Sigma    & ::= & \varnothing\mid \Sigma, \ell : (\val, \HLMath{\L})  \\
                  \STCPA{\Sigma}{\L}{\Sigma'} & \DEF & \L \subq \DOM(\Sigma) \land \L \subq \DOM(\Sigma') \, \land \, (\forall\, \ell \in \L.\, \Sigma(\ell) = \Sigma'(\ell')) \\
                  \sigma :\Sigma & \DEF & \DOM(\sigma) = \DOM(\Sigma) \, \land  \, (\forall \ell \in \DOM(\sigma). \, \exists \, \val. \Sigma(\ell) = (\val, \HLMath{\qbot}) \, \land  \, \sigma(\ell) \in \val \, \land \, \HLMath{\vallocs{\sigma(\ell)} \subq \qbot}) \\
                    \HLMath{\WFV{\Sigma}{v}}               & \DEF   & \HLMath{\vallocs{v}  \subseteq \DOM(\Sigma)}                                                                                                                                                                                                         \\
                    \HLMath{\;\WFP{p}{x}{\G}}                & \DEF    & \HLMath{\NFQ{p} \subq \DOM(\G_1) \qquad \qquad \text{where } \G = \G_1, x: \ty[p]{T}, \G_2 \text{ for some } T} \\

                    \DEPS{\sigma}{L}{\sigma'}              & = & \forall \ell \in \DOM(\sigma). \ \ell \not\in L \Rightarrow \sigma(\ell) = \sigma'(\ell)                                                                                                                                                             \\
                    \\
                    \UV{\ty{Bool}}                       & =    & \{ \UTC{H}{\Sigma}{c} \}                                                                                                                                                                                    \\

                    \UV{\TRef \ \ty[\qbot]{T}}   & =    & \{ \UTC{H}{\Sigma}{\ell} \mid \HLMath{\WFV{\Sigma}{\ell}} \, \land \, (\exists \, \val, \Sigma(\ell) = (\val, \qbot) \,\land \,
                                (\forall\, \Sigma', v. \, \HLMath{\vallocs{v} \subq \qbot} \Rightarrow \\
                                & & \quad (v \in \val \iff \UTC{H}{\Sigma'}{v} \in \UV{T}))) \}            \\

                    \\
                    \UV{(x:\ty[s]{T}) \to U^{r}} & =    & \{ \UTC{H}{\Sigma}{\cl{H'}{\lambda x. t}{\QF}} \mid  \HLMath{\WFV{\Sigma}{\cl{H'}{\lambda x. t}{\QF}}}  \, \land \,  (\forall \, v, \sigma', \Sigma'. \, \NUM{1}{\HLMath{\STCPB{\Sigma}{\varslocs{H'}{\QF}}{\Sigma'}}} \, \land \,           \\
                                                           &      & \quad  \WFS{\sigma'}{\Sigma'} \, \land  \, \UTC{H}{\Sigma'}{v}\in \UV{T}\, \land \,   \NUM{2}{\HLMath{\vallocs{v} \subq  \varslocs{H}{\NFQ{s}} \, \ccup{\QSelf \in s} \, \varslocs{H'}{\QF} \, \ccup{\QFresh \in s} \, \overline{\varslocs{H'}{\QF}}}} \, \Rightarrow \,  \\
                                                           &      & \quad \quad \quad \exists \, \sigma'', \Sigma'', v'. \, \config{H'\extends (x,v)}{\sigma'}{t} \eval \pconfig{v'}{\sigma''} \, \land  \, \WFS{\sigma''}{\Sigma''} \, \land \, \STC{\Sigma'}{\Sigma''} \, \land \,                                               \\
                                                           &      & \quad \quad \quad \quad \quad \UTC{H}{\Sigma''}{v'} \in \UV{U} \,\land \\
                                                           &      & \quad \quad \quad \quad \quad \NUM{3}{\HLMath{\vallocs{v'} \subq (\varslocs{H}{\NFQ{r}\qminus x} \cap \varslocs{H'}{\QF}) \, \ccup{x \in r} \vallocs{v} \, \ccup{\QSelf \in r}\, \varslocs{H'}{\QF} \, \ccup{\QFresh \in r} \, \frlocs{\Sigma''}{\Sigma'}}}) \, \land \, \\
                                                           &      & \quad \quad \quad \quad \quad \NUM{4}{\DEPS{\sigma'}{\HLMath{\varslocs{H'}{\QF}} \, \cup \, \vallocs{v}}{\sigma''}} \}      \\

                    \\
                    \Ut{\flt}{\ty[p]{T}}                   & =    & \{ \UTC{H}{\Sigma}{t} \mid  \forall \, \sigma. \, \WFS{\sigma}{\Sigma} \, \land  \, \exists \,  \sigma', \Sigma', v'.\,  \config{H}{\sigma}{t} \eval \pconfig{v'}{\sigma'} \, \land \, \WFS{\sigma'}{\Sigma'} \, \land \, \STC{\Sigma}{\Sigma'} \, \land                      \\
                                                           &      & \quad \quad \quad \quad \quad \, \, \UTC{H}{\Sigma'}{v'} \in \UV{T} \, \land \, \HLMath{\vallocs{v'} \subq \varslocs{H}{\flt \cap \NFQ{p}} \ccup{\QFresh\in p} \frlocs{\Sigma'}{\Sigma}} \land \, \DEPS{\sigma}{\HLMath{\varslocs{H}{\flt}}}{\sigma'} \}                                     \\
                    \\
                    \HLMath{\UG{\G[\flt]}}                 & = & \{(H, \Sigma) \mid \DOM(H) = \DOM(\G) \, \land \, \HLMath{\flt \subq \DOM(\G) \, \land \, \WFGA{H}{\Sigma}{\flt} \, \land \, \WFGB{H}{\Sigma}{\flt}} \}                                                                            \\
                    \HLMath{\WFGA{H}{\Sigma}{\flt}}        & = & \forall \, x, T, p. \, \G(x) = \ty[\HLMathSUPER{p}]{T}\, \Rightarrow  \, \HLMath{\WFP{p}{x}{\G}} \, \land \, (x \in \flt \, \Rightarrow \, \UTC{H}{\Sigma}{H(x)} \in \UV{T}) \, \land \,               \\
                                                           &      &  \quad \quad \, \, \qquad \qquad \qquad \qquad  \HLMath{(\QFresh \not\in p \,  \, \Rightarrow \vallocs{H(x)} \subq \varslocs{H}{\NFQ{p}})}  \\
                    \HLMath{\WFGB{H}{\Sigma}{\flt}}        & = & \HLMath{\forall p, p'. \, \NFQ{p} \subq \flt \, \land \, \NFQ{p'} \ \subq \flt  \, \Rightarrow \varslocs{H}{\NFQ{p}} \cap \varslocs{H}{\NFQ{p'}} \subq \varslocs{H}{\qsat{\NFQ{p}} \cap \qsat{\NFQ{p'}}}}
                \end{array}%
        \end{mathpar}%
    \end{mdframed}
    \caption{The value interpretation of types, terms and typing context interpretation for \X{} . Differences from \langb{} (in \figref{fig:lambda_lr}) are highlighted.
       Qualifiers $s$ and $r$ may include the self-references marker $\QSelf$, and are used to specify function domain and codomain. The notation $\STCPB{\Sigma}{L}{\Sigma'}$ means the RST introduced in \secref{sec:motiv}.
     }\label{fig:unary_base}
    \vspace{-4ex} 
\end{figure*}

 We refine the definition of store typings from \figref{fig:lambda_lr} by adding reachable locations to each entry. 
Now, a store typing $\Sigma$ maps from locations to pairs of semantic types and reachable locations, as shown in \figref{fig:unary_base}.
A well-formed store $\sigma$ \wrt store typing $\Sigma$ demands that 
each location $\ell$ in $\sigma$ stores well-typed values (\wrt $\Sigma$), which reach the empty set of locations, \ie $\HLMath{\vallocs{\sigma(\ell)} \subq \qbot}$.
This condition reflects the restriction in $\X{}$, 
where reference types are limited to the form of $\ty[\qbot]{\TRef~T}$, 
\ie, the referent of a reference does not reach any locations. 
This restriction is relaxed by $\maybelang$ in \secref{sec:unary_nested}.

Our logical definition is established on a well-defined logical program state,
which is defined in the semantic typing context $\UG{\G[\flt]}$ shown in
\figref{fig:unary_base} (bottom).
The logical program state specifies what can be observed by the observation $\flt$,
as anything outside the observation is irrelevant. %
In terms of encapsulation safety, the observation can also be considered as a boundary to another world. 
The semantic typing context defines two reachability invariants: $\WFQA$ and $\WFQB$. %
Invariant $\WFQA$ states that locations reachable from non-fresh values are bounded by those from its qualifier.

Invariant $\WFQB$ states that reachability qualifier intersection  distributes over locations \wrt qualifier saturation, which is written as $\qsat{p}$.
This invariant is critical for function applications,  where one needs to ensure an argument value is permissible to invoke the function 
by inspecting the argument's qualifier against the prescribed separation/overlapping from a function's type.
Following \citet{wei2023polymorphic}'s work, we assign minimal variable sets (called one-step reachability) 
and compute transitive closures when checking separation/overlapping.
\figref{fig:unary_base} (top) recaps their definitions of variable reachability and qualifier saturation, which compute all reachable variables transitively.
In addition, the predicate $\WFP{p}{x}{\G}$ for $x: \ty[p]{T} \in \G[\flt]$ (defined in \figref{fig:unary_base}) prohibits cycles in qualifiers, 
as the typing context is extended with the binder (in a $\lambda$ term),
where a parameter's qualifier is in the same scope as the $\lambda$ term. Thus,
cycles in qualifiers are not syntactically well-formed.

\subsection{The Logical Interpretation of Types and Terms}\label{sec:unary_lr_base}

The definitions of logical interpretation of types and terms are shown in the middle of \figref{fig:unary_base}.
They follow a similar logical structure as for $\langb{}$ in \figref{fig:lambda_lr}, 
but are enriched with the knowledge about reachability specified by the qualifiers in types (highlighted).

We use the notation $\ccup{b}$ to denote the conditional set union operator, \ie, if the predicate $b$ is true, 
the operator performs a union of the left- and right-hand operands;
otherwise, the result defaults to the left-hand operand only.
Let $p$ be a reachability qualifier that may include the freshness marker $\QFresh$ and the self-reference marker $\QSelf$.
We write $\NFQ{p}$ to denote dropping the freshness and the self-reference markers in $p$, resulting in the set of variables appearing in $p$,
\ie, $\NFQ{p} := p \qminus (\QFresh, \QSelf)$.

\bfparagraph{Interpretation of Reachability}
\figref{fig:unary_base} (top) refines function $\vallocs{v}$ that computes reachable locations from value $v$ for \X{}. 
The definition follows that for \langb{} (\figref{fig:lambda_lr}), 
except for function values. 
Here, reachable locations from a closure record $\cl{H}{\lambda x.t}{\QF}$ include those from its qualifier, \ie, $\varslocs{H}{\QF}$, 
which computes reachable locations from variables in $\QF$. 
We often write $\varslocs{H}{\QF}$ for the reachable locations from a closure record when the context is clear.

\bfparagraph{Well-Defined Reachable Locations}
To use the reachability information in logical definitions, we use the predicate $\WFV{\Sigma}{v}$ (shown in \figref{fig:unary_base}) 
to guarantee reachable locations from a value $v$ are well-defined \wrt a store typing $\Sigma$,
\ie, $\vallocs{v} \subq \DOM(\Sigma)$. 
For example, a location $\ell$ can only reach itself. Thus the predicate $\WFV{\Sigma}{\ell}$ checks $\ell \in \DOM(\Sigma)$.
For values of type $\ty{Bool}$, their reachable locations are the empty set.
So their well-defined predicates are trivially true, and thus are omitted from its value interpretation.

\bfparagraph{Reference Types} The value interpretation of reference type $\TRef~\ty[\qbot]{T}$ are store locations 
 that hold values of type $T$, and their contents reach the empty set of locations.
 Comparing with \langb{} that only allows references to hold Boolean values, here, references can store any typeable values as long as they reach the empty set of locations.

\bfparagraph{Function Types}
The logical interpretation of function type $(x:\ty[s]{\ty{T}}) \to U^{r}$ is defined in \figref{fig:unary_base}, 
where $s$, $T$ and $U$ are closed under value environment $H$.
Their closedness predicates are omitted in the figure, and in the logical definitions in the subsequent sections. 
The qualifier of the return value may include parameter $x$, as well as the freshness and self-reference markers.
The qualifiers of a function and its argument specify how the argument value communicates with its function value (formula $\circledx{2}$).
Here, we discuss several examples in~\citet{wei2023polymorphic}'s work against the formula.
Consider the following reachability-polymorphic identity function:
\begin{lstlisting}
def id(x: T$\trackfresh$): T$\trackset{x}$ = x             // : ((x: T$\trackfresh$) => T$\trackset{x}$)$\track$
\end{lstlisting}
The type specifies that \code{id} cannot reach anything from its context ($q = \qbot$ in $\circledx{2}$),
and accepts argument values that may reach separate locations from what \code{id} reaches ($s = \QFresh$ in $\circledx{2}$).
Thus, the \code{id} function accepts any values of type \code{T}.

A different parameter qualifier specifies permissible overlapping between functions and their argument values.
Consider the following variants of \code{id} which capture some variable \code{z} in the context:
\begin{lstlisting}[aboveskip=1pt,belowskip=1pt]
def id2(x: T$\trackset{\qfresh}\,\,\,\,\,$): T$\trackset{x}$ = { val u = z; x }            // : ((x: T$\trackset{\qfresh}\,\,\,\,\,$) => T$\trackset{x}$)$\trackset{z}$
def id3(x: T$\trackset{\qfresh,z}$): T$\trackset{x}$ = { val u = z; x }            // : ((x: T$\trackset{\qfresh,z}$) => T$\trackset{x}$)$\trackset{z}$
\end{lstlisting}
The qualifiers on the function types and their parameters
specify the reachability assumption that the
implementation can \emph{observe} about its context ($q = \{z\}$ in $\circledx{2}$), and about any
given argument value, respectively. 
Function \code{id2} accepts arguments reaching anything that does not (directly or transitively) reach \code{z} ($s = \QFresh$ in $\circledx{2}$),
and \code{id3} permits \code{z} in the parameter's qualifier, \ie, allowing any argument ($s = \{\QFresh, z\}$ in $\circledx{2}$).

The reachability qualifiers of return values express how they communicate with arguments and function values (formula $\circledx{3}$).
Consider the following function that mutates a captured reference cell and returns the argument (here, $q = \{c1\}$, $s = \{c1, \QFresh\}$).
We annotate that the argument \code{x} is potentially aliased with the captured argument \code{c1}
but apply the function with argument \code{c2}.
Our type system would not propagate such imprecision by tracking
one-step reachability. The return type only tracks the argument \code{x} ($r = \{x\}$ in $\circledx{3}$), meaning the return value reaches what the argument reaches.
When applying different arguments to the function, precise reachability
is retained: \looseness=-1
\begin{lstlisting}
... // c1: T$\trackset{c1}$, c2: T$\trackset{c2}$
def incr(x: T$\trackset{c1,\qfresh}$): T$\trackset{x}$ = { c1 := !c1 + 1; x }     // : ((x: T$\trackset{c1,\qfresh}$) => T$\trackset{x}$)$\trackset{c1}$
incr(c1)                                        // : T$\trackset{c1}$
incr(c2)                                        // : T$\trackset{c2}$  $\leftarrow$ precision retained
\end{lstlisting}
The logical definition precisely captures our crucial design that has the freshness marker $\QFresh$ in qualifiers
to explicitly communicate (non-)freshness, which is preserved by dependent
application and substitution~\cite{wei2023polymorphic} (see rule \textsc{t-app$\QFresh$} in \figref{fig:semantic_typing_base}). 

If the argument's qualifier includes both the self-reference and the freshness markers, any argument values of the proper pretype are allowed.
If the self-reference marker appears in the return value's qualifier, as in the  @cell@ example in \secref{sec:overview}, 
it means the return value can reach whatever the returning closure reaches.

\bfparagraph{The Term Interpretation.}
The term interpretation also adds the interpretation of reachability for its result values $v$.
If the result value $v$ is not fresh,  then we check that its reachable locations are 
bounded by those from the reachability qualifier observed by the observation $\flt$, \ie,
$\flt \cap \NFQ{p}$; otherwise, it may additionally include fresh locations $\frlocs{\Sigma'}{\Sigma}$.
The preserved store values are those not reachable from observation $\flt$.

\subsection{The Compatibility Lemmas, Fundamental Theorem and Adequacy}\label{sec:soundness_base}
\begin{figure}[t]\footnotesize
  \begin{mdframed}[innertopmargin=0pt, innerbottommargin=0pt, leftmargin=0pt, rightmargin=2pt]
    \judgement{Semantic Typing}{\BOX{\strut\G[\flt] \models t : \ty[p]{T}}} %
    \begin{minipage}[t]{.62\linewidth}\vspace{0pt}
      \begin{minipage}[t]{.45\linewidth}\vspace{0pt}
      \infrule[t-cst]{\\
      }{
        \G[\flt] \models\, c : \ty[\qbot]{\ty{Bool}}
      }
      \end{minipage}%
      \begin{minipage}[t]{.04\linewidth}
        \hspace{1pt}%
      \end{minipage}%
      \begin{minipage}[t]{.49\linewidth}\vspace{0pt}
      \infrule[t-var]{
        \G(x) =  \ty[p]{T} \qquad x \in \flt
      }{
        \G[\flt] \models\, x : \ty[x]{T}
      }
      \end{minipage}%
      \infrule[t-abs]{
        (\G, \, x: \ty[{\flt\cap(s[\QF/\QSelf]) \ccup{\QFresh \in s} \QFresh}]{T})^{\QF',x} \models\, t : \ty[{r[\QF'/\QSelf]}]{U}\ \\
        \QF' = \flt \cap \QF \qquad 
        \NFQ{s} \subq \QF' \\
      }{
        \G[\flt] \models\, (\lambda x.t)^{\QF'} : ((x:\ty[s]{T}) \to \ty[r]{U})^{\QF}
      }
      \infrule[t-app]{
      \G[\flt]\models\, t_1 : \ty[p]{\left((x:\ty[s]{T}) \to \ty[r]{U}\right)} \quad 
      \QFresh \not\in s, o \\
      \G[\flt]\models\, t_2 : \ty[o]{T} \qquad 
      \NFQ{o} \subq \NFQ{s} \subq \flt  \qquad \NFQ{r} \subq \flt,x \qquad
       x \not\in \FV(U)
      }{
      \G[\flt]\models\, t_1~t_2 : \ty[{r[o/x, p/\QSelf]}]{U}
      }
      \infrule[t-app$\QFresh$]{
        \G[\flt]\models\, t_1 : \ty[p]{\left((x:\ty[s]{T}) \to \ty[r]{U}\right)} \quad 
        \QFresh \in s \quad  
        x \not\in \FV(U) \\
        \G[\flt]\models\, t_2 : \ty[o]{T} \qquad 
       \QSelf \not\in s \Rightarrow \qsat{\NFQ{p}} \cap \qsat{\NFQ{o}} \subq \NFQ{s} \quad
       \NFQ{s} \subq \flt \quad
       \NFQ{r} \subq \flt,x   
      }{
      \G[\flt]\models\, t_1~t_2 : \ty[{r[o/x, p/\QSelf]}]{U}
      }
      \begin{minipage}[t]{.52\linewidth}
        \infrule[t-ref]{
        \G[\flt]\models\, t : \ty[\qbot]{T}
      }{
        \G[\flt]\models\, \tref~t : \ty[\QFresh]{(\TRef~\ty[\qbot]{T})}
      }
      \end{minipage}%
      \begin{minipage}[t]{.01\linewidth}
        \hspace{1pt}%
      \end{minipage}%
      \begin{minipage}{.49\linewidth}
      \infrule[t-$!$]{ %
        \G[\flt]\models\, t : \ty[p]{(\TRef~\ty[\qbot]{T})} \\
      }{
        \G[\flt]\models\, !t : \ty[\qbot]{T}
      }
      \end{minipage}%
    \end{minipage}%
    \begin{minipage}[t]{.02\linewidth}
      \hspace{1pt}%
    \end{minipage}%
    \begin{minipage}[t]{.37\linewidth}\vspace{0pt}      
      \infrule[t-$:=$]{
        \G[\flt]\models\, t_1 : \ty[{p}]{(\TRef~\ty[\qbot]{T})} \\
        \G[\flt]\models\, t_2 : \ty[\qbot]{T}
      }{
        \G[\flt]\models\, t_1 \coloneqq t_2 : \ty[\qbot]{\ty{Bool}}
      }
      \vgap
      \vgap
      \infrule[t-seq]{
        \G[\flt_1]\models\, t_1 : \ty[q_1]{\ty{Bool}}\\
        \G[\flt_2]\models\, t_2 : \ty[q_2]{\ty{Bool}} \\
        \flt_1 \subq \flt \qquad \flt_2 \subq \flt
      }
      {
        \G[\flt]\models\, t_1;t_2 : \ty[\qbot]{\ty{Bool}} \\
      }
      \vgap
      \infrule[t-sub]{ %
         \G[\flt] \models \, t: \ty[p_1]{T} \\
         \NFQ{p_1} \subq \NFQ{p_2} \subq \DOM(\G) \\
         p_3 = p_2 \ccup{\QFresh \in p_1 \lor \QFresh \in p_2} \QFresh
      }{
         \G[\flt] \models \, t: \ty[p_3]{T}
      }
      \vgap
      \infrule[t-sub-var]{ %
      \G[\flt] \models t: \ty[p]{U} \qquad  
      x \in \NFQ{p}   \\
      \G(x) = \ty[q]{T} \qquad   
      q \subq \flt \\ 
      }
      {
        \G[\flt] \models t: \ty[p{[q/x]}]{U} 
      }
    \end{minipage}%
    \\ 
    \judgement{Qualifier Substitution}{\BOX{r[p/x]}\ \BOX{r[q/\QSelf]}}\vspace{-1ex}
    $$
    \begin{array}{l@{\;}c@{\;}ll@{\quad\qquad\qquad\qquad}l@{\;}c@{\;}ll}
        r[p/x] & = & r\qminus\{x\}\qlub p & x\in r         & r[q/\QSelf] & = & r\qminus\{\QSelf\} \qlub q & \QSelf \in r \\
        r[p/x] & = & r                      & x\notin r      & r[q/\QSelf] & = & r         & \QSelf \not\in r
      \end{array}$$ \vspace{-1ex}\\
    \textbf{\textsf{Qualifier Shorthands}} \qquad
    $\small p,q := p \qlub q\qquad x := \{x\} \qquad \QFresh :=\{\QFresh\}\qquad  \QSelf := \{\QSelf\} \qquad \fstarred{p} := \{\qfresh\} \qlub\{\QSelf\} \qlub p$ \\
    \phantom{\textbf{\textsf{Qualifier Shorthands}} \qquad} $p \ccup{b} q := b = \ttrue~?~ p,q : p \qquad\qquad\qquad\quad\phantom{1} \NFQ{p} := p \qminus (\QFresh, \QSelf)$ \\
  \end{mdframed}
  \caption{Semantic typing rules for \X{}. Qualifiers $s$ and $r$ may include the self-references marker $\QSelf$, and are used to specify function domain and codomain.} \label{fig:semantic_typing_base}
  \vspace{-3ex}
\end{figure}

The semantic typing judgment is defined as $\G[\flt] \models t: \ty[p]{T} \DEF \forall\, (H, \Sigma) \in \UG{\G[\flt]}. \, \UTC{H}{\Sigma}{t} \in \Ut{\flt}{\ty[p]{T}}$,
meaning that term $t$ inhabits the term interpretation of type $\ty[p]{T}$
under any value environment $H$ and store typing $\Sigma$ that satisfies the
semantic interpretation of typing context $\UG{\G[\flt]}$.

\figref{fig:semantic_typing_base} shows the semantic typing rules.
For readability, we often drop the set notation for qualifiers and write them down as comma-separated lists of atoms.
Each of the rules is a compatibility lemma, and has been proved in Rocq. 
The syntactic typing judgments have the form $\G[\flt] \ts t: \ty[p]{T}$.
The syntactic typing rules have the same shape as the semantic one, but turning the symbol $\models$ (in \figref{fig:semantic_typing_base}) into the symbol $\ts$.

We have proved the compatibility lemmas, fundamental theorem and adequacy for \X{}.
The fundamental theorem follows immediately from the compatibility lemmas, and 
adequacy, semantic type safety, and termination follow from the fundamental theorem and the definition of semantic term interpretation in \figref{fig:unary_base}.

\bfparagraph{Encapsulation} Now, we study encapsulation property demanded for rule \textsc{$\beta$-equiv} in \secref{sec:equiv}.
The typing judgment $\G[\qbot] \ts t : \ty[\qbot]{T}$ expresses \emph{encapsulated computations}, 
which are logically characterized by the following lemma:
\begin{lemma}[Encapsulated Computations (\X{})]\label{lem:unary_base_encap}
        If $\G[\qbot] \ts t : \ty[\qbot]{T}$, 
        then $\forall \, (H, \Sigma) \in \UG{\G[\qbot]}.$ 
        $\forall \, \Sigma'. \,
        \STCPB{\Sigma}{\qbot}{\Sigma'} \Rightarrow      
        (H, \Sigma', t) \in \Ut{\qbot}{\ty[\qbot]{T}}$. 
\end{lemma}
\noindent Note that the notation $\STCPB{\Sigma}{\qbot}{\Sigma'}$ means the RST introduced in \secref{sec:motiv}. The lemma expresses that expression $t$ is valid at type $\ty[\qbot]{T}$ \wrt arbitrary store typing, including the empty one.
Note that term $t$ can be an implementation of an abstract data type, 
which may internally allocate new store locations. 
The empty qualifier (in $\ty[\qbot]{T}$) dictates that $t$'s evaluation result cannot be tracked, 
hence are encapsulated.

\subsection{Subtyping}\label{sec:subtyping}
\begin{figure}[t]\footnotesize
\begin{mdframed}
  \judgement{Semantic Interpretation of Subqualifiers, Subpretypes and Subtypes}{\BOX{\Xsub{}}} \\\vspace{-2.5ex}
  \begin{footnotesize}
    \begin{mathpar}
            \begin{array}{@{}r@{\hspace{1ex}}l@{\hspace{1ex}} l}
              \G \models q_1 <: q_2 & \DEF &  \forall (H, \Sigma) \in  \UG{\G[\qbot]}. \, \varslocs{H}{q_1} \subq \varslocs{H}{q_2} \\
              \G \models \ty{T_1} <: \ty{T_2} & \DEF & \forall (H, \Sigma) \in  \UG{\G[\qbot]}. \, \forall v. \, \UTC{H}{\Sigma}{v} \in \UV{\ty{T_1}} \Rightarrow  \UTC{H}{\Sigma}{v} \in \UV{\ty{T_2}} \\
              \G[\flt] \models \ty[p_1]{T_1} <: \ty[p_2]{T_2} & \DEF & \forall (H, \Sigma) \in  \UG{\G[\flt]}. \, \forall t. \, \UTC{H}{\Sigma}{t} \in \Ut{\flt}{\ty[p_1]{T_1}} \Rightarrow \UTC{H}{\Sigma}{t} \in \Ut{\flt}{\ty[p_2]{T_2}}
            \end{array}
          \end{mathpar}
    \end{footnotesize}\\
\judgement{Semantic Typing}\BOX{\strut\G[\flt] \models t : \ty[p]{T}}\vspace{-1ex}\\
\begin{minipage}[t]{0.3\linewidth}
 \qquad \qquad \qquad \qquad \qquad $\cdots$ 
\end{minipage}%
\begin{minipage}[t]{.02\linewidth}
  \hspace{1pt}%
\end{minipage}%
\begin{minipage}[t]{0.58\linewidth}
   \infrule[t-sub-stp]{
\G[\flt] \models t: \ty[p_1]{T_1} \qquad \G[\flt] \models \ty[p_1]{T_1} <: \ty[p_2]{T_2} \qquad
 \NFQ{p_2} \subq \flt
}{
\G[\flt] \models t: \ty[p_2]{T_2}
}
\end{minipage}%
\\
\judgement{Semantic Subtyping}{\BOX{\G \models p <: p}\ \BOX{\G \models s <: s}\BOX{\G \models\ty{T} <: \ty{T}}\ \BOX{\G[\flt] \models\ty[p]{T} <: \ty[o]{U}}}\\
\begin{minipage}[t]{.53\linewidth}\vspace{7pt}
\begin{minipage}[t]{.48\linewidth}\vspace{2pt}
  \infrule[q-sub]{
  p_1\subq p_2\subq \starred{\DOM(\G)}
}{
  \G \models p_1 <: p_2
}
\end{minipage}%
\begin{minipage}[t]{.02\linewidth}
  \hspace{1pt}
\end{minipage}%
\begin{minipage}[t]{.51\linewidth}\vspace{2pt}
\infrule[q-sub-\QSelf]{
  s_1 \subq s_2 \subq \fstarred{\DOM(\G)}
}
{
\G \models s_1 <: s_2
} 
\end{minipage}%
\vspace{6pt}
\infrule[q-var]{
        \G(x) = \ty[q_2]{T} \quad x \in q_1 \quad q_1 \subq \DOM(\G)
      }
      {
        \G \models \starred{q_1} <: \starred{q_1[q_2/x]} 
      }      
\begin{minipage}[t]{.66\linewidth}\vspace{0.3pt}      
\infrule[q-trans]{
  \G \models p_1 <: p_2 \quad
  \G \models p_2 <: p_3  
}{
  \G \models p_1 <: p_3
}
\end{minipage}%
\begin{minipage}[t]{.01\linewidth}
  \hspace{1pt}
\end{minipage}%
\begin{minipage}[t]{.36\linewidth}\vspace{0.3pt}     
  \infrule[s-refl]{ \G \ts T \, \mbox{\text{wf}}} {   
  \G\models\ty{T} <: \ty{T}
}
\end{minipage}%
\end{minipage}%
\begin{minipage}[t]{.01\linewidth}
  \hspace{1pt}
\end{minipage}%
\begin{minipage}[t]{.47\linewidth}\vspace{0pt}
\infrule[s-ref]{
  \G \models \ty{U} <: \ty{S} \quad
  \G \models \ty{S} <: \ty{U} \\ 
  \G \models p_1 <: p_2  \quad
  \G \models p_2 <: p_1
}{
  \G \models \ty{\TRef~\ty[p_1]{U}} <: \ty{\TRef~\ty[p_2]{S}}
}
\vgap
   \infrule[s-fun]{
    \G \models \ty{T_2} <: \ty{T_1} \quad
    \G \models \ty{U_1} <: \ty{U_2} \quad 
    \QSelf \in s_1 \Rightarrow \QFresh \in s_1 \\ 
    \G \models s_2 <: s_1 \quad 
    \G, x: \ty[s_2]{T_2} \models r_1 <: r_2 \\
  }{
    \G \models\ty{\ty[s_1]{T_1} \to \ty[r_1]{U_1}} <: \ty{\ty[s_2]{T_2} \to \ty[r_2]{U_2}}
  }
\infrule[sq-sub]{
  \G \models\ty{S} <: \ty{T}\quad\quad \G \models p <: q 
}{
  \G[\flt] \models\ty[p]{S} <: \ty[q]{T}
}

\end{minipage}%
\end{mdframed}
\caption{Semantic subtyping rules for \Xsub{}. }
\label{fig:subtyping_base}
\vspace{-3ex}
\end{figure}

 To keep the exposition consistent with prior work~\cite{wei2023polymorphic},
we extend \X{} with subtyping, resulting in \Xsub{}.
\figref{fig:subtyping_base} shows the interpretation of subqualifiers, subpretypes and subtypes,
as well as the semantic subtyping rules. 
Qualifier subtyping includes the subset relation (\textsc{q-sub} and \textsc{q-sub-$\QSelf$}), 
and transitivity (\textsc{q-trans}). 
Rule \textsc{q-var} is contextual and allows assigning the qualifier to a type in the upper level of the reachability chain. 
Subtyping for reference type (\textsc{s-ref}) and function type (\textsc{s-fun}) are standard modulo qualifiers.
Rule \textsc{s-fun} additionally demands that if the argument qualifier includes the self-reference marker, then it must be fresh.
Subtyping for types (\textsc{sq-sub}) forwards to qualifier subtyping and ordinary type subtyping judgments. 
Rule \textsc{t-sub-stp} is the subsumption rule, allowing a term with a subtype to be safely used in a context where a supertype is expected.

The semantic subtyping notion provided by the logical relations for \Xsub{} provides an ideal vehicle to identify subtyping rules in a specific scenario, which is not present in prior work~\cite{wei2023polymorphic}.
Here we discuss one of the key technical points with
the following example, demonstrating how the local variable \code{x} from the result type's qualifier can be replaced by the self-reference marker $\QSelf$,
by using stepwise subtyping rules.
\[
\setlength{\abovedisplayskip}{0pt}
\setlength{\belowdisplayskip}{0pt}    
    ((x: \ty[\QFresh,\QSelf]{T})  \to  \ty[x]{U})^{\QF} \RSTEP{1} ((x: \ty[q']{T})  \to  \ty[x]{U})^{\QF} \RSTEP{2} ((x: \ty[q']{T})  \to  \ty[q']{U})^{\QF} \RSTEP{3} ((x: \ty[q']{T})  \to  \ty[\QSelf]{U})^{\QF}
\]
\noindent Here we assume $q' <: q$.
The step $\rect{1}$ is established by contravariance on the argument type, which is always valid as the qualifier $\{\QFresh,\QSelf\}$ permits any argument values of type $T$.
The step $\rect{2}$ is established by covariance on the result type, provided  $\{x\} <: q'$ by
the rule \textsc{q-var}.
The step $\rect{3}$ is valid only in the context such that $q' <: q$, meaning what is reachable from $q'$ is a subset of those reachable from the function's qualifier $\QF$.
This kind of stepwise reasoning enables a sound way of introducing self-references via subtyping to type escaping values without term-level coercions.
See \citet{jia2024escape}'s work for more detailed elaboration.

\bfparagraph{The Compatibility Lemmas, Fundamental Theorem and Adequacy}
The syntactic subqualifiers, subpretyping and subtyping judgments and 
those syntactic rules 
have the same form as the semantic ones, 
but turning the symbol $\models$ (in \figref{fig:subtyping_base}) into the symbol $\ts$.
\begin{theorem}[Fundamental Theorem of Subqualifiers]\label{them:qsp_fun}
    Every syntactically well-typed subqualifier relation is semantically well-typed, \ie, 
    if $\G \ts p_1 <: p_2$, then $\G \models p_1 <: p_2$, and 
    if $\G \ts s_1 <: s_2$, then $\G \models s_1 <: s_2$
\end{theorem}

\begin{theorem}[Fundamental Theorem of Subpretypes]\label{them:stp_fun}
    Every syntactically well-typed subpretype relation is semantically well-typed, \ie, 
    if $\G \ts T_1 <: T_2$, then $\G \models T_1 <: T_2$
\end{theorem}

\begin{theorem}[Fundamental Theorem of Subtypes]\label{them:qstp_fun}
    Every syntactically well-typed subtype relation is semantically well-typed, \ie, 
    if $\G[\flt] \ts \ty[p_1]{T_1} <: \ty[p_2]{T_2}$, then $\G[\flt] \models \ty[p_1]{T_1} <: \ty[p_2]{T_2}$
\end{theorem}
The definition of semantic typing for \Xsub{} follows what is defined for \X{}, \ie, $\G[\flt] \models t: \ty[p]{T} \DEF \forall\, (H, \Sigma) \in \UG{\G[\flt]}. \, \UTC{H}{\Sigma}{t} \in \Ut{\flt}{\ty[p]{T}}$,
meaning that term $t$ inhabits the term interpretation of type $\ty[p]{T}$
under any value environment $H$ and store typing $\Sigma$ that satisfies the
semantic interpretation of typing context $\UG{\G[\flt]}$.

We have proved the compatibility lemmas, fundamental theorem and adequacy for \Xsub{}.
The fundamental theorem follows immediately from the compatibility lemmas, and 
adequacy and semantic type safety follow from the fundamental theorem and the definition of semantic term interpretation.

\subsection{Extension with Higher-Order Mutable References (\maybelang{})}
\label{sec:unary_nested}
As presented so far, \X{} only supports
\emph{first-order} mutable references, \ie, mutable references cannot hold values
that contain references to other store locations. 
In this section, we introduce \maybelang{}, which extends \X{} to support higher-order mutable references.
We outline the major differences from \X{} here. 

\bfparagraph{The \maybelang{} Language}
We relax the restriction on referent type $\ty[\qbot]{T}$ 
for mutable references, \ie, 
a referent's type can now carry a reachability qualifier $\ty[q]{T}$, 
indicating that the referent is a non-fresh value of type $T$, 
and may reach locations specified by $q$.

Following $\X{}$, a store typing maps from locations to pairs of semantic types and reachable locations.
Let $\ell$ be a location in store typing $\Sigma$, such that $\Sigma(\ell) = (\val, \L)$, for some $\val$ and $\L$, where $\val$ is its semantic type, 
and $\L$ prescribes the set of locations reachable from the value stored in location $\ell$. 
As location $\ell$ can indirectly reach some locations in a store,
following prior work~\cite{wei2023polymorphic}, we use location saturation, 
written as $\lls{\Sigma}{\L}$, 
to transitively compute reachable sets, and demand that the saturated reachable locations from the content is a subset of
the saturated ones from $\L$. 
In addition, following the syntactic typing rules that disallow cyclic references~\cite{wei2023polymorphic}, our logical definition 
requires that earlier allocated locations cannot reach those allocated later, which is defined in $\sigma :\Sigma$, \ie, $\forall \ell' \in \L. \, \ell' < \ell $. 
Here, store locations are ordered by their allocation time.

\bfparagraph{The Logical Interpretation of Types and Terms}\label{sec:unary_lr_nested}
\begin{figure*}[t]\footnotesize
    \begin{mdframed}
        \judgement{Location Saturation}{\BOX{$\maybelang{}$}}\vspace{-1ex} \\
    $
      \begin{array}{l@{\ \,}c@{\ \,}l@{\qquad\qquad\qquad\ \ }l}
        \ell \xreaches{\Sigma} \ell' \Leftrightarrow \Sigma(l) =(\val, \L \cup \{ \ell' \}) \qquad
        \lls{\Sigma}{\ell} \Leftrightarrow \{\ell' \mid \ell \xreaches{\Sigma}^{\ast} \ell' \} \qquad
        \lls{\Sigma}{\L}:= \bigcup_{\ell \in \L} \lls{\Sigma}{\ell}
      \end{array}
    $ \\ \\
        \judgement{Unary Logical Relations}{} \vspace{-1ex}
        \begin{mathpar}
                \begin{array}{@{\hspace{-0.5ex}}r@{\hspace{0.5ex}}l@{\hspace{0.5ex}} l}

                    \WFV{\Sigma}{v}                                      & =    & \HLMath{\lls{\Sigma}{\vallocs{v}}}  \subseteq \DOM(\Sigma)                                                                                                                                                                                                                    \\
                    \Sigma    & ::= & \varnothing\mid \Sigma, \ell : (\val, \L)  \\
                    \sigma :\Sigma & \DEF & \DOM(\sigma) = \DOM(\Sigma) \, \land  \, (\forall \ell \in \DOM(\sigma). \, \exists \, \val, \L. \, \Sigma(\ell) = (\val, \L) \, \land \, \sigma(\ell) \in \val \, \land \, \\
                                 &   &    \NUM{1}{\HLMath{\lls{\Sigma}{\vallocs{\sigma(\ell)}} \subq \lls{\Sigma}{\L}}} \, \land \, \NUM{2}{\HLMath{(\forall \ell' \in \L. \, \ell' < \ell )}}) \\

                    \\
                    \UV{\ty{Bool}}                                     & =    & \{ \UTC{H}{\Sigma}{c} \}                                                                                                                                                                                                             \\

                    \\
                    \UV{\HLMath{\TRef \ \ty[q]{T}}}                      & =    & \{ \UTC{H}{\Sigma}{\ell} \mid  \WFV{\Sigma}{\ell}  \, \land \, (\exists\, \val, \L. \, \Sigma(\ell) = (\val, \L) \, \land \,  \HLMath{\L =  \varslocs{H}{q}} \, \land \, \HLMath{\lls{\Sigma}{\L} \subseteq \lls{\Sigma}{\ell}} \, \land \\
                                                                         &      & \qquad (\forall \, \Sigma', v. \STCPB{\Sigma}{\HLMath{\lls{\Sigma}{\L}}}{\Sigma'} \land  \HLMath{\lls{\Sigma'}{\vallocs{v}} \subq \lls{\Sigma'}{\L}}  \Rightarrow  (v \in \val \iff \UTC{H}{\Sigma'}{v} \in \UV{T})))\} \\

                    \\
                    \UV{(x:\ty[s]{T}) \to \ty[r]{U}}             & =    & \{ \UTC{H}{\Sigma}{\cl{H'}{\lambda f(x). t}{\QF}} \mid  \WFV{\Sigma}{\cl{H'}{\lambda f(x). t}{\QF}}  \, \land \,  (\forall \, v, \sigma', \Sigma'. \,  \HLMath{\STCPB{\Sigma}{\llsvars{H'}{\Sigma}{\QF}}{\Sigma'}} \, \land       \\
                                                                         &      & \qquad  \, \WFS{\sigma'}{\Sigma'} \, \land \, \UTC{H}{\Sigma'}{v}\in \UV{T}\, \land \, \HLMath{\lls{\Sigma'}{\vallocs{v}} \subq \llsvars{H'}{\Sigma'}{\NFQ{s}} \, \ccup{\QSelf \in s} \, \llsvars{H'}{\Sigma'}{q} \, \ccup{\QFresh \in s} \overline{\llsvars{H'}{\Sigma'}{q}}}  \Rightarrow           \\
                                                                         &      & \qquad \exists \, \Sigma'', \sigma'', v'. \, \config{H'\extends (x,v)}{\sigma'}{t} \eval \pconfig{v'}{\sigma''}   \, \land  \, \STC{\Sigma'}{\Sigma''} \, \land \, \WFS{\sigma''}{\Sigma''} \, \land \, \\
                                                                         &      & \qquad \qquad \UTC{H}{\Sigma''}{v'} \in \UV{U} \, \land \\
                                                                         &      & \qquad \qquad \HLMath{\lls{\Sigma''}{\vallocs{v'}} \subq (\llsvars{H'}{\Sigma''}{\NFQ{r}\qminus x} \cap \llsvars{H'}{\Sigma''}{\QF}) \, \ccup{x \in r} \, \lls{\Sigma''}{\vallocs{v}} \, \ccup{\QSelf \in r} \llsvars{H'}{\Sigma''}{\QF}} \, \ccup{\QFresh \in r} \, \frlocs{\Sigma''}{\Sigma'} \, \land \\
                                                                         &      & \qquad \qquad \DEPS{\sigma'}{\HLMath{\llsvars{H'}{\Sigma'}{\QF} \, \cup \, \lls{\Sigma'}{\vallocs{v}}}}{\sigma''} )\} \\

\\

                    \Ut{\varphi}{\ty[p]{T}}                              & =    & \{ \UTC{H}{\Sigma}{t} \mid \forall \, \sigma: \Sigma. \,  \exists. \,\Sigma', \sigma', v'.\,  \config{H}{\sigma}{t} \eval \pconfig{v'}{\sigma'} \,  \land \, \STC{\Sigma}{\Sigma'} \, \land \, \sigma': \Sigma' \,  \land                                                     \\
                                                                         &      & \qquad \UTC{H}{\Sigma'}{v'} \in \UV{T}  \, \land \, \lls{\Sigma'}{\vallocs{v}} \subq \llsvars{H}{\Sigma'}{\flt \cap \NFQ{p}} \ccup{\QFresh\in p} \frlocs{\Sigma'}{\Sigma} \, \land \, \DEPS{\sigma}{\HLMath{\llsvars{H}{\Sigma}{\flt}}}{\sigma'} \} \\
                                                                         
                    \\

                    \UG{\G[\flt]}                                        & = & \{(H, \Sigma) \mid \DOM(H) = \DOM(\G) \, \land \, \flt \subq \DOM(\G) \, \land \, \WFGA{\Sigma}{H}{\flt} \, \land \, \WFGB{\Sigma}{H}{\flt} \}                                                                                                                                \\
                    \WFGA{\Sigma}{H}{\flt}                               & = & \forall \, x, T, p. \, \G(x) = \ty[p]{T} \Rightarrow  \,  \WFP{p}{x}{\G} \, \land \,                                                                                                                                                                 \\
                                                                         &      & \qquad \qquad (x \in \flt \Rightarrow  \UTC{H}{\Sigma}{H(x)} \in \UV{T}) \, \land \, (\QFresh \not\in p \, \land \, \HLMath{\NFQ{p} \subq \flt} \Rightarrow \HLMath{\lls{\Sigma}{\vallocs{H(x)}}} \subq \HLMath{\llsvars{H}{\Sigma}{\NFQ{p}}}) \\
                                                                         
                    \WFGB{\Sigma}{H}{\flt}                               & = & \forall p, p'. \, \NFQ{p} \subq \flt \, \land \, \NFQ{p'}  \subq \flt \, \land \,  \HLMath{\qsat{\NFQ{p}} {\cap} \qsat{\NFQ{p'}} \subq \flt} \Rightarrow \HLMath{\llsvars{H}{\Sigma}{\NFQ{p}}} \cap \HLMath{\llsvars{H}{\Sigma}{\NFQ{p'}}} \subq \HLMath{\llsvars{H}{\Sigma}{\qsat{\NFQ{p}} {\cap} \qsat{\NFQ{p'}}}}.
                \end{array}
        \end{mathpar}
        \\
        \textsf{\textbf{Semantic Typing Judgment}} \qquad  $\G[\flt] \models t: \ty[p]{T} \quad \DEF \quad \forall\, (H, \Sigma) \in \UG{\G[\flt]}. \, \UTC{H}{\Sigma}{t} \in \Ut{\flt}{\ty[p]{T}}$ \\
        \\
        \textsf{\textbf{Semantic Typing}} \\ \vspace{-2pt}
        \begin{minipage}[t]{.31\linewidth}\vspace{-3pt}
          \infrule[t-ref]{
        \G[\flt]\models t : \ty[\HLMathSUPER{q}]{T} \qquad \HLMath{q \subq \flt} 
          }{
        \G[\flt]\models \tref~t : \ty[\HLMath{\starred{q}}]{(\TRef~\ty[\HLMathSUPER{q}]{T})}
        }
          \vgap
        \end{minipage}%
        \begin{minipage}[t]{.02\linewidth}
          \hspace{0.9pt}%
        \end{minipage}%
        \begin{minipage}[t]{.33\linewidth}\vspace{-8pt}
          \vgap
          \infrule[t-$!$]{ %
            \G[\flt]\models t : \ty[p]{(\TRef~\ty[\HLMathSUPER{q}]{T})} \qquad
            q \subq \flt 
          }{
            \G[\flt]\models !t : \ty[\HLMath{q}]{T}
          }
        \end{minipage}%
        \begin{minipage}[t]{.02\linewidth}
          \hspace{0.9pt}%
        \end{minipage}%
        \begin{minipage}[t]{0.3\linewidth}\vspace{-12pt}
          \infrule[t-$:=$]{
            \G[\flt]\models t_1 : \ty[p]{(\TRef~\ty[\HLMathSUPER{p_1}]{T})} \\
            \G[\flt]\models t_2 : \ty[\HLMathSUPER{\NFQ{p_1}}]{T} \qquad 
            \HLMath{\NFQ{p_1} \subq \flt}
          }{
            \G[\flt]\models t_1 \coloneqq t_2 : \ty[\qbot]{Bool}
          } 
        \end{minipage}%
    \end{mdframed}
    \caption{The value interpretation of types, terms, typing context interpretation, semantic typing judgment and refined semantic typing rules for \maybelang{}. Differences from \X{} are highlighted. 
    }
    \label{fig:unary_nested}
    \vspace{-3ex}
\end{figure*}%

 The definitions of the value interpretation of types, terms, and typing contexts (shown in \figref{fig:unary_nested})
follow \X{}, but taking location saturation into consideration. 
For example, the predicate $\WFV{\Sigma}{v}$ is refined to 
ensure all the indirectly reachable locations are well-defined \wrt a store typing, 
\ie, $\WFV{\Sigma}{v}  = \lls{\Sigma}{\vallocs{v}}  \subseteq \DOM(\Sigma)$.                                                                                                                                                                                                                    

Similarly, in the definition of value interpretation of Boolean and function types, and term interpretation,
the reachability properties defined by $\vallocs{v}$ and $\varslocs{H}{\QF}$ for \X{} are now
defined using location saturation, \ie, $\lls{\Sigma}{\vallocs{v}}$ and $\llsvars{H}{\Sigma}{\QF}$ for some store typing $\Sigma$.
For example, formula $\circledx{2}$ in \figref{fig:unary_base} is refined to $\lls{\Sigma'}{\vallocs{v}} \subq \llsvars{H'}{\Sigma'}{\NFQ{s}} \, \ccup{\QSelf \in s} \, \llsvars{H'}{\Sigma'}{q} \, \ccup{\QFresh \in s} \overline{\llsvars{H'}{\Sigma'}{q}}$.

Values of reference type $\ty[q]{\TRef~T}$ are store locations $\ell$, where its content may change over time,
but must always be of type $\ty{T}$.
The reachability qualifier $q$ allows the reasoning to time travel back and forth between two store typings as long as they agree on 
the reachable locations from the referent. 
Thus, the value interpretation of a reference type \figref{fig:unary_base} is refined to
$(\forall \, \Sigma', v. \, \STCPB{\Sigma}{\lls{\Sigma}{\L}}{\Sigma'} \land  \lls{\Sigma'}{\vallocs{v}} \subq \lls{\Sigma'}{\L}  \Rightarrow  (v \in \val \iff \UTC{H}{\Sigma'}{v} \in \UV{T}))$, 
where $\lls{\Sigma}{\L}$ is the logical interpretation of $q$, \ie, $\L = \varslocs{H}{q}$.

\bfparagraph{The Time Travelling Property} The time travelling lemma (\lemref{lem:valt_store_change} in \secref{sec:motiv}) is refined with saturated locations as follows: %
\begin{lemma}[Time Travelling ($\maybelang$)]\label{lem:valt_store_change_nested}
    If $(H, \Sigma, v) \in \UV{T}$, 
    and $\STCPB{\Sigma}{\lls{\Sigma}{\vallocs{v}}}{\Sigma'}$,
    then $(H, \Sigma', v) \in \UV{T}$. 
\end{lemma}
\noindent The lemma allows the validity of the value $v$ at type $T$ to be preserved from store typing $\Sigma$ to 
another $\Sigma'$, as long as they agree on the saturated locations reachable from value $v$.

\bfparagraph{The Compatibility Lemmas, Fundamental Theorem and Adequacy}
The definition of semantic typing judgment follows what was defined for $\X{}$ in \figref{fig:unary_base}.
\figref{fig:unary_nested} shows the selected semantic typing rules.
Following prior work~\cite{wei2023polymorphic}, the type system supports a restricted form of nested references:
only non-fresh values can be stored/read from/written into in a reference (\textsc{t-ref}, \textsc{t-!} and \textsc{t-:=}).
As referent qualifiers are invariant, store typing is monotonic.
Each of the semantic typing rules is a compatibility lemma, and has been proven in Rocq. 

We have proved the compatibility lemmas, fundamental theorem and adequacy for \maybelang{}.
The fundamental theorem follows immediately from the compatibility lemmas, and 
adequacy, semantic type safety, and termination follow from the fundamental theorem and the definition of semantic term interpretation.

\bfparagraph{Encapsulation} The encapsulated computation lemma for \X{} (\lemref{lem:unary_base_encap}) 
remains valid for \maybelang. Its formalization is similar, and thus is omitted.

\bfparagraph{Discussion 1: Modeling Key Features in Prior Work} 
Our logical relations for $\maybelang$ models a major subset of \citet{wei2023polymorphic}'s $\maybelang$ system.
We summarize the differences as follows: 
\begin{enumerate}[label=\textbf{(\arabic*)}, leftmargin=*, wide]
\item \textbf{Self-References:}  As mentioned previously, we use a self-reference marker $\QSelf$, instead of formalizing self-references 
$\lambda f$ as a binder of a $\lambda$ term as in \citet{wei2023polymorphic}'s work.
However, this simplified formalization retains their key feature, \ie, the lightweight form of qualifier polymorphism  
which is achieved by rule \textsc{t-app$\QFresh$} in both works. 

\item \textbf{Deep Substitution:} As mentioned previously, to simplify the formalization, our \textsc{t-app} rule does not consider deep substitution as theirs does. 
This restriction is purely technical, and does not appear to impact expressiveness in any significant way~\cite{jia2024escape} as deep references can also be modeled indirectly via chains of argument and self-references.

\item \textbf{Reference Rules:} In addition, our references rules require that a reference's qualifier be bounded 
by observation $\flt$. This property is not explicitly specified in their system, 
but can be inferred from their tying rules.
Thus, our reference rules are semantically equivalent to theirs.
\end{enumerate}

\bfparagraph{Discussion 2: Termination}
Our semantic is parameterized over a fuel value that bounds the steps that the execution is allowed;
if it runs out of fuel, \text{timeout} will be returned.
In this setting, we prove termination by providing enough fuel for the execution.
In addition, proving termination in the presence of higher-order mutable references requires some constraints:  
the qualifiers attached to a store location prohibit storing any values
that may reach the location itself, thus excluding cycles.
This is a limitation in prior work~\cite{wei2023polymorphic} we discovered,
leading to programs like Landin`s knot~\cite{DBLP:journals/cj/Landin64} not being typeable. 
In particular, the introduction and assignment rules for mutable references in ~\citet{wei2023polymorphic}'s work prohibit creating cycles through the store. 
Specifically, rules \textsc{t-ref} and \textsc{t-assgn} (in their work) ensure that referent qualifiers are treated as invariant, so only values with equivalent qualifiers can be assigned. 
Consider a typical version of Landin`s knot~\cite{DBLP:journals/cj/Landin64}:
\begin{lstlisting}
 ($\lambda$ x. (x := ($\lambda$ z. !x z); !x))(ref ($\lambda$ y y))
\end{lstlisting}
In this example, the reference type for the argument cannot capture any reachable variable, 
but the function body updates the reference with a value that captures variable \code{x}. 
Thus, the example is not typable in their system.

\bfparagraph{Discussion 3: Strong Type Soundness} Prior works~\cite{DBLP:journals/pacmpl/BaoWBJHR21,wei2023polymorphic} proved syntactic type soundness 
using small-step semantics, where reachability properties are checked for intermediate values through their preservation theorems. 
In contrast, our work extends the big-step semantics $\eval$ to a total evaluation function,
making a distinction between \text{timeout}, \text{errors}, and normal values, leading to strong soundness results~\cite{DBLP:conf/popl/AminR17}.
In particular, we define reachability properties as invariants in the value interpretation of types and terms in the logical framework. 
The fundamental theorem is proven by induction on the type derivation ($\G[\flt] \ts t: \ty[q]{T}$),
where each case is the corresponding compatibility lemma (\ie, semantic typing rule).
Therefore, the reachability properties are also verified for each sub-expression.

\subsection{Write Effect System Extension (\maybelange{})}
\label{sec:effects}
In the previously presented frameworks, 
any reachable location could be modified while evaluating an expression. 
This section presents \maybelange{} that adds observable write effect qualifiers 
to track which locations are modified, as opposed to just read or, \eg,
passed as argument without being dereferenced.
This gives us a stronger store value preservation property than previous frameworks.

\begin{figure*}\footnotesize
    \begin{mdframed}
        \judgement{Unary Logical Relations}{\BOX{$\maybelange{}$}}\vspace{-8pt}
        \begin{mathpar}
            \begin{footnotesize}
                \begin{array}{@{}r@{\hspace{1ex}}l@{\hspace{1ex}} l}

                    \UV{(x:\ty[s]{\ty{T}}) \stackrel{\EPS}{\to} U^{r}} & = & \{ \UTC{H}{\Sigma}{\cl{H'}{\lambda x. t}{\QF}} \mid  \WFV{\Sigma}{\cl{H'}{\lambda x. t}{\QF}}  \, \land \,  (\forall \ v, \sigma', \Sigma'. \, \STCPB{\Sigma}{\llsvars{H'}{\Sigma}{\QF}}{\Sigma'} \ \land \ \WFS{\sigma'}{\Sigma'} \ \land                                \\
                                                                &   & \qquad  \ \UTC{H}{\Sigma'}{v}\in \UV{T}\ \land \  \lls{\Sigma'}{\vallocs{v}} \subq \llsvars{H}{\Sigma'}{\NFQ{s}} \, \ccup{\QSelf \in s} \, \llsvars{H'}{\Sigma}{\QF} \, \ccup{\QFresh \in s} \, \overline{\llsvars{H'}{\Sigma}{\QF}}   \Rightarrow \\
                                                                &   & \qquad \qquad  \exists \, \sigma'', \Sigma'', v'. \,   \config{H'\extends (x,v)}{\sigma'}{t} \eval \pconfig{v'}{\sigma''} \, \land  \, \STC{\Sigma'}{\Sigma''} \, \land \, \WFS{\sigma''}{\Sigma''} \, \land \\
                                                                &   & \qquad \qquad \qquad \UTC{H}{\Sigma''}{v'} \in \UV{U} \, \land \\
                                                                &   & \qquad \qquad \qquad \lls{\Sigma''}{\vallocs{v'}} \subq (\llsvars{H'}{\Sigma''}{\NFQ{r}\qminus x} \cap \llsvars{H'}{\Sigma''}{\QF}) \, \ccup{x\in r} \lls{\Sigma''}{\vallocs{v}} \, \ccup{\QSelf \in r}\ \llsvars{H'}{\Sigma''}{\QF} \\
                                                                &   & \qquad \qquad \qquad \qquad \qquad  \ccup{\QFresh \in r} \, \frlocs{\Sigma''}{\Sigma'} \, \land \\
                                                                &   & \qquad \qquad \qquad \HLMath{\DEPS{\sigma'}{\llsvars{H}{\Sigma'}{\EPS\qminus x} \, \ccup{x \in \EPS} \, \lls{\Sigma'}{v} \ccup{\QSelf \in \EPS} \, \llsvars{H'}{\Sigma'}{\QF} }{\sigma''}}  )\} \\

                  \\

                    \Ut{\varphi}{\ty[p]{T}\ \EPS}               & = & \{ \UTC{H}{\Sigma}{t} \mid \forall \, \WFS{\sigma}{\Sigma}. \,  \exists \, \Sigma', \sigma', v'.\,  \config{t}{H}{\sigma} \eval \pconfig{v'}{\sigma'} \,  \land \, \STC{\Sigma}{\Sigma'} \, \land \, \sigma': \Sigma' \,  \land                            \\
                                                                &   & \qquad \UTC{H}{\Sigma'}{v'} \in \UV{T}  \, \land \, \lls{\Sigma'}{\vallocs{v}} \subq \llsvars{H}{\Sigma'}{\flt \cap \NFQ{p}} \ccup{\QFresh\in p} \frlocs{\Sigma'}{\Sigma} \, \land \, \HLMath{\DEPS{\sigma}{\llsvars{H}{\Sigma}{\flt \cap \EPS}}{\sigma'}} \} \\
                    
                \end{array}
            \end{footnotesize}
        \end{mathpar}%
        \\ \\
        \textbf{\textsf{Syntax}} \quad $\EPS \in \mathcal{P}_{\mathsf{fin}}(\Var\uplus \{ {\QSelf }\})$  \qquad \qquad \textbf{\textsf{Flow-Insensitive Sequential Effects Composition}} \qquad
        $\EPS[1] \EFFSEQ \EPS[2] := \EPS[1] \cup \EPS[2]$ %
    \end{mdframed}
    \caption{Effect interpretation and selected typing rules for \maybelange{}. Differences from \maybelang{} are highlighted.
    Effect qualifiers ($\EPS$) are finite set of variables, and may include the self-reference marker $\QSelf$. The effect composition (\eg, $\EPS[1] \EFFSEQ \EPS[2]$) is defined as set union (\ie, $\EPS[1] \cup \EPS[2]$).}
    \label{fig:unary_nested_effs}
    \vspace{-3ex}
\end{figure*}

 \bfparagraph{Effect Types}
The syntax of effect qualifiers are shown in \figref{fig:unary_nested_effs} (bottom).
Effect qualifiers ($\EPS$) are finite sets of variables, and may include the self-reference marker $\QSelf$,
denoting the locations (in the pre-state) that may be
modified during computation.
Function types carry effect qualifiers $\EPS$ for their latent effects.

\bfparagraph{Logical Interpretation of Effects}
\figref{fig:unary_nested_effs} shows
the interpretation of function types and terms that includes effects, which are highlighted in the figure.
Consider the following example:
\begin{lstlisting}
... // c1: T$\trackvar{c1}$, c2: T$\trackvar{c2}$
def f (x: T$\trackfresh$) = { c1 := !x + 1 }   // : ((x:T$\trackfresh$) =>${\trackvar{\textcolor{eff}{c1}}}$ ())$\trackset{c1}$
f(c2)                             // : () $\texttt{\textcolor{eff}{c1}}$          
\end{lstlisting}
The function @f@ has its latent effect on the variable @c1@, meaning that its body can only modify the locations reachable from @c1@, 
as shown in the highlighted formula in \figref{fig:unary_nested_effs}.
When the function is invoked, \ie, @f(c2)@, the effect is just @c1@ as well, according to rule \textsc{e-app-$\QFresh$}.
In contrast, without the effect extension, $\maybelang$ approximates the write effect to what the argument and function reaches.
As a result, @f(c2)@ would induce effect on both @c1@ and @c2@, leading to a weaker value preservation property, as illustrated by formula $\circledx{4}$ and term interpretation in \figref{fig:unary_base}.

Other definitions follow \maybelang{},  and are omitted in \figref{fig:unary_nested_effs}. %

\bfparagraph{Semantic Typing Judgment and Rules} 
\begin{figure}[t]\footnotesize
  \begin{mdframed}
    \typicallabel{t-abs}
    \judgement{Semantic Typing}{\BOX{\strut\G[\flt] \models t : \ty[p]{T}\ \EPS}} %
    \begin{minipage}[t]{.65\linewidth}\vspace{0pt}
      \begin{minipage}[t]{.45\linewidth}\vspace{0pt}
      \infrule[e-cst]{
        \\
      }{
        \G[\flt] \models c : \ty[\qbot]{\ty{Bool}}\ \PURE
      }
      \end{minipage}%
      \begin{minipage}[t]{.02\linewidth}
        \hspace{1pt}
      \end{minipage}%
      \begin{minipage}[t]{.5\linewidth}\vspace{0pt}
      \infrule[e-var]{
        \G(x) = x : \ty[p]{T} \quad\quad x \in \flt
      }{
        \G[\flt] \models x : \ty[x]{T}\ \PURE
      }
      \end{minipage}%
      \infrule[e-abs]{
        (\G, \, x:  \ty[{\flt\cap(s[\QF/\QSelf]) \ccup{\QFresh \in s} \QFresh}]{T})^{\QF',x} \models t : \ty[{r[\QF'/\QSelf]}]{U}\ \EPS{\FX{[\QF'/\QSelf]}} \\
        \QF' = \flt\cap \QF \qquad
        \NFQ{s} \subq \QF' \\
      }{
        \G[\flt] \vdash \tabs{\lambda x.t}{\QF'} : ((x: \ty[s]{T}) \stackrel{\EPS}{\to} \ty[r]{U})^{\QF}\ \PURE
      }
      \infrule[e-app]{
      \G[\flt]\models t_1 : \ty[p]{\left((x: \ty[s]{T}) \stackrel{\EPS[3]}{\to} \ty[r]{U}\right)}\ \EPS[1] \qquad 
      \QFresh \not\in s, o \\
      \G[\flt]\models t_2 : \ty[o]{T}\ \EPS[2] \qquad
      \NFQ{o} \subq \NFQ{s} \subq \flt \qquad
      \NFQ{r} \subq \flt,x \qquad 
      x \not\in \FV(U)\\
      \EPS[3] \subq \flt,x \qquad
      \theta = [o/x, p/\QSelf] \qquad
      \theta' = [\NFQ{o}/x, \NFQ{p}/\QSelf]
      }{
      \G[\flt]\models t_1~t_2 : \ty[r{\theta}]{U}\ \FX{(\EPS[1]\EFFSEQ\EPS[2]\EFFSEQ\EPS[3])\theta'}
      }
      \infrule[e-app-$\QFresh$]{
      \G[\flt]\models t_1 : \ty[p]{\left(x: \ty[s]{T} \stackrel{\EPS[3]}{\to} \ty[r]{U}\right)}\ \EPS[1] \qquad
      \theta = [o/x, p/\QSelf] \\
      \G[\flt]\models t_2 : \ty[o]{T}\ \EPS[2] \qquad
      \QSelf \not\in s \Rightarrow  \qsat{p} \cap \qsat{\NFQ{o}} \subq \NFQ{s} \qquad  
      \QFresh \in s \qquad \NFQ{s} \subq \flt \\      
      x \not\in \FV(U) \qquad
      \NFQ{\EPS[3]} \subq \flt,x \qquad
      \NFQ{r} \subq \flt, x \qquad 
      \theta' = [\NFQ{o}/x, \NFQ{p}/\QSelf]
      }{
      \G[\flt]\models t_1~t_2 : \ty[r\theta]{U}\ \FX{(\EPS[1]\EFFSEQ\EPS[2]\EFFSEQ\EPS[3])\theta'}
      }
      \infrule[e-$:=$]{
        \G[\flt]\models t_1 : \ty[p]{(\TRef~\ty[p_1]{T})}\ \EPS[1] \qquad 
        \G[\flt]\models t_2 : \ty[\NFQ{p_1}]{T}\ \EPS[2] \qquad 
        \NFQ{p_1} \subq \flt
        }{
        \G[\flt]\models t_1 \coloneqq t_2 : \ty[\qbot]{\ty{Bool}}\ \EPS[1]\EFFSEQ\EPS[2]\EFFSEQ\FX{\NFQ{p}}     }
          
    \end{minipage}%
    \begin{minipage}[t]{.01\linewidth}
      \hspace{0.5pt}%
    \end{minipage}%
    \begin{minipage}[t]{.35\linewidth}\vspace{0pt}
      \infrule[e-ref]{
          \G[\flt]\models t : \ty[q]{T}\ \EPS \qquad q \subq \flt 
        }{
          \G[\flt]\models \tref~t : \ty[\starred{q}]{(\TRef~\ty[q]{T})}\ \EPS
        }
        \vgap
      \infrule[e-$!$]{ %
        \G[\flt]\models t : \ty[p]{(\TRef~\ty{T^{q}})}\ \EPS \\
        q \subq \flt
      }{
        \G[\flt]\models !t : \ty[q]{T}\ \EPS
      }
      \vgap
      \infrule[e-seq]{
        \G[\flt_1]\models t_1 : \ty[q]{\ty{Bool}}\ \EPS[1]\\
        \G[\flt_2]\models t_2 : \ty[p]{\ty{Bool}}\ \EPS[2] \\
        \flt_1 \subq \flt \qquad
        \flt_2 \subq \flt
      }
      {
        \G[\flt]\models t_1;t_2 : \ty[\qbot]{\ty{Bool}}\ \EPS[1]\EFFSEQ\EPS[2] \\
      }
      \vgap
      \infrule[e-sub]{ %
        \G[\flt] \models t : \ty[p_1]{T}\ \EPS[1] \\
        \NFQ{p_1} \subq \NFQ{p_2} \subq \DOM(\G) \\ 
        \NFQ{\EPS[1]} \subq \NFQ{\EPS[2]} \subq \DOM(\G) \\
        p_3 = p_2 \ccup{\QFresh \in p_1 \lor \QFresh \in p_2} \QFresh
      }{
        \G[\flt]\models t : \ty[p_3]{T}\ \EPS[2]
      }
      \vgap
      \infrule[e-sub-var]{
        \G[\flt]\models t: \ty[p]{U}\ \EPS \qquad x \in \NFQ{p} \\
        \G(x) = \ty[q]{T}  \qquad  
        q \subq \flt
      }
      {
        \G[\flt]\models t: \ty[p{[q/x]}]{U}\ \EPS
      }
    \end{minipage}%
    \\ \\
    \textbf{\textsf{Effects Shorthands}} \qquad
    $\EPS[1] \EFFSEQ \EPS[2] := \EPS[1] \cup \EPS[2]$ %
  \end{mdframed}
  \caption{Semantic typing rules for \maybelange{}.}
    \label{fig:typing_effs}
    \vspace{-3ex}
\end{figure}

 The definition of the semantic typing judgment is defined as: {\small$~\G[\flt] \models t: \ty[p]{T}\ \EPS  \DEF \forall\, (H, \Sigma) \in \UG{\G[\flt]}. \, \UTC{H}{\Sigma}{t} \in \Ut{\flt}{\ty[p]{T}\ \EPS}$}. 
It means that term $t$ inhabits the term interpretation of type $\ty[p]{T}\ \EPS$
under any value environment $H$ and store typing $\Sigma$ that satisfies the
semantic interpretation of the typing context $\UG{\G[\flt]}$, 
which follows \maybelang{}.%

\figref{fig:typing_effs} shows the semantic typing rules for \maybelange{}, 
extending those of \maybelang{} with observable write effect qualifiers ($\EPS$),
denoting the set of locations (in the pre-state) that may be modified during the execution.
The effect rules are straightforward:
the final effect of a compound term combines the effects of
sub-terms with the intrinsic effect of this term. 
For example, the effects of assignments (rule \textsc{E:=}) consist of two parts: (1) the effects $\EPS[1]$ and $\EPS[2]$,
which are those of sub-terms, and (2) the effect $\FX{\NFQ{p}}$, which denotes the reachable locations from variables in $\NFQ{p}$
being modified. The final effects are sequential composition of those effects.
In a flow-insensitive effect system like $\maybelange$, 
the composition %
is defined as set union 
.~\footnote{ 
One could also incorporate a flow-sensitive effect system, as suggested by~\citet{DBLP:journals/pacmpl/BaoWBJHR21}, 
to model more complex effects.}

\bfparagraph{Fundamental Theorem and Adequacy}
The syntactic typing judgments have the form $\G[\flt] \ts t: \ty[p]{T}\ \EPS$, 
and the typing rules have the same shape as the semantic ones, but turning the symbol $\models$ into the symbol $\ts$.
We have proved the compatibility lemmas, fundamental theorem and adequacy for \maybelange{}.
The fundamental theorem follows immediately from the compatibility lemmas, and 
adequacy, semantic type safety, and termination follow from the fundamental theorem and the definition of semantic term interpretation.

\bfparagraph{Encapsulation} 
The encapsulated computation lemma for \maybelange demand that an encapsulated computation does not induce observable effects. 
\begin{lemma}[Encapsulated Computation (\maybelange{})]\label{lem:unary_effs_encap}
     If $\G[\qbot] \ts t : \ty[\qbot]{T} \ \PURE$,
     then $\forall \, (H, \Sigma) \in \UG{\G[\qbot]}.$ 
     $\forall \, \Sigma'. \,
        \STCPB{\Sigma}{\qbot}{\Sigma'} \Rightarrow      
     (H, \Sigma', t) \in \Ut{\qbot}{\ty[\qbot]{T} \ \PURE}$. 
\end{lemma}

\section{Contextual Equivalence via Binary Logical Relations}
\label{sec:direct-lr}

We now switch from modeling soundness, termination, and 
other properties of a \emph{single} expression to properties concerning \emph{pairs}
of expressions, specifically notions of \emph{observational equivalence}. 
Thus, we extend our unary logical relations for $\maybelang{}$ and $\maybelange{}$
to binary logical relations, with the intention that semantically equivalent values
and terms are exactly the ones that will be logically related. 
We formalize their judgments into one, as $\maybelange{}$ only adds conditions regarding effects. 
For example, the typing judgment is written as $\G[\flt] \models \ty[p]{T}\ \op{\EPS}$, 
where formula $\op{\EPS}$ applies only to $\maybelange{}$.

\subsection{Contextual Equivalence}
\label{sec:direct_context}
Two programs are contextually equivalent if a well-typed program context cannot distinguish between them,
\ie, they have the same observable behavior.
Here, we define contextual equivalence to verify equational rules (\secref{sec:equiv}).
A program $t_1$ is said to be \emph{contextually equivalent} to another program $t_2$, written as $\G[\flt] \models t_1 \equiva t_2 : \ty[p]{T}\ \op{\EPS}$,
if for any program context $C$ with a hole of type $\ty[p]{T}\ \op{\EPS}$,
if $C[t_1]$ has some (observable) behavior, then so does $C[t_2]$.
The definition of context $C$ is shown soon. 
Following the approach of \citet{logical-approach} and related prior work
\cite{DBLP:conf/popl/AhmedDR09}, we define
a judgement for logical equivalence using binary logical relations,
written as $\G[\flt] \models t_1 \equivlog t_2 : \ty[p]{T}\ \op{\EPS}$.

\begin{figure}[t]\footnotesize
	\begin{mdframed}
		\judgement{Context for Contextual Equivalence}{\BOX{\maybelang{}/\maybelange{}}}\vspace{-8pt}
		\[\begin{array}{l@{\ \ }c@{\ \ }l@{\qquad\qquad\ }l@{\ \ }c@{\ \ }l}
				{C} & ::= & \square \mid C\ t \mid t\ C \mid  \tabs{\lambda x. C}{\QF} \mid \tref ~C \mid\ !~{C} \mid {C} := {t} \mid {t} := {C} \mid t;~{C} \mid {C};~t &  &  & \\
			\end{array}\]
		\judgement{Context Typing Rules}{
			\BOX{C : (\G[\flt]; \ty[q]{T}\ \op{\EPS}) \carrow (\GP[\fltp]; \ty[q']{T'}\ \op{\EPSPR})}
		}
		\begin{minipage}[t]{.32\linewidth}\vspace{2pt}
			\infrule[c-$\square$]{\\\\
			}{
				\square: (\G[\flt]; \ty[q]{{T}}\ \op{\EPS}) \carrow (\G[\flt]; \ty[q]{{T}}\ \op{\EPS})
			}
		\end{minipage}%
		\begin{minipage}[t]{.01\linewidth}
			\hspace{1pt}
		  \end{minipage}%
		\begin{minipage}[t]{.66\linewidth}\vspace{0pt}
			\infrule[c-$\lambda$]
			{  C : (\G[\flt]; \ty[p]{{S}}\ \op{\EPS}) \carrow ( (\G'\ ,\ x: \ty[{\fltp\cap(s[\QF/\QSelf])\ccup{\QFresh \in s} \QFresh}]{T})^{\QF', x}; \ty[{r[\QF'/\QSelf]}]{U} \ \op{\EPSPR}) \\
			  \QF' = \fltp \cap \QF \quad \NFQ{s} \subq \QF'
			}
			{
				\tabs{\lambda x. C}{\QF}: (\G[\flt]; \ty[p]{{S}} \ \op{\EPS}) \carrow (\GP[\fltp]; ((x: \ty[s]{{T}})  \stackrel{\op{\EPSPR}}{\to} \ty[r]{{U}})^{q} \ \PURE)
			}
		\end{minipage}%
		\vspace{1ex}\\
		\begin{minipage}[t]{1.0\linewidth}\vspace{0pt}
			\infrule[c-app-$\QFresh$-1]
			{
			C: (\G[\flt]; \ty[p]{S} \ \op{\EPS[1]}) \carrow
			(\GP[\fltp]; ((x: \ty[s']{{T'}}) \stackrel{\op{\EPS[3]}}{\to}  \ty[r']{{U'}})^{p'} \ \op{\EPS[4]})
			\quad
			\GP[\fltp] \ts t_2: \ty[o']{{T'}} \ \op{\EPS[2]} \quad
			\QFresh \in s' \quad 
			\NFQ{s'} \subq \fltp \quad \\
			\QSelf \not\in s' \Rightarrow \qsat{\NFQ{\QF'}} \cap \qsat{\NFQ{o'}} \subq \NFQ{s'} \quad
			\NFQ{r'} \subq \fltp, x \quad 
			 \op{\EPS[3] \subq \fltp,x} \quad 
			 x\notin\FV(U') \quad
			 \theta = [o'/x, \NFQ{p'}/\QSelf]
			}
			{
			C \ t_2: (\G[\flt]; \ty[p]{S} \ \EPS[1]) \carrow (\GP[\fltp]; \ty[r' \theta]{U'} \ (\EPS[4] \EFFSEQ \EPS[2] \EFFSEQ \EPS[3])\theta)
			}
			\vgap
			\infrule[c-app-$\QFresh$-2]
			{
			\GP[\fltp] \ts t_1: ((x: \ty[s']{T'}) \stackrel{\op{\EPS[4]}}{\to}  \ty[r']{{U'}})^{p'} \ \op{\EPS[2]}
			\quad
			C: (\G[\flt]; \ty[p]{{S}} \ \EPS[1]) \carrow (\GP[\fltp]; \ty[o]{{T'}}  \ \op{\EPS[3]}) \quad
            \QFresh \in s' \quad 
			\NFQ{s'} \subq \flt \\ 
			\QSelf \not\in s' \Rightarrow \qsat{\NFQ{\QF'}} \cap \qsat{\NFQ{o}} \subq \NFQ{s'} \qquad
			 \NFQ{r'} \subq \fltp, x \qquad 
			 \op{\EPS[3] \subq \fltp,x} \qquad 
			 x\notin\FV(U') \qquad
			 \theta = [o/x, \NFQ{p'}/\QSelf] 
			}
			{
			t_1 \ C: (\G[\flt]; \ty[p]{S} \ \op{\EPS[1]}) \carrow (\GP[\fltp]; \ty[r'\theta]{U'} \ \op{(\EPS[2] \EFFSEQ \EPS[3] \EFFSEQ \EPS[4])\theta})			
			}
			\vgap

		\end{minipage}%
	\end{mdframed}
	\caption{Selected context typing rules for the \maybelang{}- and \maybelange{} calculus.}
	\label{fig:context}
	\vspace{-3ex}
\end{figure}

 Unlike reduction contexts, contexts $C$ for reasoning about equivalence allow a ``hole'' to appear in any place.
We write $C: (\G[\flt]; \ty[p]{T}\ \op{\EPS}) \carrow (\GP[\fltp]; \ty[p']{T'}\ \op{\EPSPR})$ to 
mean that the context $C$ is a program of type 
$\ty[p']{T'} \ \op{\EPSPR}$ (closed under $\GP[\fltp]$) with a hole that can be filled 
with any program of type $\ty[p]{T}\ \op{\EPS}$ (closed under $\G[\flt]$).
The typing rules for well-typed contexts imply that if
$\G[\flt] \ts t : \ty[p]{T}\ \op{\EPS}$ 
and $C:  (\G[\flt]; \ty[p]{T}\ \op{\EPS})\carrow (\GP[\fltp]; \ty[p']{T'}\ \op{\EPSPR})$ hold, 
then $\GP[\fltp] \ts C[t] : \ty[p']{T'}\ \op{\EPSPR}$.
\figref{fig:context} shows the selected typing rules for well-typed contexts.

Two well-typed terms, $t_1$ and $t_2$, under typing context $\G[\flt]$, are \emph{contextually equivalent} if any occurrences of the first term in a closed term can be replaced by the second term without affecting the \emph{observable results} of reducing the program, which is formally defined as follows:
\begin{definition}[Contextual Equivalence]\label{def:standard_equiv} Term $t_1$ is \emph{contextually equivalent} to $t_2$,
  written $\G[\flt] \models t_1 \equiva t_2: \ty[p]{T}\ \op{\EPS}$, if $\G[\flt] \ts t_1: \ty[p]{T}\ \op{\EPS}$, and $\G[\flt] \ts t_2: \ty[p]{T}\ \op{\EPS}$, and 
 $\forall \, C: (\G[\flt];  \ty[p]{T}\ \op{\EPS}) \carrow ( \emptyset; \ty[\qbot]{\TUnit}\ \op{\PURE}). \
        C[t_1]\downarrow \ \Longleftrightarrow \ C[t_2] \downarrow$.
\end{definition}
\noindent We write $t\downarrow$ to mean term $t$ terminates, if
$\config{t}{\emptyset}{\emptyset} \, \eval \, \pconfig{v}{\sigma}$,
for some value $v$ and final store $\sigma$.

The above definition is standard \cite{DBLP:conf/popl/AhmedDR09} and defines a partial program equivalence.
However, since we focus on a total fragment of the systems here, program termination can not be used as an observer for program equivalence.
We thus rely on a refined definition of contextual equivalence using
Boolean contexts.
\begin{align*}
    \small
    \setlength{\abovedisplayskip}{0pt}
    \setlength{\belowdisplayskip}{0pt}
     & \forall \, C: (\G[\flt];  \ty[p]{T}\ \op{\EPS}) \carrow ( \emptyset; \ty[\qbot]{\ty{Bool}}\ \op{\PURE}). \, \exists \ \sigma, \sigma', v. \,\config{C[t_1]}{\emptyset}{\emptyset}\, \eval \, \pconfig{v}{\sigma} \; \wedge\; \config{C[t_2]}{\emptyset}{\emptyset} \, \eval \, \pconfig{v}{\sigma'}.
\end{align*}
That is to say, we consider two terms contextually equivalent if they yield the same answer value in all Boolean contexts.

\subsection{The World Model}
\begin{figure*}[t]\footnotesize
    \begin{mdframed}[innertopmargin=0pt, innerbottommargin=2pt, leftmargin=1pt, rightmargin=2pt]
       \judgement{Relational Worlds, Well-Defined Relations and Store Typing}{
       \BOX{\maybelang{}/\maybelange{}}} \vspace{-2ex}
        \begin{mathpar}
                \begin{array}{@{\hspace{-2ex}}r@{\hspace{1ex}}l@{\hspace{1ex}} l}
                 \Sigma &:= &\qbot \mid \Sigma, \ell: \L \\
                 \sigma : \Sigma  & \DEF & \DOM(\sigma) = \DOM(\Sigma) \, \land \, (\forall \ell \in \DOM(\Sigma). \lls{\Sigma}{\vallocs{\sigma(\ell)}} \subq \lls{\Sigma}{\Sigma(\ell)} \, \land \, (\forall \ell' \in \Sigma(\ell). \, \ell' < \ell))\\
                 \\
                 \STF{\W}{\L}{\L'}       & \DEF & (\L, \L') \subq \DOM(\W) \, \land \, (\forall \ell_1, \ell_2, \, (\ell_1, \ell_2) \in f \Rightarrow (\ell_1 \in \L \iff \ell_2 \in \L')) \\
            \BSTCP{\W}{\L}{\L'}{W'} & \DEF & \STF{\W}{\L}{\L'} \, \land \, \STF{\W'}{\L}{\L'} \, \land \, \STCPB{\W_1}{\L}{\W'_1} \, \land \, \STCPB{\W_2}{\L'}{\W'_2} \, \land    \\
                                  &   & (\forall \, \ell_1 \in \L, \, \ell_2 \in \L'. \, \W(\ell_1, \ell_2) \Rightarrow \W_{\val}(\ell_1, \ell_2) = \W'_{\val}(\ell_1, \ell_2)) \, \land     \\
                                  &      & (\forall \, \ell_1 \in \L, \, \ell_2 \in  \L'.\, (\ell_1, \ell_2) \in \W \leftrightarrow  (\ell_1, \ell_2) \in \W')                         \\
                                  
                 (\sigma_1, \sigma_2) : \W & \DEF & \sigma_1: \W_1 \, \land \, \sigma_2 : \W_2 \, \land \, (\forall \ell_1, \ell_2,\, \W(\ell_1, \ell_2) \Rightarrow (\sigma(\ell_1), \sigma(\ell_2)) \in \W_{\val}(\ell_1, \ell_2) \, \land \\
                                         &    & \STF{\W}{\lls{\W_1}{\sigma_1(\ell_1)}}{\lls{\W_2}{\sigma_2(\ell_2)}}  ) \\
                \end{array}
        \end{mathpar}
    \end{mdframed}
    \caption{Definitions of relational worlds, well-defined relations and store typing for binary logical relations.}
    \label{fig:worlds}
    \vspace{-3ex}
\end{figure*}%

 Following other prior works~\cite{DBLP:conf/icfp/ThamsborgB11,DBLP:conf/ppdp/BentonKBH07,ahmed2004semantics},
we apply Kripke logical relations to our systems.
Our logical relations are indexed by types and store layouts via \emph{worlds},
generalized store typings relating two stores.
This allows us to interpret $\TRef \ \ty[q]{T}$ as an allocated location that holds values of type $\ty[q]{T}$.
The invariant that all allocated locations hold well-typed values \wrt the world must hold in the pre-state and be re-established in the post-state of a computation.
The world may grow as more locations may be allocated.
It is important that this invariant must hold in future worlds, which is commonly referred to as \emph{monotonicity}.

\bfparagraph{World, Relational Worlds \& World Extension} As discussed in \secref{sec:unary_nested},
our store layout is specified by the logical interpretation of reachability qualifiers in a store typing, and is free of cycles.
The notion of world for our systems is defined as: %
\begin{definition}[World]\label{def:world} A world $\W$ is a tuple $(\Sigma_1, \Sigma_2, f, \bval)$, where
    \begin{itemize}
        \item $\Sigma_1$ and $\Sigma_2$ are store typings defined in \figref{fig:worlds},

        \item $ f \subseteq (\DOM(\Sigma_1) \times \DOM(\Sigma_2))$ is a partial bijection.
        \item $ \bval \subseteq (\DOM(\Sigma_1) \times \DOM(\Sigma_2) \times \val \times \val)$.
    \end{itemize}
\end{definition}
A world is meant to define relational stores.
The partial bijection captures the fact that a relation holds under permutation of store locations.
The notation $\bval$ maps from pairs of locations to semantic types. 
A semantic type is a set of values (\val).

If $\W = (\Sigma_1, \Sigma_2, f, \bval)$ is a world, we refer to its components as follows: \vspace{-5pt}

\[\small
\begin{array}{@{}r@{\hspace{1ex}}l@{\hspace{1ex}} l}
    \W(\ell_1, \ell_2) & = & \begin{cases}
                                 (\ell_1, \ell_2) \in f & \ell_1 \in \DOM(\Sigma_1) \text{ and } \ell_2 \in \DOM(\Sigma_2) \text{ and when defined} \\
                                 \bot              & \text{otherwise}
                             \end{cases} 
\end{array}%
\]
\vspace{-2ex}
\[\small
\begin{array}{@{}r@{\hspace{1ex}}l@{\quad\qquad}r@{\hspace{1ex}}l@{\quad\qquad}r@{\hspace{1ex}}l@{\quad\qquad}r@{\hspace{1ex}}l@{}}
    \DOM_1(\W)   & = \DOM(\Sigma_1) & \DOM_2(\W)   & = \DOM(\Sigma_2) & \DOM(\W)   & = (\DOM(\Sigma_1), \DOM(\Sigma_2)) \\
    \W_1 & = \Sigma_1 & \W_2 & = \Sigma_2 & \W_{\val} & = \bval
\end{array}
\]

If $\W$ and $\W'$ are worlds, such that
$
    \DOM_1(\W) \cap \DOM_1(\W') = \DOM_2(\W) \cap \DOM_2(\W') = \emptyset
$,
then $\W$ and $\W'$ are called disjoint, and we write $\W \extends \W'$ to mean extending $\W$ with a disjoint world $\W'$. %
Let $\sigma_1$ and $\sigma_2$ be two stores. We write $\WFRS{\sigma_1}{\sigma_2}{\W}$ to mean the stores are well-defined \wrt the world $\W$, which is formally defined in \figref{fig:worlds}.

Similar to the unary versions, we define relational worlds, written as $\BSTCP{\W}{\L}{\L'}{\W'}$
to mean that types and relations are preserved over two related worlds at a pair of locations $(\L, \L')$.
Note that the pair of locations can not be arbitrary. Consider related locations at world $\W$, \eg, $(\ell_1, \ell_2) \in f$.
If $\ell_1 \in \L$, but $\ell_2 \not\in \L'$, then the relation is broken, resulting in ill-defined relational worlds.
Thus, we use the predicate $\STFN$ (defined in \figref{fig:worlds}) to make sure the relation is defined in both directions.
In other words, $\ell_1$ must be related to some location, and it cannot be related to more than one location.
\figref{fig:worlds} summarizes the definitions of these notations.
We write $\STC{\W}{\W'}$ to mean $\BSTCP{\W}{\DOM_1(\W)}{\DOM_2(\W)}{\W'}$.
The definition of world extension satisfies reflexivity and transitivity.

\subsection{Binary Logical Relations for \maybelang{} and  \maybelange{}}
\label{sec:binary}
\begin{figure*}[t]\footnotesize
    \begin{mdframed}[innertopmargin=0pt, innerbottommargin=2pt, leftmargin=1pt, rightmargin=2pt]
       \judgement{Binary Value Interpretation of Types and Terms}{
       \BOX{\maybelang{}/\maybelange{}}} \vspace{-2ex}
        \begin{mathpar}
                \begin{array}{@{\hspace{-2ex}}r@{\hspace{1ex}}l@{\hspace{1ex}} l}
                    \BVT{\ty{Bool}}{\ty{Bool}}                                        & =             & \{ \BTC{\VE}{\W}{c}{c} \}                                                                                                                                                              \\

                    \BVT{\TRef \ \ty[q_1]{T_1}}{\TRef \ \ty[q_2]{T_2}}                    & =             & \{ \BTC{\VE}{\W}{\ell_1}{\ell_2} \mid  \HLMath{\W(\ell_1, \ell_2)} \, \land \, \NUM{1}{\HLMath{\STF{\W}{\{\ell_1\}}{\{\ell_2\}}}} \, \land (\W(\ell_i) = \varslocs{\VE_i}{q_i}  \, \land \,\\ 
                                                                                          &               &  \quad \lls{\W_i}{\W_i(\ell_i)} \subq \lls{\W_i}{\ell_i} \land (\forall \, \sigma_1', \sigma_2', \W', v_1, v_2. \, \WFRS{\sigma_1'}{\sigma_2'}{\W'} \land  \BSTCP{\W}{\lls{\W_1}{\W_1(\ell_1)}}{\lls{\W_2}{\W_2(\ell_2)}}{\W'} \land\\
                                                                                          &               &  \quad \lls{\W_i'}{\vallocs{v_i}} \subq \lls{\W_i'}{\W_i(\ell_i)} \, \land \, \NUM{2}{\HLMath{\STF{\W'}{\lls{\W_1'}{\vallocs{\sigma_1'(\ell_1)}}}{\lls{\W_2'}{\vallocs{\sigma_2'(\ell_2)}}} \Rightarrow}} \\ 
                                                                                          &               &  \quad \quad \quad \HLMath{((v_1, v_2) \in \W_{\val}(\ell_1, \ell_2) \iff \BTC{\VE}{\W'}{v_1}{v_2} \in \BVT{T_1}{T_2})})) \} \\
                    \\                                                                                                                                                                                                               
                    
                    \mathcal{V}\lbrack\!\lbrack(x:\ty[s_1]{T_1}) \LEFFS{\op{\EPS[1]}} \ty[r_1]{U_1}, \phantom{?}    & =             & \{ \BTC{\VE}{\W}{\cl{H_1}{\lambda x. t_1}{\QF_1}}{\cl{H_2}{\lambda x. t_2}{\QF_2}} \mid \NUM{3}{\HLMath{\STF{\W}{\llsvars{\W_1}{H_1}{\QF_1}}{\llsvars{\W_2}{H_2}{\QF_2}}}} \land \\
                          (x:\ty[s_2]{T_2}) \LEFFS{\op{\EPS[2]}} \ty[r_2]{U_2}\rbrack\!\rbrack   &                &    \,     \, (\forall v_1, v_2, \W', \sigma_1', \sigma_2'. \, \WFRS{\sigma_1'}{\sigma_2'}{\W'}  \, \land  \, \BSTCP{\W}{\llsvars{\W_1}{H_1'}{\QF_1}}{\llsvars{\W_2}{H_2'}{\QF_2}}{W'} \, \land \\
                                                                                          &               &   \quad \NUM{4}{\HLMath{\STF{\W'}{\DOM_1(\W')}{\DOM_2(\W')}}} \,  \land  \, \HLMath{\BTC{\VE}{\W'}{v_1}{v_2}\in \BVT{T_1}{T_2}}  \land \\
                                                                                          &               &   \quad \lls{\W_i'}{\vallocs{v_i}} \subq \llsvars{\W_i'}{\VE_i}{\NFQ{s_i}}  \ccup{\QSelf \in s_i}  \llsvars{H_i}{\W_i'}{\QF_i} \ccup{\QFresh \in s_i} \overline{\llsvars{H_i}{\W_i'}{\QF_i}} \, \Rightarrow \\
                                                                                          &               &  \quad \quad \quad   (\exists \, \W'', \sigma_1'', \sigma_2'', v_1', v_2'. \, \config{t_1}{H_1;(x, v_1)}{\sigma_1'} \, \eval \, \pconfig{v_1'}{\sigma_1''}  \land  \config{t_2}{H_2;(x, v_2)}{\sigma_2'} \, \eval \, \pconfig{v_2'} {\sigma_2''} \, \land \\
                                                                                          &               &  \quad \quad \quad \quad \quad \quad  \STC{\W'}{\W''} \, \land \, \HLMath{\STF{\W''}{\DOM_1(\W'')}{\DOM_2(\W'')}} \, \land \, \WFRS{\sigma_1''}{\sigma_2''}{\W''}  \, \land \\
                                                                                          &               &  \quad \quad \quad \quad \quad \quad  \HLMath{\BTC{\VE}{\W''}{v_1'}{v_2'} \in \BVT{U_1}{U_2}} \, \land                                     \\
                                                                                          &               &  \quad \quad \quad \quad \quad \quad  \lls{\W''_i}{\vallocs{v_i'}} \subq (\llsvars{\VE_i}{\W''_i}{\NFQ{r_i}\qminus x} \cap \llsvars{H_i}{\W''_i}{\QF_i}) \ccup{x \in r_i}  \lls{\W''_i}{\vallocs{v_i'}} \ccup{\QSelf \in r_i} \llsvars{H_i}{\W''_i}{\QF_i} \\
                                                                                          &               &  \quad \quad \quad \quad \quad \quad \quad \quad  \quad \quad \quad \ccup{\QFresh \in r_i} \frlocs{\W''_i}{\W'_i} \, \land \\
                                                                                          &               &  \quad \quad \quad \quad \quad \quad  \HLMathP{\DEPS{\sigma_i'}{\llsvars{\W_i'}{H_i}{\QF_i} \, \cup \, \lls{\W_i'}{\vallocs{v_i}}}{\sigma_i''}}   \qquad \HLMathG{\DEPS{\sigma_i'}{\llsvars{\W_i'}{H_i}{\EPS[i]\qminus x} \, \ccup{x \in \EPS[i]} \, \lls{\W_i'}{v_i} \ccup{\QSelf \in \EPS[i]} \, \llsvars{H'_i}{\W'_i}{\QF_i}}{\sigma_i''}}))  \,  \} \\

                    \\
                    \BMt{\flt}{\ty[p]{T}\ \op{\EPS}}                                           & =        & \{ \BTC{\VE}{\W}{t_1}{t_2} \mid \forall \, \sigma_1, \sigma_2. \WFRS{\sigma_1}{\sigma_2}{\W} \, \land \, \exists \, \W', \sigma_1', \sigma_2', v_1, v_2.\,  \config{t_1}{\VE_1}{\sigma_1} \, \eval \, \pconfig{v_1}{\sigma_1'} \, \land            \\
                                                                                          &               & \quad \config{t_2}{\VE_2}{\sigma_2} \, \eval \, \pconfig{v_2}{\sigma_2'} \, \land \, \STC{\W}{\W'} \, \land \, \HLMath{\STF{\W'}{\DOM_1(\W')}{\DOM_2(\W')}} \, \land \, \WFRS{\sigma_1'}{\sigma_2'}{\W'} \, \land                                  \\
                                                                                          &               & \quad \HLMath{\BTC{\VE}{\W'}{v_1}{v_2} \in \BVT{T}{T}} \, \land \, \lls{\W_i'}{\vallocs{v_i}} \subq \llsvars{\W_i}{\VE_i}{\flt \cap p} \ccup{\QFresh \in p} \frlocs{\W_i'}{\W_i} \, \land \,                                                                                                                  \\
                                                                                          &               & \quad \HLMathP{(\DEPS{\sigma_1}{\llsvars{\W_1}{\VE_1}{\flt}}{\sigma_1'} \, \land \, \DEPS{\sigma_2}{\llsvars{\W_2}{\VE_2}{\flt}}{\sigma_2'})} \,  \qquad \, \HLMathG{(\DEPS{\sigma_1}{\FX{\llsvars{\W_1}{\VE_1}{\flt \cap \EPS}}}{\sigma_1'} \, \land \, \DEPS{\sigma_2}{\FX{\llsvars{\W_2}{\VE_2}{\flt \cap \EPS}}}{\sigma_2'})} \}      \\
                    \\
                    \UG{\G[\flt]}                                                         & =             & \{(\VE, \W) \, \mid \, \DOM_i(\VE) = \DOM(\G) \, \land \, \flt \subq \DOM(\G) \, \land \, \WFGA{\sigma}{\VE}{\flt} \, \land \, \WFGB{\W}{\VE}{\flt} \}                                                                                                                   \\
                    
                    \WFGA{\W}{\VE}{\flt}                               & = & \forall \, x, T, p. \, \G(x) = \ty[p]{T} \Rightarrow  \, \WFP{p}{x}{\G} \, \land \, (\HLMath{x \in \flt \Rightarrow}  \HLMath{\BTC{\VE}{\W}{\VE_1(x)}{\VE_2(x)} \in \BVT{T}{T}}) \, \land \,                 \\
                                        &      & (\QFresh \not\in p \, \land \, \NFQ{p} \subq \flt \, \land \, x \in \flt \Rightarrow \lls{\W_i}{\vallocs{\VE_i(x)}} \subq \llsvars{\VE_i}{\W_i}{\NFQ{p}}) \\
                    \WFGB{\W}{\VE}{\flt}                               & = & \forall p, p'. \, \NFQ{p} \subq \flt \, \land \, \NFQ{p'}  \subq \flt \, \land \,  \qsat{\NFQ{p}} {\cap} \qsat{\NFQ{p'}} \subq \flt \Rightarrow \llsvars{\VE_i}{\W_i}{\NFQ{p}} \cap \llsvars{\VE_i}{\W_i}{\NFQ{p'}} \subq \llsvars{\VE_i}{\W_i}{\qsat{\NFQ{p}} {\cap} \qsat{\NFQ{p'}}}.
                    
                \end{array}
        \end{mathpar}
    \end{mdframed}
    \caption{The binary interpretation of types and terms, and semantic context interpretation. The highlighted formulas in \textcolor{teal}{teal} are assertions on well-formed relations, and related values.
    Formulas in \textcolor{pink}{pink} and in \textcolor{gray}{gray} are exclusively for \maybelang{} and \maybelange{}.}
        \label{fig:binary}
    \vspace{-5ex}
\end{figure*}

 This section presents the definition of binary logical relations for our systems, as shown in \figref{fig:binary}.
In the figure, we use the subscript $i$ in formulas to refer to both programs when the context is clear, where $i = 1$ or $2$. 
For example, in the value interpretation of reference types, we write $\W(\ell_i) = \varslocs{\VE_i}{q_i}$ to mean $\W(\ell_1) = \varslocs{\VE_1}{q_1} \, \land \, \W(\ell_2) = \varslocs{\VE_2}{q_2}$.

In the figure, the highlighted formulas in \textcolor{teal}{teal} are assertions on well-formed relations, and related values.
Formulas in \textcolor{pink}{pink} and in \textcolor{gray}{gray} are exclusively for \maybelang{} and \maybelange{} respectively.
As in their unary counterparts, \maybelang{} interprets observations as an approximation of write effects, 
and \maybelange{} establishes a stronger value preservation property with explicit write effect annotations.

\bfparagraph{The Binary Value and Term Interpretation}
As our systems use dependent types, we consider two related values at two equivalent but not identical types.
For example, in the proof of rule \textsc{$\beta$-equiv} in \secref{sec:beta}, we relate two types across substitutions.
Thus, our relational interpretation of type ${T}$ is written as $\BVT{T}{T}$, which is a set of tuples of form $\BTC{\VE}{\W}{v_1}{v_2}$,
where $\VE$ is a pair of relational value environments (defined in \figref{fig:binary}, bottom), $\W$ is a world,
and $v_1$ and $v_2$ are values.

A pair of Boolean values are related if they are both \text{true} or \text{false}.
A pair of locations are related if they store related values, in addition to their unary counterpart.
Two closure records are related
if applying their enclosing $\lambda$ terms with argument values that  are related at the domain type, 
their reduction results are related at the codomain type.
In addition to the unary counterpart, we use predicate $\STFN$ to ensure that 
the relations on the locations reachable from related values are well-defined, \eg, $\circledx{1}$, $\circled{2}$, $\circledx{3}$ and $\circledx{4}$ in \figref{fig:binary}.

Two terms are related if their reduction results are related at the given types.

\bfparagraph{Example: Re-ordering (part 1)}In the following example, we assume three distinct Boolean references $c_1$, $c_2$ and $c_3$. 
The two sub-programs $t_1$ and $t_2$ update references $c_1$ and $c_2$ respectively, 
and are typed in the $\maybelange$ system that tracks observable write effects:~\footnote{Conjunction can be encoded by our sequence terms.}
\[\small
\begin{array}{@{}r@{\hspace{1ex}}l@{\hspace{1ex}}l@{\hspace{5ex}}l@{\hspace{1ex}}l@{\hspace{1ex}}l@{\hspace{1ex}}l@{\hspace{1ex}}l@{\hspace{1ex}}l@{\hspace{1ex}}l@{\hspace{1ex}}l}
\multicolumn{3}{l}{\!\!\G[c_1, c_2, c_3] \ts c_1 : (\TRef ~\ty{Bool})^{c_1}} & & \multicolumn{3}{l}{\!\!\G[c_1, c_2, c_3] \ts c_2 : (\TRef ~\ty{Bool})^{c_2}} & & \multicolumn{3}{l}{\!\!\G[c_1, c_2, c_3] \ts c_3 : (\TRef ~\ty{Bool})^{c_3}}\\
t_1 & \DEF & c_1 \ := \ !c_3\ \wedge\ !c_1    & & \G[c_1, c_3] \ts t_1 : Bool\ \FX{c_1} \\
t_2 & \DEF & c_2 \ := \ !c_3\ \wedge\ !c_2    & & \G[c_2, c_3] \ts t_2 : Bool\ \FX{c_2}
\end{array}
\]
The goal is to show that $t_1;t_2$ and $t_2;t_1$ are logically equivalent:
$
\G[c_1, c_2, c_3] \models t_1;t_2 \equivlog t_2;t_1 : Bool \ \FX{c_1,c_2}.
$
Let $(\VE, \W) \in \UG{\G[c_1, c_2, c_3]}$.
By the definition of typing context interpretation in \figref{fig:binary}, 
we know that $(\VE, \W, \VE_1(c_i), \VE_2(c_i)) \in \BVT{\TRef\ \ty{Bool}}{\TRef \ \ty{Bool}}$, where $i \in \{1, 2, 3\}$.
By the definition of value interpretation of reference types, 
we know that there exists three pairs of locations (referred by the three references),
which are related in the world $\W$ (formula $\circledx{1}$). 
Thus, each pair of the related locations stores a pair of equivalent Boolean values (formula $\circledx{2}$ in \figref{fig:binary}),
where all the referents (boolean values) reach the empty location, satisfying the premise of formula $\circledx{2}$ trivially.
We can prove the equivalence by applying \thmref{thm:reordering} in \secref{sec:reorder}.

\bfparagraph{The Time Travelling Property} Similar to the unary versions, our relational reasoning can time travel back and forth from a past world to another (possible future) one 
as long as they agree on pairs of reachable locations from the given values: 
\begin{lemma}[Time Travelling for Binaries]\label{lem:valt_store_change_binary}
    If $(\sigma_1, \sigma_2): \W$, 
    and $(\VE, \W, v_1, v_2) \in \BVT{T}{T'}$,
    and $\BSTCP{\W}{\lls{\W_1}{\vallocs{v_1}}}{\lls{\W_2}{\vallocs{v_2}}}{\W'}$,
    then $(\VE, \W', v_1, v_2) \in \BVT{T}{T'}$.
\end{lemma}
\noindent This lemma allows us to prove the validity of a pair of values preserved \wrt a constructed world in the proof of rules 
\textsc{re-order-$\maybelang$}, \textsc{re-order-$\maybelange$} and 
\textsc{$\beta$-equiv} in \secref{sec:equiv}.

\subsection{The Compatibility Lemmas, Fundamental Theorem and Soundness}
\label{sec:binary_soundness}

\figref{fig:binary} (bottom) defines the interpretation of typing contexts. %
In the definition, $\VE$ ranges over a pair of relational value environments that are finite maps from variables $x$ to pairs of values $(v_1, v_2)$.
If $\VE(x) = (v_1, v_2)$, then $\VE_1(x)$ denotes $v_1$ and $\VE_2(x)$ denotes $v_2$.
We write $\DOM_1(\VE)$ and $\DOM_2(\VE)$ to mean the domain of the first and second value environment respectively.
The binary semantic typing judgment is defined as: $\G[\flt] \models t_1 \equivlog t_2: \ty[p]{T}\ \op{\EPS} \DEF \forall (\VE, \W) \in \UG{\G[\flt]}. (\VE, \W, t_1, t_2) \in \BMt{\flt}{\ty[p]{T}\ \op{\EPS}}$.

\begin{theorem}[Fundamental Property]\label{thm:binary_fp} If $\G[\flt] \ts t: \ty[p]{T}\ \op{\EPS}$, then $\G[\flt] \models t \equivlog t : \ty[p]{T}\ \op{\EPS}$.
\end{theorem}

\begin{lemma}[Congruency of Binary Logical Relations]\label{lem:direct_congruence} The binary logical relation is closed under well-typed program contexts,
    \ie, if $\G[\flt] \models t_1 \equivlog t_2: \ty[p]{T}\ \op{\EPS}$,
    and $C:(\G[\flt]; \ty[p]{T}\ \op{\EPS}) \carrow (\GP[\flt']; \ty[p']{T'}\ \op{\EPSPR})$, then $\GP[\flt'] \models C[t_1] \equivlog C[t_2]: \ty[p']{T'} \ \op{\EPSPR}$.
\end{lemma}

\begin{lemma}[Adequacy of the binary logical relations]\label{lem:direct_adequacy}
    The binary logical relation preserves termination, \ie, if $\emptyset \models t_1 \equivlog t_2: \ty[\qbot]{T}\ \op{\PURE}$,
    then $\exists \ \sigma, \sigma', v_1, v_2. \ \config{t_1}{\emptyset}{\emptyset} \, \eval \, \pconfig{v_1}{\sigma} \wedge \config{t_2}{\emptyset}{\emptyset} \, \eval \, \pconfig{v_2}{\sigma'}$.
\end{lemma}

\begin{theorem}[Soundness of Binary Logical Relations]\label{thm:binary_lr_soundness} The binary logical relation is sound w.r.t. contextually equivalence, \ie,
    if $\G[\flt] \ts t_1: \ty[p]{T}\ \op{\EPS}$ and $\G[\flt] \ts t_2: \ty[p]{T}\ \op{\EPS}$, then
    $\G[\flt] \models t_1 \equivlog t_2: \ty[p]{T}\ \op{\EPS}$ implies $\G[\flt] \models t_1 \equiva t_2: \ty[p]{T}\ \op{\EPS}$.
\end{theorem}

The formalization of the encapsulated computation lemma for the binary logical relations is omitted here. 
We will show its use in the proof of rule \textsc{$\beta$-equiv} (\ie, \lemref{lem:inv3}) in \secref{sec:beta}.

\bfparagraph{Key Properties} Now, we present key properties of our binary logical relations.
The following lemma allows us to tighten a given semantic typing context to the one specified by a sub-observation. 
It is used  in the proof of the compatibility lemma for a sequence term, and rules \textsc{re-order-$\maybelang$} and \textsc{re-order-$\maybelange$} in \secref{sec:equiv}.
\begin{lemma}[Semantic Typing Context Tightening]\label{lem:envt_tighten}
    Let $(\VE, \W) \in  \UG{\G[\flt]}$. For all $\flt'$, such that $\flt' \subq \flt$, we have $(\VE, \W) \in  \UG{\G[\flt']}$.
\end{lemma}

\noindent The following lemma allows us to obtain a subset of related locations that are well-defined: %
\begin{lemma}[Relation Tightening]\label{lem:envt_filter_deep} Let $(\VE, \W) \in \UG{\G[\flt]}$, 
    and $(\sigma_1, \sigma_2) : \W$,
    and $\flt' \subq \flt$,
    then $\STF{\W}{\llsvars{\VE_1}{\W_1}{\flt'}}{\llsvars{\VE_2}{\W_2}{\flt'}}$
\end{lemma}

The following lemma allows the reasoning to focus on a world $\W'$, where 
only types and the relation on locations $(\L_1', \L_2')$ are preserved.
\begin{lemma}[Relational Worlds Tightening]\label{lem:stchain_tighten} Let $\W$ and $\W'$ be two worlds, and $\L_1$, $\L_2$, $\L_1'$ and $\L_2'$ be four set of locations,
    such that  $\BSTCP{\W}{\L_1}{\L_2}{\W'}$, and $\L_1 \subq \L_1'$,
    and $\L_2 \subq \L_2'$, and $\STF{\W'}{\L_1'}{\L_2'}$.
    Then $\BSTCP{\W}{\L_1'}{\L_2'}{\W'}$.
\end{lemma}

The above tightening lemmas are commonly used together in the proofs of the compatibility lemmas (\eg, for the $\lambda$- and $\tref$ cases).
In those proofs, we establish relational worlds for the value interpretation of function types and reference types respectively 
by those tightening lemmas, where a  world extension is assumed from a given term interpretation. 
Also in the proofs of rules \textsc{re-order-$\maybelang$} and \textsc{re-order-$\maybelange$} in \secref{sec:reorder},
those tightening lemmas allow the reasoning to focus on locations specified by the pair of sub-observations.

\bfparagraph{Discussion: Example in \figref{fig:redis}} Now we provide informal justification for the safe use of unsafe features 
(\ie, the assertion statement) in the example of \figref{fig:redis},  
using the binary logical relation for $\maybelange$.
We reduce the problem by showing the program where the assertion is substituted with 
\code{!x == 1} is observationally equivalent to the one where the assertion is substituted with 
\code{true}.
This allows us to establish the safety of the assertion without extending $\maybelange$.
The reasoning is straightforward: by the fundamental property, 
the two programs right before the substituted statements are observationally equivalent,  
where their two local references are related. 
Moreover, since the effect of function \code{f} is variable $y$, which reaches locations separate from variable $x$.
Thus, the referent of $x$ remains unchanged, and hence, the two programs are observationally equivalent.

\section{Equational Theory}\label{sec:equiv}
\begin{figure*}[t]\footnotesize
    \begin{mdframed}[innertopmargin=0pt, innerbottommargin=2pt, leftmargin=0pt, rightmargin=2pt]
        \judgement{Equational Rules}{\BOX{\maybelang{}/\maybelange{}}}\vspace{-2ex}
        \begin{mathpar}
                \inferrule*[left=re-order-$\maybelang{}$]
                {\G[\flt_1] \ts t_1 : \ty[p_1]{\ty{Bool}} \ \quad
                 \G[\flt_2] \ts t_2 : \ty[p_2]{\ty{Bool}} \ \quad
                 \flt_1 \subq \flt \quad
                 \flt_2 \subq \flt \quad
                 \qsat{\flt_1}\,{\cap}\, \qsat{\flt_2} = \qbot 
                }
                {\G[\flt] \models t_1;t_2 \equivlog t_2;t_1 : \ty[\qbot]{\ty{Bool}}} 
                \\
                \inferrule*[left=re-order-$\maybelange{}$]
                {\G[\flt_1] \ts t_1 : \ty[p_1]{\ty{Bool}} \ \EPS[1]\quad
                 \G[\flt_2] \ts t_2 : \ty[p_2]{\ty{Bool}} \ \EPS[2]\quad
                 \flt_1 \subq \flt \quad
                 \flt_2 \subq \flt \quad
                 \qsat{\flt_1}\,{\cap}\, \qsat{\EPS[2]} = \qbot \quad 
                 \qsat{\flt_2}\,{\cap}\, \qsat{\EPS[1]} = \qbot
                }
                {\G[\flt] \models t_1;t_2 \equivlog t_2;t_1 : \ty[\qbot]{\ty{Bool}} \ \EPS[1]\EFFSEQ\EPS[2]}
                \\
                \inferrule*[left=$\beta$-equiv]{
                    (\G, \, x:  \ty[{\flt\cap(s[\QF/\QSelf]) \ccup{\QFresh \in s} \QFresh}]{T})^{\QF',x} \ts t_2 : \ty[{r[\QF'/\QSelf]}]{U}\ \EPS{\FX{[\QF'/\QSelf]}} \\
                    \G[\qbot] \ts t_1: \ty[\qbot]{T} \ \PURE \\\\
                    \QF' = \flt\cap \QF \\
                    \NFQ{s} \subq \QF'\\
                    \theta = [\qbot/x, \QF/\QSelf]                }
                {\G[\flt] \models (\lambda x.t_2)^{\QF'}~t_1 \equivlog t_2[t_1/x]: \ty[r\theta]{U} \ \EPS \FX{\theta} }
        \end{mathpar}
    \end{mdframed}
    \caption{The re-ordering and $\beta$-inlining rules for \maybelang{}/\maybelange{}.}
    \label{fig:equiv}
    \vspace{-3ex}
\end{figure*} This section shows applications of our binary logical relations by proving two re-ordering  and  $\beta$-equivalence rules.
We outline the key ideas of the proofs, primarily by presenting key lemmas and suitable worlds definitions used in the proof.
Full proofs can be found in the Rocq artifacts.

\subsection{Re-ordering}\label{sec:reorder}
\usetikzlibrary{arrows.meta, positioning, patterns}
\usetikzlibrary{decorations.pathmorphing}

\tikzset{fontscale/.style = {font=small}
}
\begin{figure}[t]
\begin{minipage}{0.50\linewidth}
\begin{tikzpicture}[every node/.style={font=\small}]
    \node (PS) at (-1,1.5) {\texttt{\textbf{Pre-State:} $\W = (\Sigma_1, \Sigma_2, f, \bval)$}};
    \node (ST) at (0,1) {\texttt{$(\sigma_1, \sigma_2): \W$}};
    \node (V1) at (0,0) {
       \begin{tikzpicture}[scale=0.4]
	                       \begin{scope} %
                            \draw[draw = ForestGreen] (-0.8, 1) circle (1);
                            \draw[draw = ForestGreen] (0.8, 1) circle (1);
                            \draw[draw = red] (-1, 1) circle (0.4);
                            \draw[draw = red] (1, 1) circle (0.4);
                            \node at (-2.5,1) {\textcolor{ForestGreen}{$\flt_1$}};
                            \node at (2.5,1) {\textcolor{ForestGreen}{$\flt_2$}};
                          \end{scope}
                       \end{tikzpicture}
    };
    \node (E1) at (-1.8, -1.5) { $t_1, \VE_1, \sigma_1 \eval  v_a, \sigma_1'$ };
    \node (E2) at (1.8, -1.5) { $t_2, \VE_2, \sigma_2  \eval v_b, \sigma_2'$ };
    \node (E21) at (-1.8, -3) { $t_2, \VE_1, \sigma_1' \eval v_b', \sigma_1''$ };
    \node (E11) at (1.8, -3) { $t_1, \VE_2, \sigma_2' \eval v_a', \sigma_2''$ };
    \node (V4) at (0, -4.5) {
      \begin{tikzpicture}[scale=0.4]
	\begin{scope} [pattern color = ForestGreen, pattern = north east lines]
    \filldraw[draw = ForestGreen] (-0.8, 1) circle (1);
    \filldraw[draw = ForestGreen] (0.8, 1) circle (1);
    \end{scope}
    \begin{scope} %
    \draw[fill = red, draw = red] (-1, 1) circle (0.4);
    \draw[fill = red, draw = red] (1, 1) circle (0.4);
    \node at (-2.5 ,1) {\textcolor{ForestGreen}{$\flt_1$}};
    \node at (2.5,1) {\textcolor{ForestGreen}{$\flt_2$}};
    \end{scope}
\end{tikzpicture}
    };
    \node (XP1) at (-2, -1) {\texttt{$\mathbf{t_1;t_2}$}};
    \node (XP1) at (2, -1) {\texttt{$\mathbf{t_2;t_1}$}};
    \node (P1) at (-2.0, -0.6) {\texttt{\textbf{$\prog{1}$:}}};
    \node (P2) at (2.0, -0.6) {\texttt{\textbf{$\prog{2}$:}}};
    \node (POS) at (-2.2, -4.5) {\texttt{\textbf{Post-State:}}};
    \node (ST2) at (0, -5.5) {$(\sigma_1'', \sigma_2''): \W^R$};
    \draw (V1) edge [->, >= angle 90] node[left] {} (E1);
    \draw (V1) edge [->] node[left] {} (E2);
    \draw (E21) edge [->, >= angle 90] node[left] {} (V4);
    \draw (E11) edge [->, >= angle 90] node[right] {} (V4);
    \draw (E1) edge [->, >= angle 90] node[left, xshift=2.6pt,] {
      \begin{tikzpicture}[scale=0.3]
	\begin{scope} [pattern color = ForestGreen, pattern = north east lines]
    \filldraw[draw = ForestGreen] (-0.8, 1) circle (1);
    \end{scope}
    \begin{scope} %
    \draw[draw = ForestGreen] (0.8, 1) circle (1);
    \draw[fill = red, draw = red] (-1, 1) circle (0.4);
    \draw[draw = red] (1, 1) circle (0.4);
    \node at (-1.8 ,1) {\textcolor{ForestGreen}{$\flt_1$}};
    \node at (3.5,1) {\textcolor{ForestGreen}{$\flt_2$}};
    \end{scope}
\end{tikzpicture}
    } (E21);
    \draw (E2) edge [->, >= angle 90] node[right, xshift=-4pt] {
    \begin{tikzpicture}[scale=0.3]
	\begin{scope} [pattern color = ForestGreen, pattern = north east lines]
    \filldraw[draw = ForestGreen] (0.8, 1) circle (1);
    \end{scope}
    \begin{scope} %
    \draw[draw = ForestGreen] (-0.8, 1) circle (1);
    \draw[fill = red, draw = red] (1, 1) circle (0.4);
    \draw[draw = red] (-1, 1) circle (0.4);
    \node at (-3.4, 1) {\textcolor{ForestGreen}{$\flt_1$}};
    \node at (1.9,1) {\textcolor{ForestGreen}{$\flt_2$}};
    \end{scope}
\end{tikzpicture}
    } (E11);
    \draw (E1) edge [<->, ForestGreen, line join=round, decorate, decoration={zigzag, segment length=4, amplitude=.9, pre=lineto, post=lineto, pre length = 2pt, post length=2pt}] node[pos=0.1, xshift=-3pt, yshift=-8pt] {\textcolor{purple}{$O_1:~\equiv_{\flt_1}$}} (E11);
    \draw (E2) edge [<->, ForestGreen, line join=round, decorate, decoration={zigzag, segment length=4, amplitude=.9, pre=lineto, post=lineto, pre length = 2pt, post length=2pt}] node[pos=0.1, xshift=6pt, yshift=-8pt] {\textcolor{purple}{$O_2:~ \equiv_{\flt_2}$}} (E21);
    \draw (POS) edge [->, draw=none] node[left] {} (V4);
    \draw (P1) edge [->, draw=none] node[left] {} (V1);
\end{tikzpicture}
\end{minipage}%
\begin{minipage}{0.54\linewidth}
  \begin{tikzpicture}[every node/.style={font=\small}]
    \node (T) at (-1.7, 4.8) {\textcolor{ForestGreen}{\texttt{\textbf{Store locations:}}}};
    \node (V) at (-2,3.5){
      \begin{tikzpicture}[scale=0.5]
        \begin{scope} %
          \draw[draw = ForestGreen] (0, 1) circle (2);
          \draw[draw = ForestGreen] (-0.8, 1) circle (1);
          \draw[draw = ForestGreen] (0.8, 1) circle (1);
          \draw[draw = red] (-1, 1) circle (0.4);
          \draw[draw = red] (1, 1) circle (0.4);
          \node at (-1,2.2) {\textcolor{ForestGreen}{$\flt_1$}};
          \node at (1,2.2) {\textcolor{ForestGreen}{$\flt_2$}};
          \node at (-1,0.3) {\textcolor{red}{$\EPS_1$}};
          \node at (1,0.3) {\textcolor{red}{$\EPS_2$}};
          \end{scope}
      \end{tikzpicture}
    };
    \node (P1) at (1, 4.3) {\textcolor{ForestGreen}{\texttt{$\flt_1, \flt_2$: readable by $t_1$, $t_2$}}};
    \node (P2) at (1, 3.8) {\textcolor{red}{\texttt{$\EPS_1, \EPS_2$: writable by $t_1$, $t_2$}}};
    \node (P3) at (1.2, 3.3) {
    \begin{tikzpicture}[scale = 0.3]
      \begin{scope} [pattern color = ForestGreen, pattern = north east lines]
      \filldraw[draw=ForestGreen](-1.1,-2) circle (0.35);
      \end{scope}
    \end{tikzpicture}%
    {\texttt{ and }}
    \begin{tikzpicture}[scale = 0.3]
      \begin{scope} [pattern color =red, pattern = north east lines]
      \filldraw[draw=red](-0.5,-2) circle (0.35);
      \end{scope}
    \end{tikzpicture}%
    {\texttt{ to indicate \textcolor{ForestGreen}{reads}}}
    };
    \node (P4) at (1.2, 2.8) {\textcolor{black}{\texttt{and \textcolor{red}{writes} have taken place}}};
    \node (O) at (-1.8,2.0) {\texttt{\textbf{Obligations:}}};
    \node (ST1) at (0.5,1.8) {\texttt{$(\sigma_1, \sigma_2):\W$}};
    \node (ST2) at (-1.2,0.3) {\texttt{$(\sigma_1, \sigma_2'):\W^1$}};
    \node (ST3) at (2.2,0.3) {\texttt{$(\sigma_1', \sigma_2):\W^2$}};
    \node (ST4) at (-1.5, -0.8) {\texttt{$(\sigma_1', \sigma_2''):\W^3$}};
    \node (ST42) at (-1.5,-1.2) {\texttt{$\W^3 = (\Sigma_1''', \Sigma_2'', f^3, \bval^3)$}};
    \node (ST5) at (2.1,-0.8) {\texttt{$(\sigma_1'', \sigma_2'):\W^4$}};
    \node (ST52) at (2.4, -1.2) {\texttt{$\W^4=(\Sigma_1'', \Sigma_2''', f^4, \bval^4)$}};
    \node (ST6) at (0.5,-2.3) {\texttt{$(\sigma_1'', \sigma_2''):\W^R$}};
    \node[rotate=270] (E1) at (-1, -0.3) {\textcolor{purple}{$\sqsubseteq$}};
    \node (E11) at (-1.5, -0.3){\textcolor{purple}{$O_5: $}};
    \node[rotate=270] (E2) at (2.5, -0.3) {\textcolor{purple}{$\sqsubseteq$}};
    \node (E22) at (2.0, -0.3){\textcolor{purple}{$O_6: $}};
    \draw (ST1) edge [->, ForestGreen] node[left] {\textcolor{purple}{$O_3: ~ \equiv_{\flt_1}$}} (ST2);
    \draw (ST1) edge [->, ForestGreen] node[right] {\textcolor{purple}{$O_4: ~ \equiv_{\flt_2}$}} (ST3);
    \node (O7) at (0.9, -1.5) {\textcolor{purple}{$O_7:$ }};
    \draw (ST1) edge [->, ForestGreen] node[pos=0.8,right,sloped, allow upside down, xshift=1pt, yshift=24pt] {\textcolor{purple}{$\sqsubseteq$}} (ST6);
    \draw (ST42) edge [dashed, ->, ForestGreen] node[right] {} (ST6);
    \draw (ST52) edge [dashed, ->, ForestGreen] node[right] {} (ST6);
 \end{tikzpicture}
\end{minipage}%
\\
\begin{minipage}{\linewidth}
\begin{tikzpicture}[every node/.style={font=\small}]
\node (N) at (0, 0) {$\qquad \qquad \qquad \qquad \W^R = (\Sigma_1'', \Sigma_2'', f_{\overline{\EPS[1], \EPS[2]}};f^{4}_{\EPS[2]};f^{4}_{\fr};f^{3}_{\EPS[1]};f^{3}_{\fr}; \bval_{\overline{\EPS[1],\EPS[2]}}; \bval^4_{\EPS[2]};\bval^4_{\fr};\bval^3_{\EPS[1]};\bval^3_{\fr})$};
\end{tikzpicture}
\end{minipage}
\caption{Illustrated proof that $\prog{1} = t_1;t_2$ and $\prog{2} = t_2;t_1$ are logically equivalent at world $\W$. The key obligations are (1) establishing equivalence of subterms $(t_1, t_1)$ and $(t_2, t_2)$ 
at worlds $\W^{1}$ and $\W^{2}$ respectively (\textcolor{purple}{$O_1 ~ \& ~ O_2$}); and (2) showing 
the final store pairs $(\sigma_1'', \sigma_2'')$ are well-formed \wrt world $\W^R$, an extension of the inital one $\W$ (\textcolor{purple}{$O_7$}).}
\label{fig:reorder}
\vspace{-3ex}
\end{figure} Rules \textsc{re-order-$\maybelang{}$} and \textsc{re-order-\maybelange{}} in \figref{fig:equiv} permit two computations to commute.
The former is applied to $\maybelang{}$, and requires the observations of the two terms to be separate; 
the latter is applied to $\maybelange{}$, and demands the observation of one term to be separate from the effects of the other.
Here we outline the proof of rule \textsc{re-order-\maybelange{}}. Rule \textsc{re-order-$\maybelang{}$} can be proved analogously.

In \figref{fig:reorder}, the left side shows the definitions of the initial world $\W$ (top)
and final world $\W^R$ (bottom), 
the semantics of programs $\prog{1} = t_1;t_2$ and $\prog{2} = t_2;t_1$ (following the two vertical arrows),
and the major proof objections (\textcolor{purple}{$O_1-O_7$}).
The bottom right shows world relations.
We write $(\sigma_1, \sigma_2) :\W \equiv_{\flt_1} (\sigma_1, \sigma_2') : \W^1$ to mean 
that stores $(\sigma_1, \sigma_2)$ and $(\sigma_1, \sigma_2')$ are well-formed \wrt worlds $\W$ and $\W^1$ respectively,
and worlds $\W^1$ and $\W$ are related \wrt to $\flt_1$.
The notation $(\sigma_1, \sigma_2'):\W^1 \sqsubseteq (\sigma_1', \sigma_2''):\W^3$ 
is defined similarly, but for world extension.
We need to show that programs $\prog{1}$ and $\prog{2}$ are logically equivalent at world $\W$,  
where $(\VE, \W) \in \UG{\G[\flt]}$.

As shown at the top right of \figref{fig:reorder},
the entire set of store locations consists of three parts:
those not observed by either the execution of term $t_1$ or $t_2$, \ie, \textcolor{ForestGreen}{$\overline{\flt_1, \flt_2}$},
and those needed by the execution of term $t_1$ and $t_2$ respectively, \ie, \textcolor{ForestGreen}{$\flt_1$} and \textcolor{ForestGreen}{$\flt_2$}.
In addition, $t_1$ and $t_2$ exclusively modify locations in $\EPS[1]$ and $\EPS[2]$, respectively.

\bfparagraph{Key Idea: Term Equivalence Preservation (Obligations 1 \& 2)}

From the fundamental property (\thmref{thm:binary_fp}), we know that $(t_2, t_2)$/$(t_1, t_1)$ are logically equivalent at world $\W$.
Observing the semantics in \figref{fig:reorder}, term $t_1$/$t_2$ is evaluated first in $\prog{1}$/$\prog{2}$. 
The key is to show that after evaluating $t_1$/$t_2$, the equivalence $(t_2, t_2)$/$(t_1, t_1)$ still holds at some world.
Informally, the observation $\flt_2$/ $\flt_1$ specifies what is needed for maintaining the equivalence
of $(t_2, t_2)$/$(t_1, t_1)$ at world $\W$ respectively.
As the observation of one term is separate from the effects of the other, 
their equivalences are preserved.
\begin{lemma}[Term Equivalence Preservation]\label{lem:inv1}
    If $\G[\flt] \ts t : \ty[q]{T} \ \EPS $,
    and $(\VE, \W) \in \UG{\G^{\flt}}$,
    and $\W = (\Sigma_1, \Sigma_2, f, \bval)$,
    and $\WFRS{\sigma_1}{\sigma_2}{\W}$,
    and $\forall \ell_1, \ell_2. \,  \W(\ell_1, \ell_2) \Rightarrow \ell_1 \not\in \L$
    and $\DEPS{\sigma_1}{\L}{\sigma_1'}$,
    and $\Sigma_1 \sqsubseteq \Sigma_1'$,
    and $\sigma_1':\Sigma_1'$,
    and let $\W' = (\Sigma_1', \Sigma_2, f, \bval)$,
    then $\BTC{\VE}{\W'}{t}{t} \in  \BMt{\flt}{\ty[q]{T}\ \EPS}$.
\end{lemma}

We strengthen the conclusion of the lemma by saying that $(t_2, t_2)$  are equivalent at a world, such that 
their relatedness is irrelevant to anything \emph{not reachable} from $t_2$'s observation $\flt_2$,
or the fresh locations during the course of evaluating $t_1$ in $\prog{1}$. 

\begin{lemma}[Term Equivalence Preservation $\&$ Reachability]\label{lem:inv2}
    If $\G[\flt] \ts t : \ty[q]{T} \ \EPS $,
    and $(\VE, \W) \in \UG{\G^{\flt}}$,
    and $\W = (\Sigma_1, \Sigma_2, f, \bval)$,
    and $\WFRS{\sigma_1}{\sigma_2}{\W}$,
    and $\DEPS{\sigma_1}{\L}{\sigma_1'}$,
    and $\Sigma_1 \sqsubseteq \Sigma_1'$,
    and $\sigma_1':\Sigma_1'$,
    and $\L \cap \llsvars{\VE_1}{\W_1}{\flt}$ = \qbot,
    and let $\W' = (\Sigma_1', \Sigma_2, f', \bval')$,
    and $f' = \{(\ell_1, \ell_2) \in f \mid  \ell_1 \not\in \L \land (\overline{\llsvars{\VE_1}{\W_1}{\flt}} \cup \frlocs{\Sigma'}{\Sigma}) \cap \lls{\W_1}{\ell_1} = \qbot\}$,
    and $\bval' = \{(\ell_1, \ell_2, v_1, v_2) \in \bval \mid  \ell_1 \not\in L \land  (\overline{\llsvars{\VE_1}{\W_1}{\flt}} \cup \frlocs{\Sigma'}{\Sigma}) \cap \lls{\W_1}{\ell_1} = \qbot\}$,
    then $\BTC{\VE}{\W'}{t}{t} \in  \BMt{\flt}{\ty[q]{T}\ \EPS}$.
\end{lemma}

\bfparagraph{Well-Formed Relational Worlds (Obligations 3 \& 4)} %
We know $\W \sqsubseteq_{\flt_2} \W^2$ and $\W \sqsubseteq_{\flt_1} \W^1$ by the definition of relational worlds in \figref{fig:binary}.

\bfparagraph{Well-Formed World Extensions (Obligations 5 \& 6)} These can be discharged by the definition of term interpretation in \figref{fig:binary}.

\bfparagraph{Well-Formed Final World (Obligation 7)} We discuss the key technical points that show world $\W^{R}$ is a valid extension of the initial world $\W$, 
and the pair of final stores are well-formed \wrt $\W^{R}$. 

We first introduce the notations $f_{\overline{\EPS[1], \EPS[2]}}$, $f^{4}_{\EPS[2]}$, and $f^{3}_{\EPS[1]}$ shown in \figref{fig:reorder}.
Without loss of generalization, we assume $(\VE, \W) \in \UG{\G[\flt]}$, and $\W = (\Sigma_1, \Sigma_2, f, \bval)$, and the effect $\EPS$ is closed at $\G$.
The notation $f_{\EPS}$ denotes that weakening the relation to the part only covered by the interpreation of $\EPS$,
\ie, $f_{\EPS} = \{(\ell_1, \ell_2). \ (\ell_1, \ell_2) \in f \land \ell_1 \in \llsvars{\VE_1}{\W_1}{\flt \cap \EPS} \land \ell_2 \in  \llsvars{\VE_2}{\W_2}{\flt \cap \EPS} \}$.
For example, $\W_{\overline{\EPS[1];\EPS[2]}}$ denotes the relations that are unchanged during the computation. %

Now we show that $f^3_{\EPS[1]} = f_{\EPS[1]}$ and $f^4_{\EPS[2]} = f_{\EPS[2]}$. 
This is established by $\BSTCPS{\W}{\flt_2}{\W^2}$ and $\STC{\W^2}{\W^4}$,
\ie,  $\W^{1}$ preserves the relation from $f_{\flt_2}$,
and $\W^{4}$ is a valid extension of $\W^{2}$.
The notation $f^{3}_{\fr}$ means the related locations that are fresh at world $\W^{3}$,
\ie, $\forall \ell_1, \ell_2, f^3(\ell_1, \ell_2) \land \ell_1 \in \fr(\W^{3}_1, \W^{1}_1) \land \ell_2 \in \fr(\W^{3}_2, \W^{1}_2)$.
The meaning of the notation $f^{4}_{\fr}$ is defined similarly.
Now, we can see that we construct the relation in a mutually disjoint way, \ie, $\W^{R}_f$ and $\W^{R}_{\bval}$ are well-defined, and  world $\W^{R}$ is a valid extension of $\W$. 
From \figref{fig:reorder}, we know that $\sigma_1'' : \W^4_1$ and $\sigma_2'': \W^{3}_2$.
Thus, the pair of final stores are well-formed \wrt world $\W_R$.

\begin{theorem}\label{thm:reordering} Rules \textsc{re-ordering-$\maybelang{}$} and \textsc{re-ordering-$\maybelange{}$}  in \figref{fig:equiv} are sound.
\end{theorem}

\bfparagraph{Example: Re-ordering (part 2)} 
The re-ordering example in \secref{sec:binary} can now be proved directly by applying \thmref{thm:reordering},
where $\flt_1 = c_1, c_3$, $\flt_2 = c_2, c_3$, $\flt = c_1, c_2, c_3$, $\EPS[1] = c_1$, and $\EPS[2] = c_2$,
and the side conditions of rule \textsc{re-order-$\maybelange$} are all satisfied.

\bfparagraph{Discussion: Stronger than Parallel Reduction} 
Prior works on reachability types~\cite{DBLP:journals/pacmpl/BaoWBJHR21,wei2023polymorphic} do not consider any notion of equivalance.
The  ``progress and preservation in parallel reduction`` property (Corollary 4.9) in prior work~\cite{wei2023polymorphic}
only shows that reduction can proceed in parallel for top-level programs, 
but does not show that commuting two reduction sequences yields equivalent results in arbitrary typed program contexts. 
Thus, the corollary in their work cannot be used to justify our re-ordering rules.

\subsection{$\beta$-Equivalence}\label{sec:beta}
Rule \textsc{$\beta$-equiv} in \figref{fig:equiv} permits replacing a function call site $t_1$ with the body of the called function,
provided that $t_1$ is observably pure. 
Let $\prog{1}$ be $(\lambda x. t_2)^{q'} t_1$, and $\prog{2}$ be $t_2[t_1/x]$. 
We need to show that programs $\prog{1}$ and $\prog{2}$ are logically equivalent at world $\W$, where 
$(\VE, \W) \in \UG{\G[\flt]}$, $\W = (\Sigma_1, \Sigma_2, f, \bval)$, and $(\sigma_1, \sigma_2):\W$.
Recall the program semantics in the followings:
\[\footnotesize
\begin{array}{l l l l l l l l l l l}
    {\prog{1}: } & \config{\tabs{\lambda x.t_2}{q'}}{\VE_1}{\sigma_1} ~ \eval ~ \pconfig{\cl{\VE_1}{\lambda x.t_2}{q'}}{\sigma_1'}  %
                 &   t_1, \VE_1, \sigma_1'  ~ \eval ~  v_a, \sigma_1''   %
                 &   t_2, \VE_1;(x:v_a), \sigma_1'' ~ \eval ~  {v_2, \textcolor{blue}{\sigma_1'''}}  \\ %
    {\prog{2}: } & t_2[t_1/x], \VE_2, \sigma_2 ~ \eval ~  v_2', \textcolor{blue}{\sigma_2'}  \\  %
\end{array}
\]

Observing the above semantics of $\prog{1}$,
we know that the $\lambda$-term gets evaluated first with final store $\sigma_1'$, such that $\sigma_1': \Sigma_1'$.
Then $t_1$ gets evaluated to value $v_a$ with final store $\sigma_1''$, such that $\sigma_1'': \Sigma_1''$.
Let $\W^1$ be $(\Sigma_1', \Sigma, f, \bval)$, and $\W^2$ be $(\Sigma_1'', \Sigma, f, \bval)$.
By the definition of world extension in \figref{fig:binary}, we know that $\STC{\W}{\W^1}$, and $\STC{\W^1}{\W^2}$.
Thus, we know $\STC{\W}{\W^2}$ by the transitivity of world extension.

Now, our goal is to prove $(\VE_1;(x:v_a), \VE_2, W^2, t_2, t_2[t_1/x]) \in \BMt{\flt}{\ty[r\theta]{U} \ \EPS \FX{\theta}}$, where $\theta = [\qbot/x, q'/\QSelf]$.
In $\prog{2}$, term $t_1$ may get evaluated any time during the course of evaluating $t_2[t_1/x]$,
where $t_2$'s effects may already occur.
Note term $t_1$'s observation is the empty set, \ie, establishing $(t_1, t_1)$ equivalence does not need any related locations in the pre-stores.
We formalize this observation in the following lemma, which is an application of encapsulated computation lemma for binaries: 
\begin{lemma}\label{lem:inv3} If\ \ $\G[\qbot] \ts t: \ty[\qbot]{T} \ \PURE$,
    and $(\VE, \W) \in \UG{\G[\qbot]}$, 
    and $\W = (\Sigma_1, \Sigma_2, f, \bval)$,
    then\\ $\BTC{\VE}{(\Sigma_1, \Sigma_2, \qbot, \qbot)}{t}{t} \in  \BMt{\qbot}{\ty[\qbot]{T}\ \PURE}$.
\end{lemma}
From the lemma above, we know that the resulting values $(v_1, v_2)$ reach the empty set of locations, \ie, $\lls{\W^c_i}{\vallocs{v_i}}$, where $\W^c$ is the resulting 
world and $i =1$ or $2$.
Now we know that $(v_1, v_2)$ are logically equivalent at any world by \lemref{lem:valt_store_change_binary}.
The following shows the top-level semantic weakening lemma that captures the fact that 
(1) $t_1$ may get evaluated any time during the course of evaluating $t_2[x/t_1]$; 
(2) $v_a$ and the resulting value from $t_1$ in $\prog{2}$ are logically equivalent at any given world.

\begin{lemma}[Semantic Weakening]\label{lem:st_weaken1} If
    $\G[\qbot] \ts t_1: \ty[\qbot]{T} \ \PURE$,     
    and $\W^{1'} = (\Sigma_1', \Sigma_2, \qbot, \qbot)$,
    and $(\VE, \W^{1'}) \in \UG{\G[\qbot]}$, 
    and $(\sigma_1', \sigma_2): \W^{1'}$,  
    and there exists $v_a$, $\sigma_1''$ and $\Sigma_1''$, such that 
    \begin{itemize}[itemsep=1pt,topsep=0pt]
    \item $t_1, H_1, \sigma_1' ~ \eval ~v_a, \sigma_1''$, and $\vallocs{v_a} \subq \qbot$, and $\DEPS{\sigma_1'}{\PURE}{\sigma_1''}$, and $(\sigma_1'': \Sigma_1'')$, and $\STC{\Sigma_1'}{\Sigma_1''}$, and
    \end{itemize}
    ( for all $H_2'$, $\sigma^b_2$, $\Sigma^b_2$, $\L$.\ $\STCPB{\Sigma_2}{\L}{\Sigma_2^b}$, 
    and  $(\sigma_2^b: \Sigma_2^b)$ implies there exists $v_b'$, $\sigma^c_2$, $\Sigma^c_2$, such that
      \begin{itemize}[itemsep=0em,topsep=0pt]
        \item $ t_1, \VE_2;H_2', \sigma^{b}_2 ~ \eval ~v_b', \sigma^{c}_2$,  and  
        $\DEPS{\sigma^{b}_2}{\PURE}{\sigma^{c}_2}$, and  $\STC{\Sigma^b_2}{\Sigma^c_2}$, and 
      \end{itemize} 
    $W^c = (\Sigma_1'', \Sigma^c_2, \qbot, \qbot)$, and $(\sigma_1'', \sigma_2^{c}): \W^c$, 
             and $((\VE_1, \VE_2;H_2'), \W^c, v_a, v_b') \in \BVT{T}{T}$, and 
             $\lls{\Sigma_1''}{\vallocs{v_a}} \subq \qbot$, and  
             $\lls{\Sigma^{c}_2}{\vallocs{v_b'}} \subq \qbot$).
\end{lemma}

The following shows the top-level semantic substitution lemma, which uses the conclusion of the semantic weakening lemma as hypothesis.
\begin{lemma}[Semantic Substitution]\label{lem:st_subst1}
    If $(\G, \, x:  \ty[{\flt\cap(s[\QF/\QSelf]) \ccup{\QFresh \in s} \QFresh}]{T})^{\QF',x} \ts t_2 : \ty[{r[\QF'/\QSelf]}]{U}\ \EPS{\FX{[\QF'/\QSelf]}}$,
    and $q' = \flt \cap q$,
    and $\NFQ{S} \subq q'$,
    and $r\qminus x \subq \flt$,
    and $\EPS \qminus x \subq \flt$,
    and $\W = (\Sigma_1, \Sigma_2, f, \bval)$,
    and $\W^2 = (\Sigma_1'', \Sigma_2, f, \bval)$,
    and $(\VE, \W^2) \in \UG{\G[\flt]}$
    and $\STF{\W}{\DOM_1(\W)}{\DOM_2(\W)}$
    and $\STF{\W}{\DOM_1(\W^2)}{\DOM_2(\W^2)}$,
    and $t_1$ satisfies semantic weakening (\lemref{lem:st_weaken1}),
then $(\VE_1;(x:v_a), \VE_2), \W^2, t_2, t_2[t_1/x] \in  \BMt{\flt}{\ty[r\theta]{U} \ \EPS \FX{\theta}}$, where $\theta = [\qbot/x, q'/\QSelf]$. 

\end{lemma}
In the proofs of \lemref{lem:st_weaken1} and \lemref{lem:st_subst1},
we give new definitions of typing context interpretation for the not-aligned value environments, and define term substitution (omitted here).
Their proofs are conducted by induction on the type derivation of $t_1$ and $t_2$ respectively.

From the conclusion of \lemref{lem:st_subst1}, we get resulting world $\W'$. Let $f'$ and $\bval'$ be the relation and the related values in $\W'$ respectively.
Now we construct the world as 
$(\textcolor{blue}{\Sigma_1'''}, \textcolor{blue}{\Sigma_2'}, f_{\overline{\flt}}; \textcolor{blue}{f'_{\flt}};\textcolor{blue}{f'_{\fr}};\bval_{\overline{\flt}};\textcolor{blue}{\bval'_{\flt}};\textcolor{blue}{\bval'_{\fr}})$,
which is a valid extension of initial world $\W$. 
The remainder of the proof follows similarly to the proof of reordering and is thus omitted.

\begin{theorem}\label{thm:beta} Rule \textsc{$\beta$-equiv} in \figref{fig:equiv} is sound.
\end{theorem}

\bfparagraph{Discussion: Extensions} 
Rule \textsc{$\beta$-equiv} restricts the argument $t_1$ to an empty observation set, 
empty reachability qualifier, and empty write effect. While effects certainly have to 
be excluded (\eg, substitution may lead to duplication of $t_1$), scenarios with 
non-empty observation should also be safe, as long as $t_1$ is in the restricted form of variables or $\lambda$-terms.
While we have not yet mechanized such 
extended cases in Rocq, a proof would need to show that the equivalence of two terms is preserved over relational worlds \wrt the reachable locations from the given values, 
and the hypothesis of \lemref{lem:st_weaken1} needs to 
change $\W^{1'}$ to $(\Sigma_1', \Sigma_2, \flt_1)$, 
capturing the fact that reduction of $t_1$ requires locations specified in $\flt_1$. 
Given the experience with different reordering rules, we are confident
that our binary logical relation can mirror  the full range of possible 
application cases, including fresh/non-fresh arguments, self-references in the function type, etc. 

\section{Related Work}

\bfparagraphX{Reachability/Capturing Types} 
Prior works on reachability types ~\cite{DBLP:journals/pacmpl/BaoWBJHR21,wei2023polymorphic} proved syntactic type soundness (progress and preservation).
This paper presents semantic models of reachability systems and establishes stonger semantic properties via logical relations. 
The models in Section~\ref{sec:unary_base} are based on \maybelang{} in \citet{wei2023polymorphic}'s work, 
and have already been extended with support for subtyping and
polymorphism in \citet{jia2024escape}'s work. 
Recently, \citet{deng2025completecyclereachabilitytypes} refined \citet{wei2023polymorphic}'s system 
to support cyclic references, addressing the 
key limitation identified in this paper. We leave it to future work 
to model cyclic store structures and non-terminating programs
using step-indexed logic relations.
In parallel to this work, \citet{bunting2025a} investigate contextual equivalence for \citet{DBLP:journals/pacmpl/BaoWBJHR21}'s system  
using operational game semantics and present a full abstraction model based on a labelled transition system.

Closely related to reachability types and also tracking variables in types, 
capturing types~\cite{DBLP:journals/toplas/BoruchGruszeckiOLLB23,xuDegreesSeparationFlexible2023} 
are a recent effort to integrate capability tracking and 
escape checking into Scala 3. While the overall goals and design are
similar, there are also important differences. Specifically, the use
of self-references for escaping values and the modeling of freshness
is unique to reachability types.

\bfparagraph{Tracking Sharing/Uniqueness/Immutability}
\cite{DBLP:journals/scp/GianniniRSZ19,DBLP:journals/tcs/GianniniSZC19} develops calculi that infer introduced sharing during an expression's execution.
and can detect capsule references and borrowing.
Those works adopt a non-standard operational model that encodes store in the language term.
In contrast, reachability types tracks sharing as well, but uses the standard operational semantics. 
\cite{DBLP:journals/pacmpl/BianchiniDGZS22} tracks sharing and mutation in a coeffect system 
that can model uniqueness and immutability of references.
Reachability types track sharing in qualifiers, and mutation is tracked via the write effect extension (see \secref{sec:effects}).
Language features, \ie, uniqueness and borrowing, can be enforced by a flow-sensitive effect system, 
which are discussed in prior works~\cite{DBLP:journals/pacmpl/BaoWBJHR21,wei2023polymorphic}. 
A recent work ~\cite{deng2025completecyclereachabilitytypes} introduces cyclic reference types with dual-component that can be used to enforce read-only references.
The logical relations presented in this work does not rely on uniqueness and immutability of references.

\bfparagraph{Ownership Types} Ownership type systems~\cite{noble1998flexible,DBLP:conf/oopsla/ClarkePN98,DBLP:series/lncs/ClarkeOSW13,DBLP:conf/oopsla/PotaninNCB06}
typically enforce strict heap invariants and selectively re-introduce sharing in a controlled way
via borrowing~\cite{hogg1991islands,
naden2012type, clebsch2015ownership}. 
The Rust type system~\cite{DBLP:conf/sigada/MatsakisK14} adopts a strong ownership model.
Its ``shared XOR mutable'' mechanism  
enforces uniqueness of mutable references while allowing sharing among immutable ones. 
\citet{DBLP:journals/pacmpl/MarshallO24}'s work connects Granule's uniqueness modality~\cite{DBLP:journals/pacmpl/OrchardLE19} and its graded model types with ideas from Rust,
allowing ownership tracking in a graded type system.
In contrast, reachability types are designed to track sharing (thus separation), rather 
than restrict it. The focus is on higher-level languages, where many common idioms 
rely on sets of closures as an interface to interact with a piece of (shared!) 
mutable state.
Uniqueness properties can be enforced via flow-sensitive move effects~\cite{DBLP:journals/pacmpl/BaoWBJHR21,jia2024escape}.
In the context of Rust, \citet{DBLP:journals/pacmpl/0002JKD18} developed lifetime logic to model lifetimes and borrowing
in a variant of Rust's MIR, where programs are expressed in continuation-passing style.
Their logical framework can verify unsafe code with safe APIs for Rust programs. 
In contrast, our work aims to verify equational rules using observational equivalence. 
We showed an informal justification of a safe use of assertion statements by reducing the problem to proving observational equivalence (\secref{sec:binary_soundness}),
but unlike~\citet{DBLP:journals/pacmpl/0002JKD18}'s work, our goal is not to verify safe use of general unsafe code.

\bfparagraph{Capabilities and Permissions.}
Capability systems~\cite{DBLP:conf/ecoop/CastegrenW16,DBLP:conf/ecoop/BoylandNR01} have been used to reason about program resources
and external calls~\cite{drossopoulou2025reasoning}.
In Pony~\cite{DBLP:phd/ethos/Clebsch17}, reference capabilities describe what other aliases are denied~\cite{DBLP:conf/esop/DoddsFPV09,DBLP:conf/agere/ClebschDBM15}, \eg, deny local/global read/write aliases.
In contrast, reachability types~\cite{DBLP:journals/pacmpl/BaoWBJHR21,wei2023polymorphic} do not distinguish read and write capabilities.
Read-only capabilities can be enforced via cyclic reference types with dual-component~\cite{deng2025completecyclereachabilitytypes}, which is not modeled in this work.
\citet{haller2010capabilities} leverage capabilities to ensure externally unique access with borrowing.
\citet{DBLP:conf/oopsla/GordonPPBD12} extend their work with immutable references for safe parallelism.
Their symmetric parallelism rule enforces constraints on read and write that are similar to our rule \textsc{re-ordering-$\maybelange$}.
In contrast, our system does not rely on the uniqueness or immutability of references, 
but instead leverages aliasing/separation and write effects tracking.
Unique mutable references can be enforced via a flow-sensitive effect system, 
which has been discussed in prior works~\cite{DBLP:journals/pacmpl/BaoWBJHR21,wei2023polymorphic}.
We leave a thorough investigation as future work.

\bfparagraph{Step-Indexed Logical Relations}
Step-indexed logical relations have been used to prove  semantic type soundness~\cite{DBLP:journals/toplas/AppelM01,ahmed2004semantics},
program termination~\cite{DBLP:journals/pacmpl/SpiesKD21,logical-approach} 
and contextual equivalence~\cite{DBLP:conf/esop/Ahmed06}, representation independence~\cite{DBLP:conf/popl/AhmedDR09,logical-approach},
monadic encapsulation of state achieved by runST using rank-2 polymorphism~\cite{DBLP:journals/pacmpl/TimanySKB18}
for languages with general recursive and polymorphic types. 
In these approaches, types are indexed by computation steps to guarantee well-founded recursions have unique fixed points.
Since all well-typed programs are guaranteed to terminate in our systems, 
our LR models do not require step-indexes. 
As future work, we plan to extend our models to support recursion and 
cyclic references, 
and we are interested in using 
the Iris logic framework~\cite{DBLP:conf/popl/JungSSSTBD15,DBLP:conf/icfp/0002KBD16,DBLP:conf/esop/Krebbers0BJDB17,DBLP:journals/jfp/JungKJBBD18} 
to avoid manual step indexes and similar proof devices.

\bfparagraph{Effect-Based Program Transformation} Close to our approach that uses effects for validating program transformations,
\citet{DBLP:conf/ppdp/BentonKBH07, DBLP:conf/popl/Benton0N14} developed denotational, semantic relational models based on a
monotonic regions and effects for a higher-order language with dynamically allocated first-order mutable stores,
and with high-order stores without dynamic allocation~\cite{DBLP:conf/aplas/BentonKHB06,DBLP:conf/ppdp/BentonKBH09}.
Our world model is adapted from \citet{DBLP:conf/ppdp/BentonKBH07}'s work, and our use of observable write effects to prove observational equivalence coincides with \citet{DBLP:conf/popl/Benton0N14}'s work.
Unlike their works, we adopt a big-step operational semantics based on closures and environments, which is closer to real language implementations.
In addition, our effect system does not track allocation effects, thus, our models assume allocation always occurs during the reduction of a term. 
This design works well, as we can leverage reachability qualifiers. For example, the absence of the freshness marker indicates values do not reach fresh locations.
In this case, freshly allocated locations cannot be reached from clients.
\citet{DBLP:conf/ppdp/Benton0N16} developed models to validate effect-dependent program transformations for concurrent programs, 
which is beyond the scope of this work and left as future work.

\bfparagraph{World Models} 
\citet{DBLP:conf/icfp/ThamsborgB11,DBLP:journals/iandc/BirkedalJST16}'s work defines a Kripke logical relation
based on a region-polymorphic type-and-effect system for an ML-like programming language with
higher-order mutable stores with dynamic allocations.
In their work, a world is the part of the entire store that the program controls,
while our world describes the entire store, 
and we use the separation of saturated reachability qualifiers/observations and observable write effects to achieve different degrees of local reasoning.
Reachability types feature lightweight reachability polymorphism via qualifier dependent function application,
and use self-references to succinctly express if returned values are fresh, without introducing explicit quantification.
\citet{wei2023polymorphic}'s work has proven syntactic type soundness for reachability and type polymorphism (\polylang{}).

\vspace{-2ex}
\section{Conclusion}
In this paper, we presented semantic models for a family of reachability type systems with increasing sets of features,
and showed how reachability types naturally lead to strong 
local reasoning properties. 
The results established, \ie, semantic type soundness, termination, effect safety, and program equivalence,
are stronger than those in prior works on reachability types. 
We also proved key equational rules such as reordering and $\beta$-equivalence, 
providing a firm foundation for %
reachability-driven program transformations.
All results are proven in Rocq.

\begin{acks}                            %
  We thank Siyuan He and Haotian Deng for related contributions
  to reachability types.
  This work was supported in part by NSF awards 2348334 and Augusta faculty startup
  package, as well as gifts from Meta, Google, Microsoft, and VMware.
\end{acks}

\bibliography{references}

\end{document}